\documentclass{aa}
\newcommand\arcdeg{\mbox{$^\circ$}\xspace} 
\usepackage{latexsym}		
\usepackage{graphicx}		
\usepackage{rotating}		
\usepackage{natbib}  
\usepackage{savesym}
\usepackage{amssymb}
\savesymbol{doublespace}
\usepackage{xspace}
\usepackage{color}
\usepackage{multicol}
\usepackage{mdframed}
\usepackage{url}
\usepackage{subfigure}
\usepackage{lscape}
\usepackage{grffile}
\usepackage{standalone}
\standalonetrue
\usepackage{import}
\usepackage[utf8]{inputenc}
\usepackage{longtable}
\usepackage{booktabs}
\usepackage[yyyymmdd,hhmmss]{datetime}
\usepackage{fancyhdr}

\newcommand{\msun}{\ensuremath{M_{\odot}}\xspace}			

\newcommand{\hh}{\ensuremath{\textrm{H}_{2}}\xspace}			

\newcommand{\formaldehyde}{\ensuremath{\textrm{H}_2\textrm{CO}}\xspace}

\newcommand{\ortho}{\ensuremath{\textrm{o-H}_2\textrm{CO}}\xspace}
\newcommand{\para}{\ensuremath{\textrm{p-H}_2\textrm{CO}}\xspace}
\newcommand{\oneone}{\ensuremath{1_{10}-1_{11}}\xspace}
\newcommand{\twotwo}{\ensuremath{2_{11}-2_{12}}\xspace}

\newcommand{\uchii}{UC\ion{H}{ii}\xspace}

\newcommand{\hii}{H~{\sc ii}\xspace}
\newcommand{\hi}{H~{\sc i}\xspace}

\newcommand{\kms}{\textrm{km~s}\ensuremath{^{-1}}\xspace}	
\newcommand{\Kkms}{\textrm{K~km~s}\ensuremath{^{-1}}\xspace}	

\newcommand{\percc}{\ensuremath{\textrm{cm}^{-3}}\xspace}
\newcommand{\perpc}{\ensuremath{\textrm{pc}^{-1}}\xspace}

\newcommand{\persc}{\ensuremath{\textrm{cm}^{-2}}\xspace}

\newcommand{\um}{\ensuremath{\mu \textrm{m}}\xspace}    

\newcommand{\twelveco}{\ensuremath{^{12}\textrm{CO}}\xspace}
\newcommand{\thirteenco}{\ensuremath{^{13}\textrm{CO}}\xspace}

\def\ee#1{\ensuremath{\times10^{#1}}}

\def\eqref#1{Equation \ref{#1}}

\renewcommand\arcdeg{\mbox{$^\circ$}\xspace} 
\renewcommand\arcmin{\mbox{$^\prime$}\xspace} 
\renewcommand\arcsec{\mbox{$^{\prime\prime}$}\xspace}

\def\Figure#1#2#3#4#5{
\begin{figure*}[htp]
\includegraphics[scale=#4,angle=#5]{#1}
\caption{#2}
\label{#3}
\end{figure*}
}

\def\FigureTwoAA#1#2#3#4#5#6{
\begin{figure*}[htp]
\subfigure[]{ \includegraphics[scale=#5,width=#6]{#1} }
\\
\subfigure[]{ \includegraphics[scale=#5,width=#6]{#2} }
\caption{#3}
\label{#4}
\end{figure*}
}

\def\FigureFourPDF#1#2#3#4#5#6{
\begin{figure*}[htp]

\subfigure[][]{\includegraphics[width=3.5in,type=pdf,ext=.pdf,read=.pdf]{#1}}  
\subfigure[][]{\includegraphics[width=3.5in,type=pdf,ext=.pdf,read=.pdf]{#2}}\\
\subfigure[][]{\includegraphics[width=3.5in,type=pdf,ext=.pdf,read=.pdf]{#3}}  
\subfigure[][]{\includegraphics[width=3.5in,type=pdf,ext=.pdf,read=.pdf]{#4}}\\

\caption{#5}
\label{#6}
\end{figure*}
}

\begin{document}

\title{The dense gas mass fraction in the W51 cloud and its protoclusters}
\titlerunning{W51 DGMF}
\authorrunning{Ginsburg et al}
\newcommand{\eso}{$^{1}$}
\newcommand{\casa}{$^{2}$}
\newcommand{\cfa}{$^{3}$}
\newcommand{\edmonton}{$^{4}$}
\newcommand{\yale}{$^{5}$}
\newcommand{\puertorico}{$^{6}$}

\author{Adam Ginsburg{\eso},
        John Bally{\casa},
        Cara Battersby{\cfa},
        Allison Youngblood{\casa},
        Jeremy Darling{\casa},
        Erik Rosolowsky{\edmonton},
        Héctor Arce{\yale},
        Mayra E. Lebrón Santos{\puertorico}
        }

\institute{
    {\eso}{\it{European Southern Observatory, Karl-Schwarzschild-Strasse 2, D-85748 Garching bei München, Germany\\
                      \email{Adam.G.Ginsburg@gmail.com}}} \\ 
    {\casa}{\it{CASA, University of Colorado, 389-UCB, Boulder, CO 80309}} \\ 
    {\cfa}{\it{Harvard-Smithsonian Center for Astrophysics, 60 Garden
    Street, Cambridge, MA 02138, USA}} \\ 
    {\edmonton}{\it{University of Alberta, Department of Physics, 4-181 CCIS, Edmonton AB T6G 2E1 Canada}} \\ 
    {\yale}{\it{Department of Astronomy, Yale University, P.O. Box 208101, New Haven, CT 06520-8101 USA}} \\ 
    {\puertorico}{\it{Department of Physical Sciences, University of Puerto Rico, P.O. Box 23323, San Juan, PR 00931}}
    }

\date{Date: \today ~~ Time: \currenttime}


\abstract
{The density structure of molecular clouds determines how they will evolve.}
{To map the velocity-resolved density structure of the most vigorously
star-forming molecular cloud in the Galactic disk, the W51 Giant Molecular
Cloud.}
{
We present new 2 cm and 6 cm maps of \formaldehyde, radio recombination lines,
and the radio continuum in the W51 star forming complex acquired with Arecibo
and the Green Bank Telescope at $\sim50\arcsec$ resolution.
We use \formaldehyde absorption to determine the relative line-of-sight
positions of molecular and ionized gas.  We measure gas densities using the
\formaldehyde densitometer, including continuous measurements of the dense gas
mass
fraction (DGMF) over the range $10^4$ $\percc< n(\hh) < 10^6$ \percc - this is
the first time a dense gas mass fraction has been measured over a range
of densities with a single data set.}
{The DGMF in W51 A is high, $f\gtrsim70\%$ above $n>10^4$
\percc, while it is low, $f<20\%$, in W51 B.
We did not detect \emph{any} \formaldehyde emission throughout the W51 GMC; all
gas dense enough to emit under normal conditions is in front of bright
continuum sources and therefore is seen in absorption instead.  
The data set has been made public at \protect\url{http://dx.doi.org/10.7910/DVN/26818}.
}
{(1) The dense gas fraction in the W51 A and B clouds shows that W51 A will
continue to form stars vigorously, while star formation has mostly ended in W51
B.  The lack of dense, star-forming gas around W51 C indicates that
collect-and-collapse is not acting or is inefficient in W51.
(2) Ongoing high-mass star formation is correlated with $n\gtrsim1\ee{5}$
\percc gas.  Gas with $n>10^4$ \percc is weakly correlated with low and
moderate mass star formation, but does not strongly correlate with high-mass
star formation.
(3) The nondetection of \formaldehyde emission implies that the  emission
detected in other galaxies, e.g. Arp 220, comes from high-density gas that is
not directly affiliated with already-formed massive stars.  Either the
non-star-forming ISM of these galaxies is very dense, implying
the star formation density threshold is higher, or \hii regions have
their emission suppressed.
}

\maketitle

\section{Introduction}

Massive star clusters, those containing $>10^4$ \msun of stars, are among the
most visually outstanding features in the night sky \citep[see review
by][]{Longmore2014a}.  In other galaxies, they
are useful probes of the star formation history and can be individually
identified and measured \citep{Bastian2008a}.  Locally, they are the essential
laboratories in which we can study the formation of massive stars
\citep{Davies2012a}.

In order to utilize these clusters as laboratories, we need to understand their
formation in detail.  Clusters are often assumed and measured to be coeval
to within a narrow range \citep[e.g. $10^5$ yr;][]{Kudryavtseva2012a},
but uncertainties remain \citep{Beccari2010a}.
In the most massive clusters, there are predictions that multiple generations
or an extended generation of stars should form prior to gas expulsion because
the gas will remain gravitationally bound \citep{Bressert2012a}.   Feedback from
and within young massive clusters is an active field of numerical study
\citep{Rogers2013a,Dale2013a,Dale2012a,Dale2008a,Dale2005a,Parker2013a,Myers2014a,Krumholz2014a}.
While
only 5-35\% of all stars form in bound clusters\footnote{Clusters bound by the
stellar mass alone, not including the birth cloud's mass.}
\citep{Kruijssen2012a}, these clusters form the basis of our understanding of
stars and stellar evolution
\citep{Kalirai2010a}, and understanding their formation is therefore crucial.

The results of cluster formation may be decided before the first stars are
formed.  The starless initial conditions of massive clusters have not yet been
definitively observed \citep{Ginsburg2012a} though there are viable candidates
such as G0.253+0.016 \citep{Longmore2012b}.  The initial conditions for star
formation on any scale are clearly turbulent.  However, there is no evidence
whether these initial conditions differ in any qualitative way from turbulence
in local, low-mass star-forming regions.

Star formation appears to occur efficiently only above a volume
density threshold in molecular gas, specifically $n(\hh)\gtrsim10^4$ \percc,
in the Galactic disk
\citep[][who advocate a
column density threshold corresponding to this density]{Lada2010a}; note
however that the `threshold' is smooth, not  a step function as implied by the
term \citep{Padoan2013a}.
The `dense
gas mass fraction' has been measured for various definitions of `dense'
\citep{Battisti2014a,Wu2005a}, but these definitions don't always correspond to
the associated threshold
\citep{Kauffmann2010a,Parmentier2011a,Parmentier2011b}.  
Often, the `dense gas' threshold is observationally defined as the threshold
to see a given molecule, which is sometimes incorrectly assumed to correspond
to a fixed `critical density' for the molecule.
The existence of a universal 
density threshold for star formation is contentious
\citep{Burkert2013a,Clark2013b}; the star formation threshold likely
varies with the turbulent properties of clouds
\citep{Longmore2013b,Kruijssen2014c,Hennebelle2011a,Hennebelle2013a,Padoan2011b,Krumholz2005a,Federrath2012a}.
Star formation thresholds can only be evaluated when measurements of density
and column density are simultaneously available.  

\subsection{W51}
The W51 cloud complex (Figure \ref{fig:w51large}), containing the two 
massive protocluster candidates W51 Main and W51 IRS 2 from
\citet{Ginsburg2012a}, is located at
$\ell\sim49, b\sim-0.3$, very near the Galactic midplane\footnote{The midplane
at $d=5.1$ kpc is offset approximately -0.22 to -0.33 degrees from $b=0$
depending on our solar system's height above the midplane, $Z_\odot = 20$ pc or
30 pc, respectively \citep{Reed2006a,Joshi2007a}.} at a distance of 5.1 kpc
\citep{Sato2010a}.  It is a well-known and thoroughly studied collection of
clouds massing $M>10^6 \msun$
\citep{Carpenter1998a,Bieging2010a,Kang2010a,Parsons2012a}.  The radio-bright
regions are generally known as W51 A to the east, W51 B to the west, and W51 C
for the southern component, known to trace a supernova remnant
\citep{Koo1995a,Brogan2000a,Brogan2013a}.

\Figure{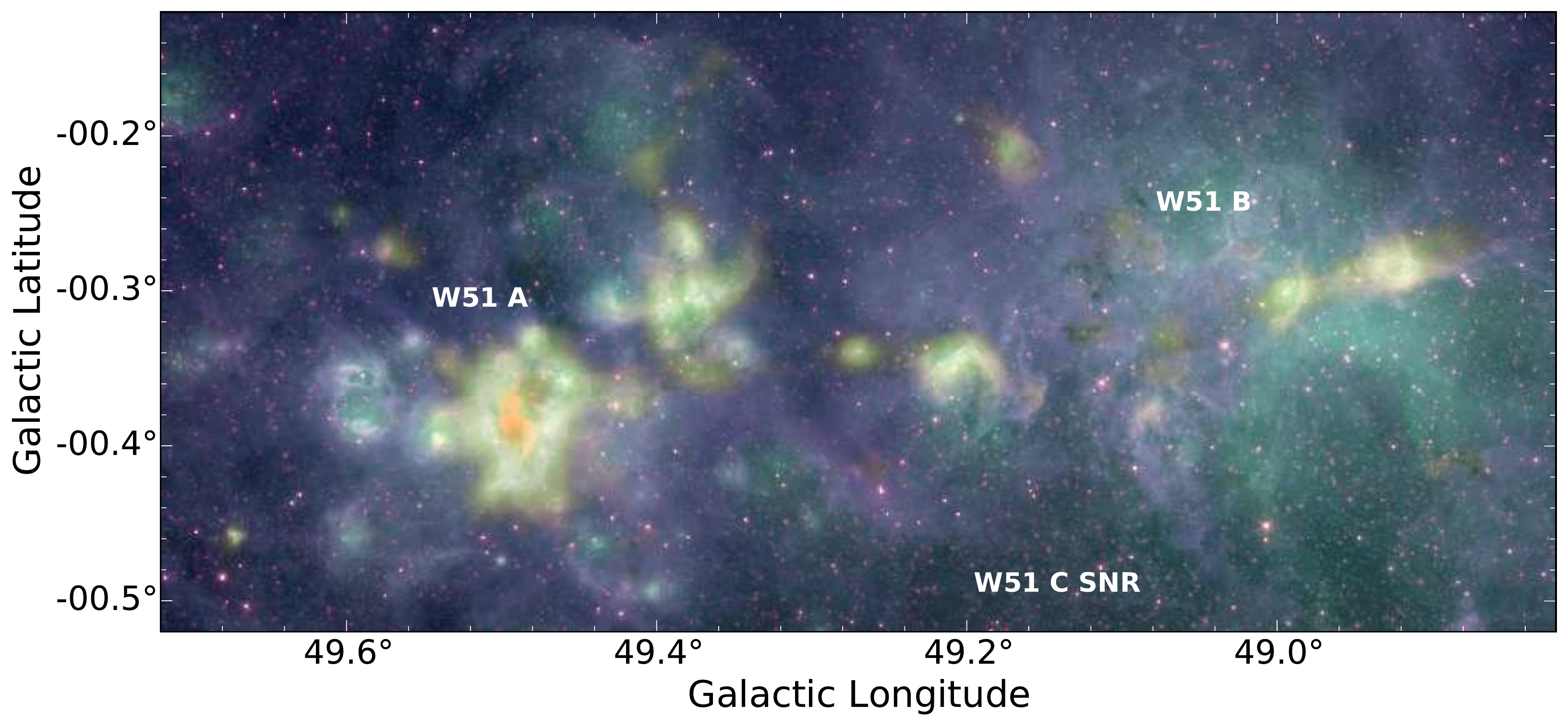}
{A color composite of the W51 region with major regions, W51 A, B, and C,
labeled.  W51 A contains the protoclusters W51 Main and W51 IRS 2; these
are blended in the light orange region around 49.5-0.38.  The
blue, green, and red colors are WISE bands 1, 3, and 4 (3.4, 12,
and 22 \um) respectively.  The yellow-orange semitransparent layer is from the
Bolocam 1.1 mm Galactic Plane Survey data \citep{Aguirre2011a,Ginsburg2013a}.
Finally, the faint whitish haze filling in most of the image is from a 90 cm
VLA image by \citet{Brogan2013a}, which primarily traces the W51 C supernova
remnant.  This haze is more easily seen in Figure \ref{fig:w51huge}.}
{fig:w51large}{0.5}{0}

\subsection{Formaldehyde}
Formaldehyde (\formaldehyde) has been recognized as a useful probe of physical
conditions in the molecular interstellar medium for decades
\citep{Mangum1993a}.  The centimeter lines, \formaldehyde \oneone (6.2 cm,
4.82966 GHz) and \twotwo (2.1 cm, 14.48848 GHz), have a peculiar excitation
process in which collisions overpopulate the lower of the two $K_c$ rotational
states, where $K_c$ is the quantum number representing the angular momentum
projected onto the long axis of the molecule.  
The overpopulated lower-energy state leads to amplified absorption, or
effective excitation temperatures less than the background temperature,
allowing \formaldehyde centimeter absorption to be seen against the cosmic
microwave background.
Because the \oneone and \twotwo
level pairs populate differently
depending on the volume density of the colliding partner (a mix of p-\hh, He,
and o-\hh), their ratio is sensitive to the local gas volume
density\footnote{Higher frequency
\formaldehyde lines show the same behavior, though the effect is progressively
weaker; see \citet{Darling2012b} Figure 2.}.

\formaldehyde \oneone has been observed in the W51 Main region with the VLA
\citep{Martin-Pintado1985a} and Westerbork \citep{Arnal1985a}, and this data
was used to gain some early constraints on the geometry of the region
\citep[e.g.][]{Carpenter1998a}.  \citet{Henkel1980a} presented observations of
the \oneone and \twotwo lines, and \citet{Martin-Pintado1985b} presented
single-dish mapping observations of the \formaldehyde \twotwo line toward the
W51 Main region, but both treated the region as a single-density structure.

\subsection{Paper Overview} 
We present a detailed examination of the dense gas in the W51 cloud complex.
Section \ref{sec:observations} presents the observations and data reduction.
Section \ref{sec:analysis} describes the analysis techniques, including
measurement of the dense gas mass fraction and derivation of the cloud
geometry.  Section \ref{sec:discussion} discusses the implications of the
density measurements for Galactic clouds and extragalactic interstellar media.
Section \ref{sec:conclusion} concludes.  There are two appendices: Appendix
\ref{sec:contrrldata} describes the radio recombination line and continuum
data.  Appendix \ref{appendix:geometry} describes details of individual
regions, with a focus on
the cloud and \hii region line-of-sight geometry.

\section{Observations \& Data Reduction}
\label{sec:observations}
The W51 survey was performed in September 2011 on the Green Bank Telescope
(GBT) and in 2012 using the Arecibo Observatory.  
The data was reduced using custom-made scripts based off of both GBTIDL's
mapping routines by Glen Langston
(\url{https://safe.nrao.edu/wiki/bin/view/Kbandfpa/KfpaPipelineHowTo}) and Phil
Perillat's AOIDL routines.  The code is available at
\url{https://github.com/keflavich/sdpy}.  The data reduction code and workflow
are included in a corresponding git repository:
\url{https://github.com/keflavich/w51_singledish_h2co_maps}.

\subsection{Arecibo 6 cm}
The Arecibo data were taken as part of project A2705 over the course of 4
nights, September 10, 11, 12, and 15 2012.  The Mock spectrometer
was used to cover the range 4.6 to 5.4 GHz with a spectral resolution $\sim$ 1
\kms, including the \ortho 4.82966 GHz line and the H107-H112$\alpha$
recombination lines. On the first night, September 10
2012, a significant fraction of the data was lost due to an internal instrument
error within the Mock spectrometer, which resulted in a loss of the
high spectral resolution component of the \formaldehyde data for that night.  As a
result, we have focused our study on the lower-resolution ($\sim1$ \kms) data.

The fields were observed with east-west maps using the C-band receiver.  No
crosshatching was performed with Arecibo.

The Arecibo data reduction process for W51
presented unique challenges: at C-band, the entire region surveyed contains
continuum emission, so no truly suitable `off' position was found within the
survey data.  Similarly, \formaldehyde is ubiquitous across the region, so it
was necessary to `mask out' the absorption lines when building an off position.
This was done by interpolating across the line-containing region of the
spectrum with a polynomial fit.  The fits were inspected interactively and
tuned to avoid over-predicting the background.

The Arecibo data were corrected to main beam brightness temperature $T_{MB}$
using a main-beam efficiency as a function of zenith angle in degrees ($za$):
$$\eta_{MB}(za) = 0.491544 + 0.00580397 za - 0.000341992 za^2$$
This is a fit to 5 years worth of calibration data acquired at Arecibo and
assembled by Phil Perrilat\footnote{The data can be retrieved from within aoIDL
with the command \texttt{n=mmgetarchive(yymmdd1,yymmdd2,mm,rcvnum=9)}, then
the $\eta_{MB}$ data are in \texttt{mm.fit.etamb}.  The calibration process is
recorded in detail in \citet{Heiles2001a}.}.

The maps were made by computing an output grid in Galactic coordinates with
15\arcsec pixels and adding each spectrum to the appropriate pixel\footnote{We
use the term `pixel' to refer to a square data element projected on the sky
with axes in Galactic coordinates.  The term `voxel' is used to indicate a cubic data
element, with two axes in galactic coordinates and a third in frequency or
velocity}.  In order
to avoid empty pixels and maximize the signal-to-noise, the spectra were added
to the grid with a weight set from a Gaussian with $FWHM=20\arcsec$, which
effectively smooths the output images from $FWHM\approx50\arcsec$ to
$\approx54\arcsec$.  See \citet{Mangum2007a} for more detail on the on-the-fly
mapping technique used here.

The Arecibo data were taken at a spectral resolution of 0.68 \kms.  The
spectra were regridded onto a velocity grid from $-50$ to 150 \kms with 1 \kms
resolution.  To achieve this, they were first Gaussian-smoothed to $FWHM=1$
\kms then downsampled appropriately.

The position-position-velocity (PPV) cubes were created with units of
brightness temperature.  The Arecibo cubes have contributions from 15-20
independent spectra in each pixel, though
this hit rate varies in a systematic striped pattern parallel to the Galactic
plane.  The small overlap regions between different maps have a significantly
higher number of samples; these regions constitute a small portion of the map.
The resulting noise level is RMS $\sim 50-60$ mK except toward the \hii
regions, where it peaks at about 400 mK.  The continuum is derived by averaging
line-free channels; its signal-to-noise peaks at $\sim900$.

The Arecibo data have smaller systematic continuum offsets than the GBT data
(Figure \ref{fig:continuum}), but they are more visually pronounced because
there is much more diffuse emission at 6 cm.  The continuum zero-point level in
the Arecibo data was set to be the 10th percentile value of each scan, which is
effectively the minimum value across each scan but with added robustness
against noise-generated false minima.  In the eastmost and westmost blocks,
this strategy was very effective, as there are clearly areas in each scan that
see no continuum.  However, in the central block, this approach resulted in a
vertical negative filament that almost certainly represents a local minimum
that should be positive.  This negative filament has values $\gtrsim-0.08$ K.
Given the excellent agreement between the three independently observed regions
in the areas that they overlap, it is clear that the continuum is reliable
above $\gtrsim0.5$ K, which is the entire regime in which it is a significant
contributor to the total background emission (at lower levels, the CMB is
dominant).

\subsection{GBT 2 cm}
The GBT data were taken as part of program AGBT10B/019.  We used the GBT
Spectrometer with 4 windows covering the \ortho 14.488479 GHz line,
H77$\alpha$ (14.12861 GHz), and two others that were not reduced targeting H$_2$CN (14.82579
GHz) and NaCl and SO (13.03606 GHz); online examination of the latter windows
suggested that we did not detect any emission or absorption.
The data presented in this paper
include sessions 10, 11, 14, 16, 17, 20, 21, and 22; the other sessions from
this project include maps of outer galaxy regions and a single-pointing survey
of Galactic plane sources that will be presented in another paper. 

Data were taken in on-the-fly mode with the GBT Ku-band dual-beam system.
Cross-hatched north-south and east-west maps were created in Galactic
coordinates. 
Each spectrum was calibrated using the first and last scans of each observation
as off positions.  The background level to be subtracted off of the
continuum was determined by linearly interpolating between these scans.

The Green Bank data have a main beam efficiency $\eta_{MB} = 0.886$, or a gain
of 1.98 K/Jy assuming a 51\arcsec beam \citep[see][for additional
discussion]{Mangum2013a}.  The GBT data were also corrected for atmospheric
opacity using Ron Maddalena's
\texttt{getForecastValues}\footnote{\url{http://www.gb.nrao.edu/~rmaddale/Weather/}}
with a typical zenith optical depth $\tau_{z}\approx0.02$; this
correction was never more than $\sim5\%$.
The GBT data were taken with a spectral resolution $\Delta v = 0.25$ \kms

Typical noise levels were $\sim10-20$ mK per 1 \kms channel; the levels vary
across the map.  

The GBT data were mapped in an orthogonal grid pattern, so the hit coverage is
more uniform on small scales than the Arecibo data, but because of the
dual-beam Ku-band system, the overall noise levels are much more patchy.
Additionally, the nights with
better weather yielded a lower noise level.  The noise ranges from $\sim7$ mK
in the W51 Main region to $\sim 20$ mK in the westmost region.  As with the
Arecibo data, the \hii region adds noise, but the peak noise towards an \hii
region is only $\sim 20$ mK.  This difference is because the diffuse \hii
region is fainter at 2 cm.  The signal-to-noise ratio in the continuum peaks at
$\sim 400$.

While the noise in the continuum is nominally quite low, there are significant
systematic effects visible in the continuum maps.  The continuum zero-point of
each GBT map was determined by assuming that the first and last scan both
observed zero continuum and that the sky background can be linearly
interpolated between the start and end of the observations.  In general, these
are good assumptions, but they leave in systematic offsets of up to
$\lesssim-0.15$ K in the maps, most likely because there is a $\sim0.15$ K
variation in the diffuse background  emission.

\FigureTwoAA{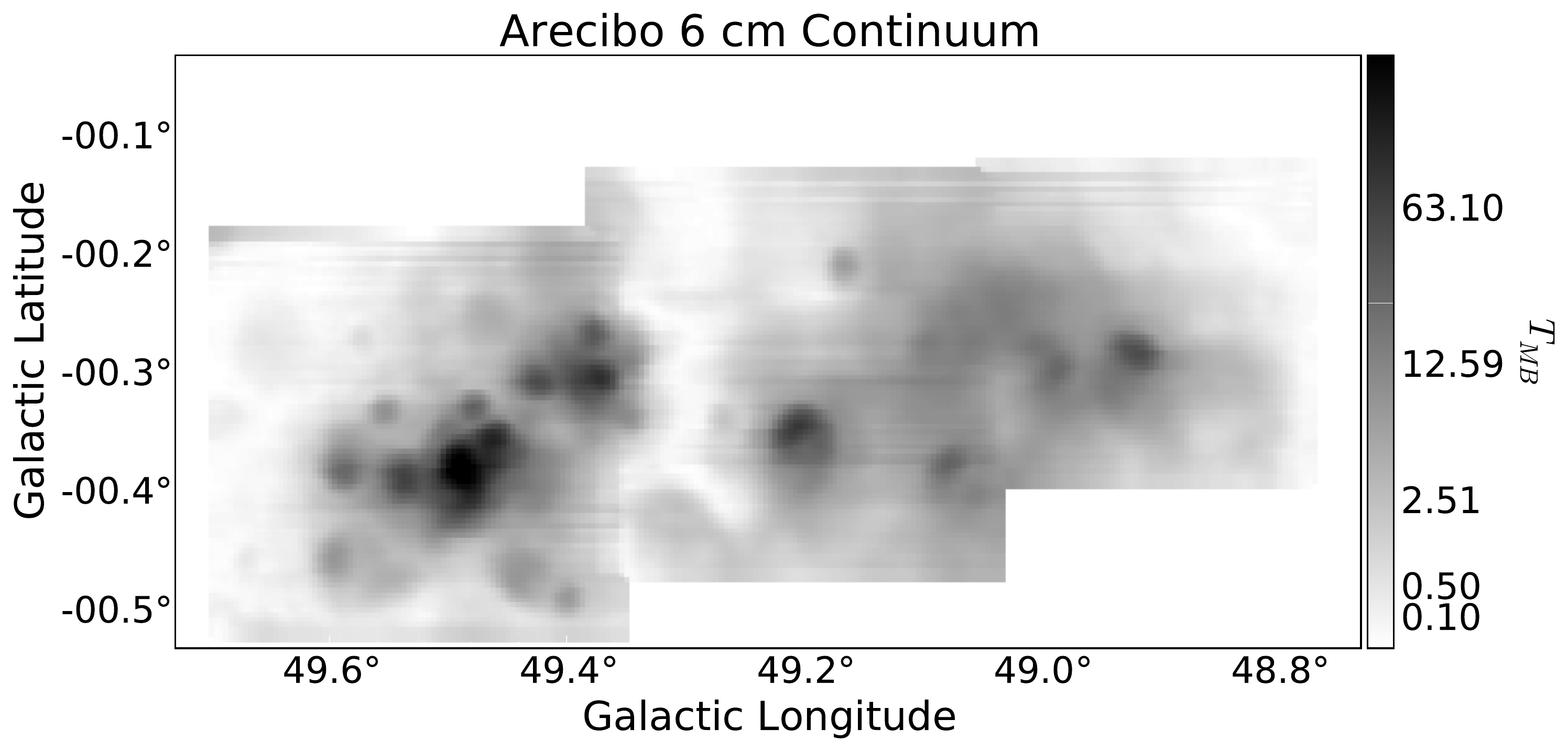}
            {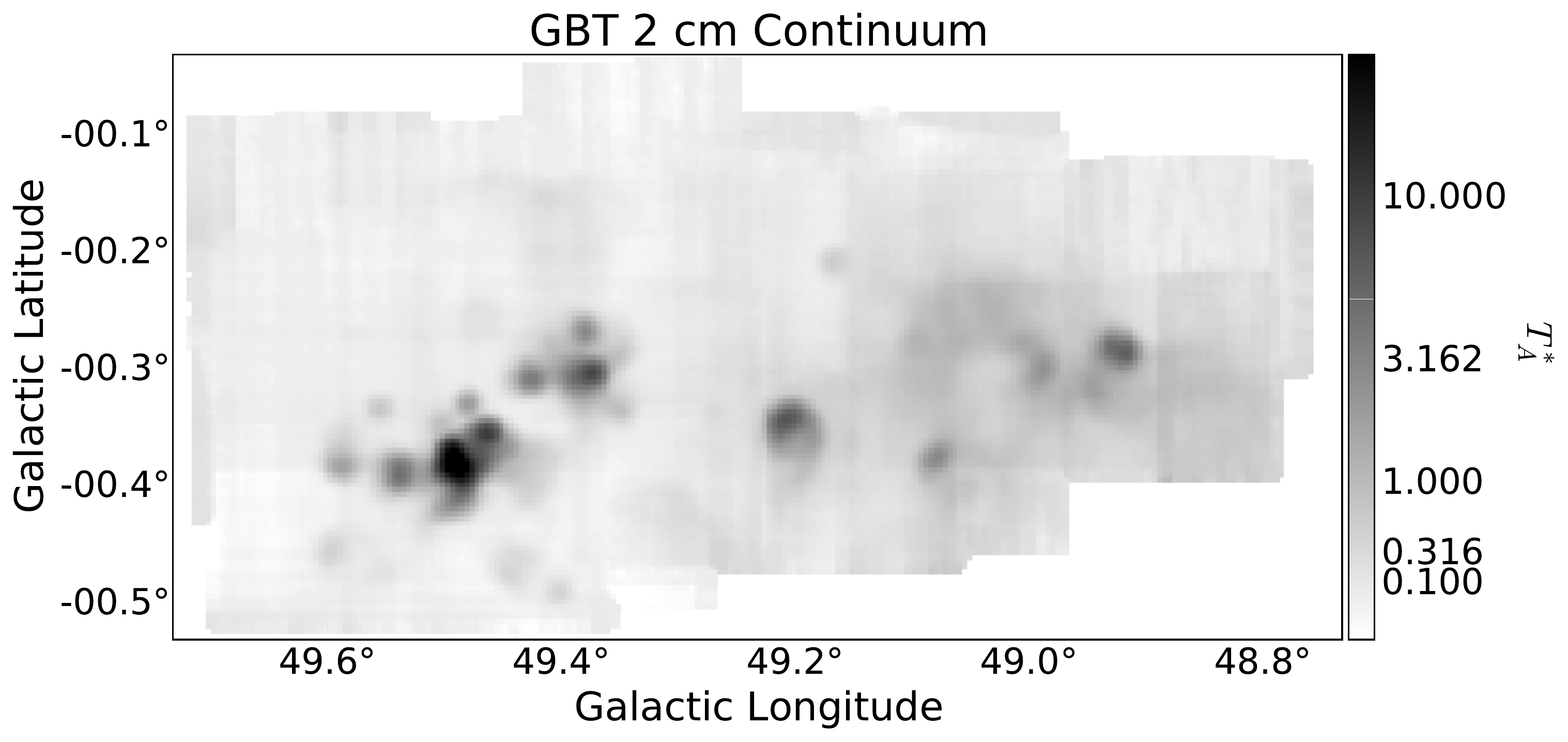}
{Continuum images of (a) the 6 cm Arecibo data and (b) 2 cm GBT
data.} {fig:continuum}{1.0}{6in}

\subsection{Optical Depth Data Cubes}

The data cubes were converted into ``optical depth'' data cubes by dividing the
integrated \formaldehyde absorption signature by the measured continuum level.
We added a fixed background of 2.73 K to the reduced images to account for the
CMB, which is absent from the images due to the background-subtraction
process.  We define an ``observer's optical depth''
\begin{equation}
    \tau_{obs} = -\ln\left[\frac{T_{MB}}{T_{bg}}\right]
\end{equation}
as opposed to the `true' optical depth, which is modeled in radiative transfer
calculations
\begin{equation}
    \tau = -\ln\left[\frac{T_{MB}-T_{ex}}{T_{bg}-T_{ex}}\right]
\end{equation}
The approximation $\tau_{obs} = \tau$ is valid for $T_{ex} << T_{bg}$, which is
true when an \hii region is the backlight but generally not when the CMB is.
Displaying the data on this scale makes regions of similar gas surface density
appear the same, rather than being enhanced where there are backlights.  The
noise is correspondingly suppressed where backlighting sources are present.

\formaldehyde absorption is ubiquitous across the map.  In the Arecibo data,
8044 of 17800 spatial pixels have peak optical depths $>5\sigma$, and 14547
have peaks $>3\sigma$, so \formaldehyde absorption is detected at $\sim80\%$ of
the observed positions.

The GBT \formaldehyde \twotwo data have lower peak signal-to-noise because the
continuum background is lower.  Additionally, the \twotwo line is expected to
trace denser gas and therefore be detected along fewer lines of sight.  The
\twotwo line is detected with a peak at $>5\sigma$ in 3497 pixels (20\%) and
$>3\sigma$ in 12254 pixels (69\%).  The high detection rate validates
\formaldehyde as an efficient dense-gas tracer.

There were no detections of \formaldehyde \oneone or \twotwo emission.  The
significance of the nondetection of emission is discussed in Section
\ref{sec:exgal}.

\subsection{A note on nondetections}
H$_2$ $^{13}$CO was not detected anywhere in the W51 complex in either the
\oneone or \twotwo lines.  The peak signal-to-noise in the \formaldehyde \oneone cube was
180 at 1 \kms resolution (corresponding to an optical depth $\tau_{obs}\sim1/5$ at
the peak continuum detection point), so we report a $3\sigma$ upper limit on the
\formaldehyde/H$_2$ $^{13}$CO ratio $R>60$, which is consistent with solar
values of the $^{12}C$/$^{13}C$
ratio.

\section{Analysis}
\label{sec:analysis}

\subsection{\formaldehyde modeling}
\label{sec:models}
The \formaldehyde line ratio can be transformed into a volume
density of hydrogen $n(\hh)$ using large velocity gradient model grids.
The column density of \ortho, the ortho-to-para-ratio of \hh, and the gas
temperature are the three main `nuisance parameters' that can be marginalized
over.

The \ortho column density is degenerate with the velocity gradient in LVG
models.  The \hh column density is degenerate with this gradient \emph{and} the
abundance of \ortho.  Precise measurements of the \formaldehyde abundance are
not generally available, but typical values of $X_{\ortho} = 10^{-10}-10^{-8}$
relative to \hh are generally assumed \citep{Mangum1993a, Ginsburg2011a,
Ginsburg2013a, Ao2013a} and found to be consistent with the observations.
Nonetheless, little is known about local variations in \ortho abundance, except
that it freezes out in cold, dense cores \citep{Young2004a}.
The abundance was left as a free parameter in the model fitting.

The model grids were generated using RADEX LVG models
\citep[python wrapper \url{https://github.com/keflavich/pyradex/}; original
code][]{van-Der-Tak2007a} and the grids were fit using
\url{https://github.com/keflavich/h2co_modeling}.  The RADEX models
assume a velocity gradient of 1 \kms \perpc.  \citet{Ginsburg2011a} and
\citet{Ginsburg2013a}
discussed the effect of a local gas density distribution on the molecular
excitation, but due to the complexity involved in accounting for these effects,
we ignore them here.  The derived physical parameters are moderately
sensitive to the input collision rates, with a 50\% error in collision rates
yielding a factor of 2 error in derived density \citep{Zeiger2010a}, but the
error on the collision rates in the low temperature regime we are modeling have
recently been improved from the previously used \citet{Green1991a} rates and
should not be a dominant factor in our calculations
\citep{Troscompt2009a,Wiesenfeld2013a}.

\subsection{\formaldehyde observables}
\label{sec:h2co}

Figure \ref{fig:peakoptdepth} shows the most important observed properties of
the \formaldehyde lines.  The figures show the peak observed optical depth
$\tau_{obs} = -\ln(T_{MB}/T_{continuum})$ in each line along with the ratio of
the \oneone to the \twotwo optical depth.  The noise is computed by measuring
the standard deviation over a signal free region (-50 to 0 \kms) along each
spatial pixel.  The cubes were masked to show significant pixels determined by:
\begin{enumerate}
    \item Selecting all voxels with $S/N > 2$ in both images or $S/N > 4$ in
        either and with at least 7 (of 26 possible) neighbors also having $S/N > 2$ 
    \item Selecting all voxels with $\ge10$ neighbors having $S/N > 2$
    \item Growing (dilating) the included mask region by 1 pixel in all
        directions
    \item Selecting all voxels with $\ge5$ neighbors marked as `included' by the
        above steps (this is a `closing' operation)
    \item (2D only) When used to mask 2D images, the selection mask is then
        collapsed along the spectral axis such that any pixel containing at
        least one voxel along the spectral axis is included
\end{enumerate}
This approach effectively includes all significant pixels and all reliably
detected regions within the data cube, though the number of neighbors used at
each step and the selected growth size are somewhat arbitrary and could be
modified with little effect.

Figures \ref{fig:peakoptdepth}-\ref{fig:upperpeakoptdepth} each contain
peak optical depth maps and two ratio maps.  The first ratio map shows the observed
optical depth ratio, while the second shows the `true' optical depth
ratio assuming an excitation temperature for each line, $T_{ex}(\oneone) = 1.0$
K and $T_{ex}(\twotwo) = 1.5$ K.  These excitation temperatures are
representative of those expected for most of the modeled parameter space in
which absorption is expected.  Fitting of individual lines-of-sight confirm
that good fits can be achieved using these assumed temperatures.

However, there are some clear outliers within the map: the clouds at G48.9-0.3
and G49.4-0.2 both show very low \oneone/\twotwo ratios over a broad area.  As
discussed in Sections \ref{sec:w51b} and \ref{sec:maus}, these two regions have
\hii regions in the foreground of the molecular gas.  The ratios displayed in
Figure \ref{fig:peakoptdepth} are therefore computed with an incorrect
background assumed; we correct for the different background in the next section.

\FigureFourPDF
{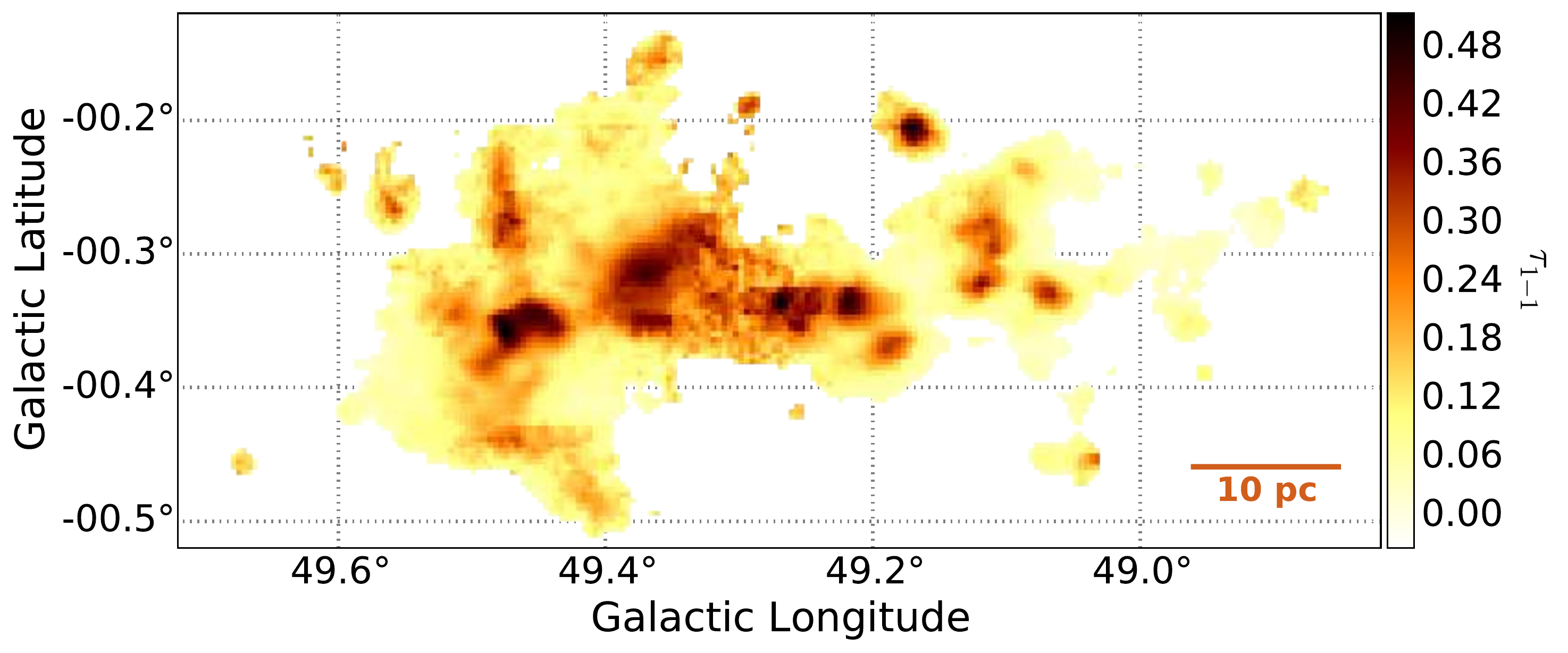}
{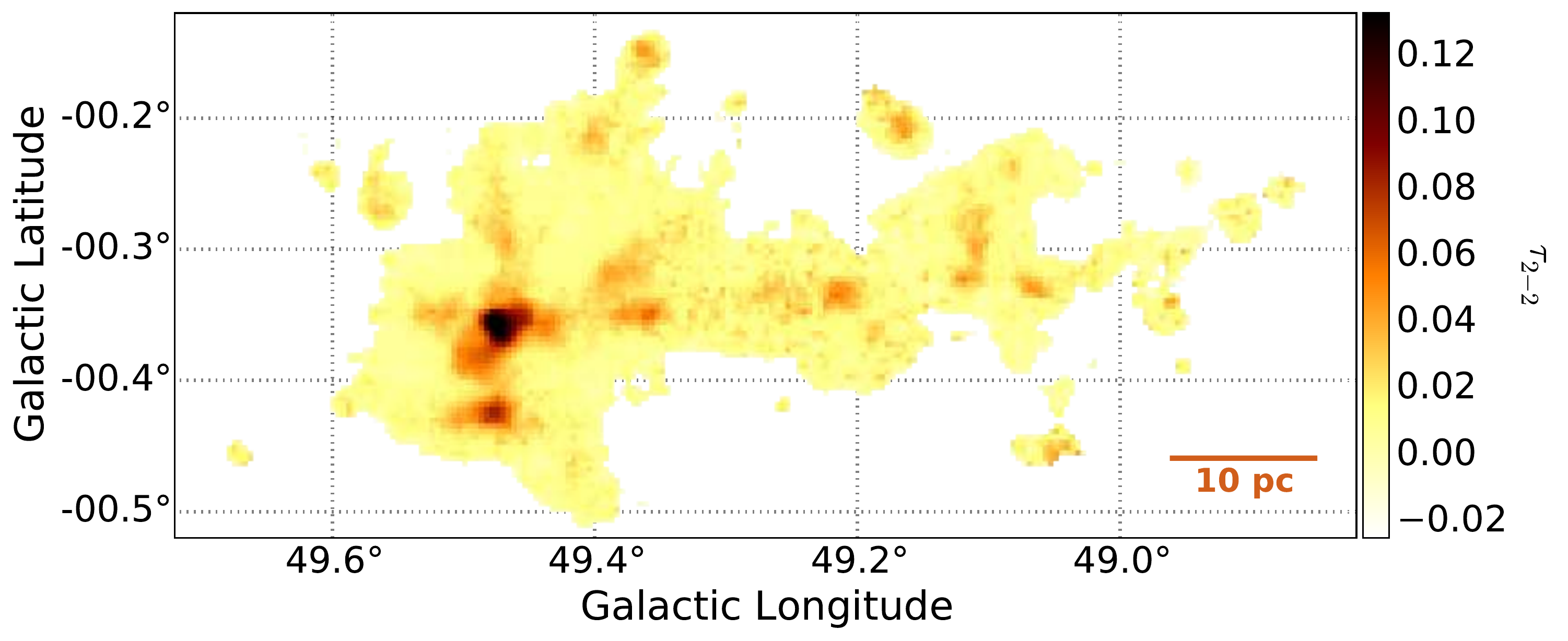}
{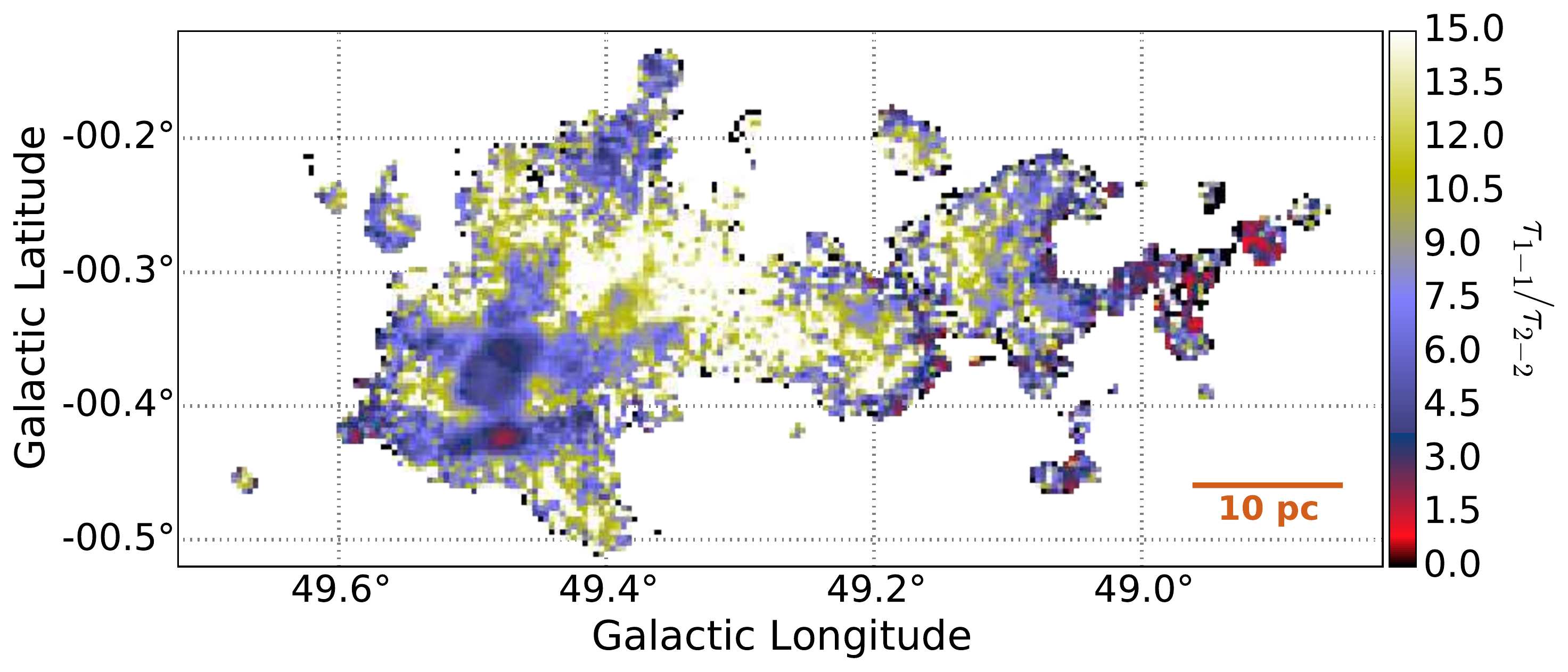}
{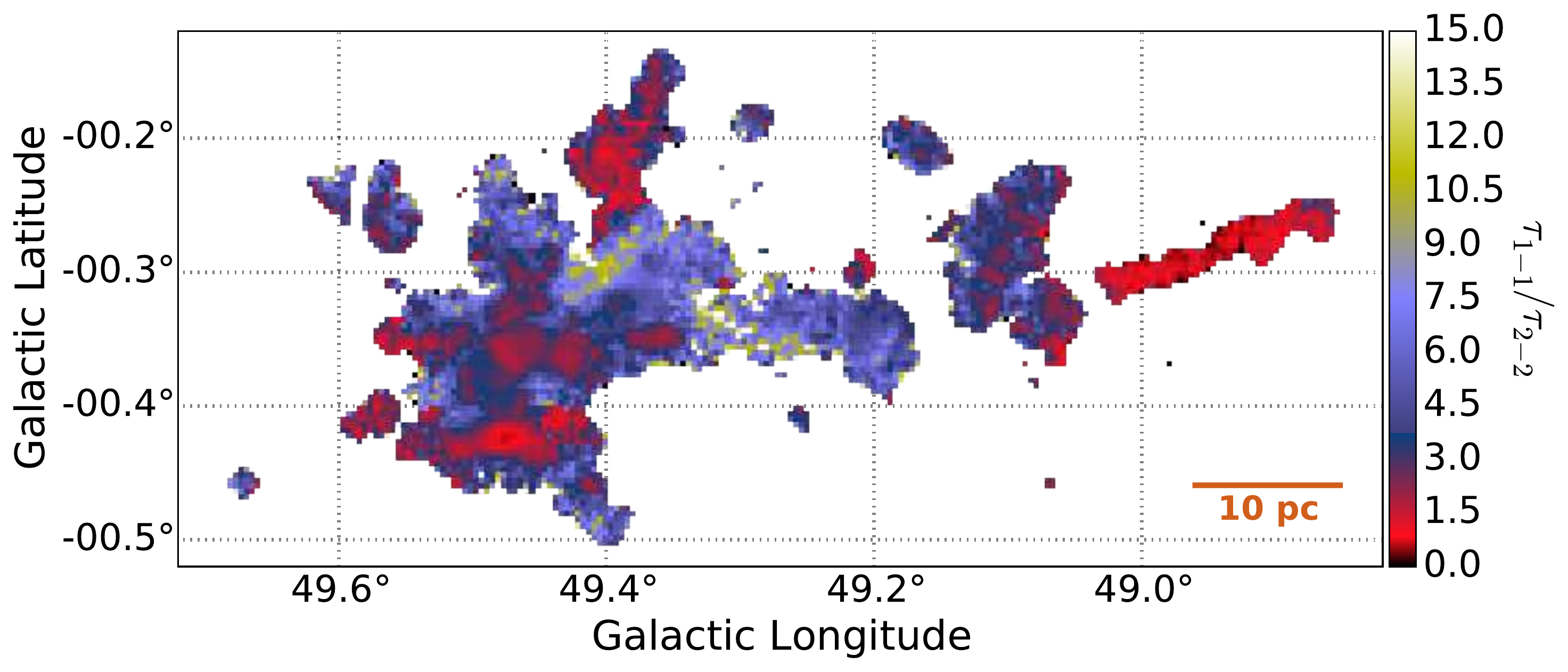}
{Plots of the peak \emph{observed} optical depth $\tau_{obs} =
-\ln(T_{MB}/T_{continuum})$ in the (a) \oneone and (b) \twotwo lines and (c)
their ratio, \oneone / \twotwo.  Figure (d) shows the `true' optical depth ratio
assuming $T_{ex}(\oneone) = 1.0$ K and $T_{ex}(\twotwo) = 1.5$ K; these are
reasonable and representative excitation temperatures but they are not fits to
the data.
The data are masked with a filter described in Section \ref{sec:h2co} and cover
the range $75 > V_{LSR} > 40$ \kms; see Figures \ref{fig:lowerpeakoptdepth} and
\ref{fig:upperpeakoptdepth} for individual velocity components.  In general,
lower (redder) ratios in figures (c) and (d) indicate higher densities, however
in the filament at 49.0-0.3, the low ratio is due to the geometry in which
$T_{continuum}$ is in the \emph{foreground} of the molecular gas.}
{fig:peakoptdepth}

\FigureFourPDF
{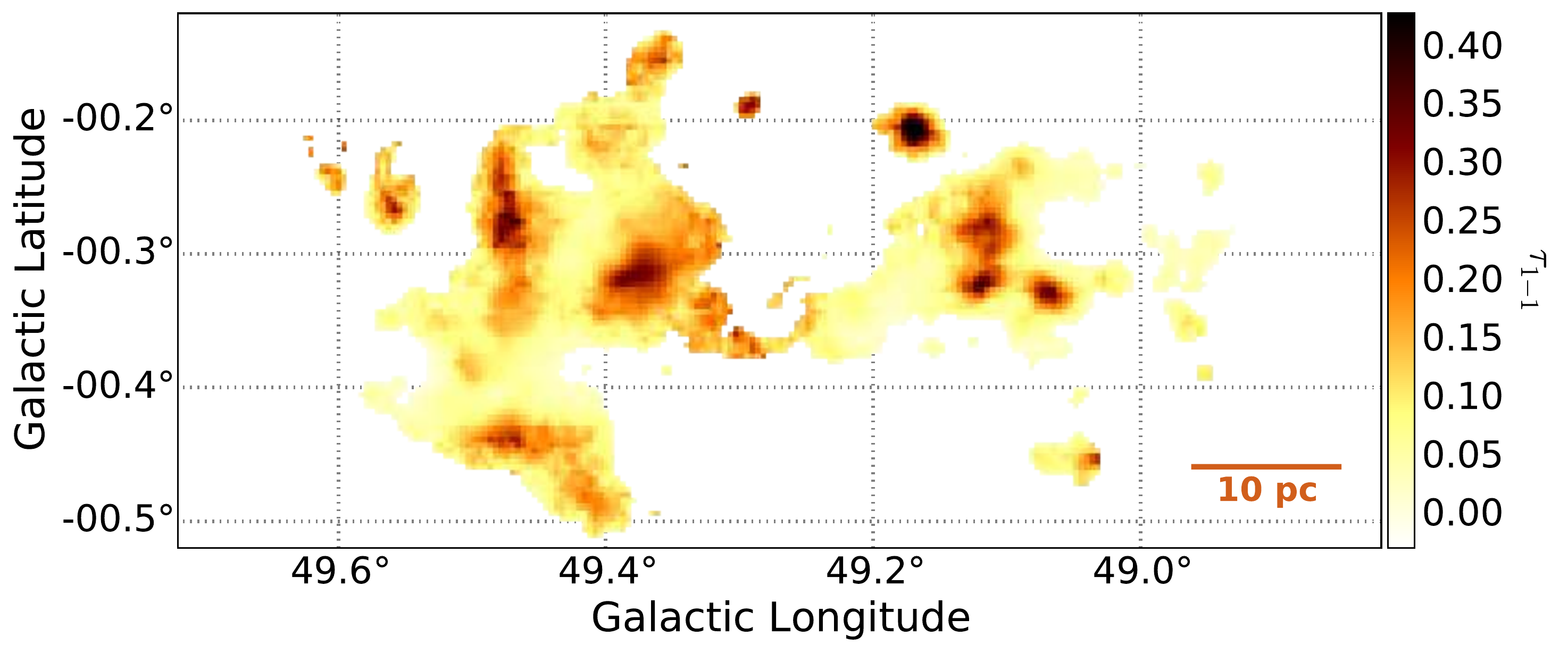}
{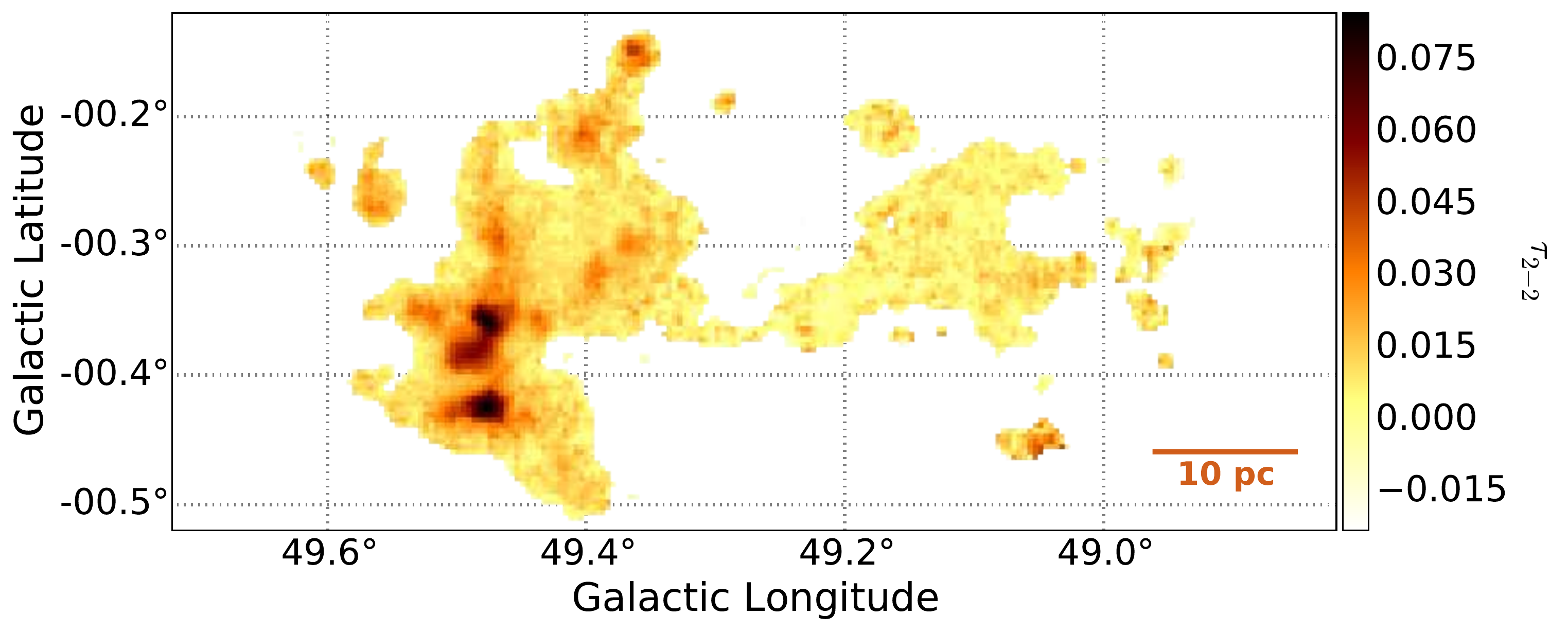}
{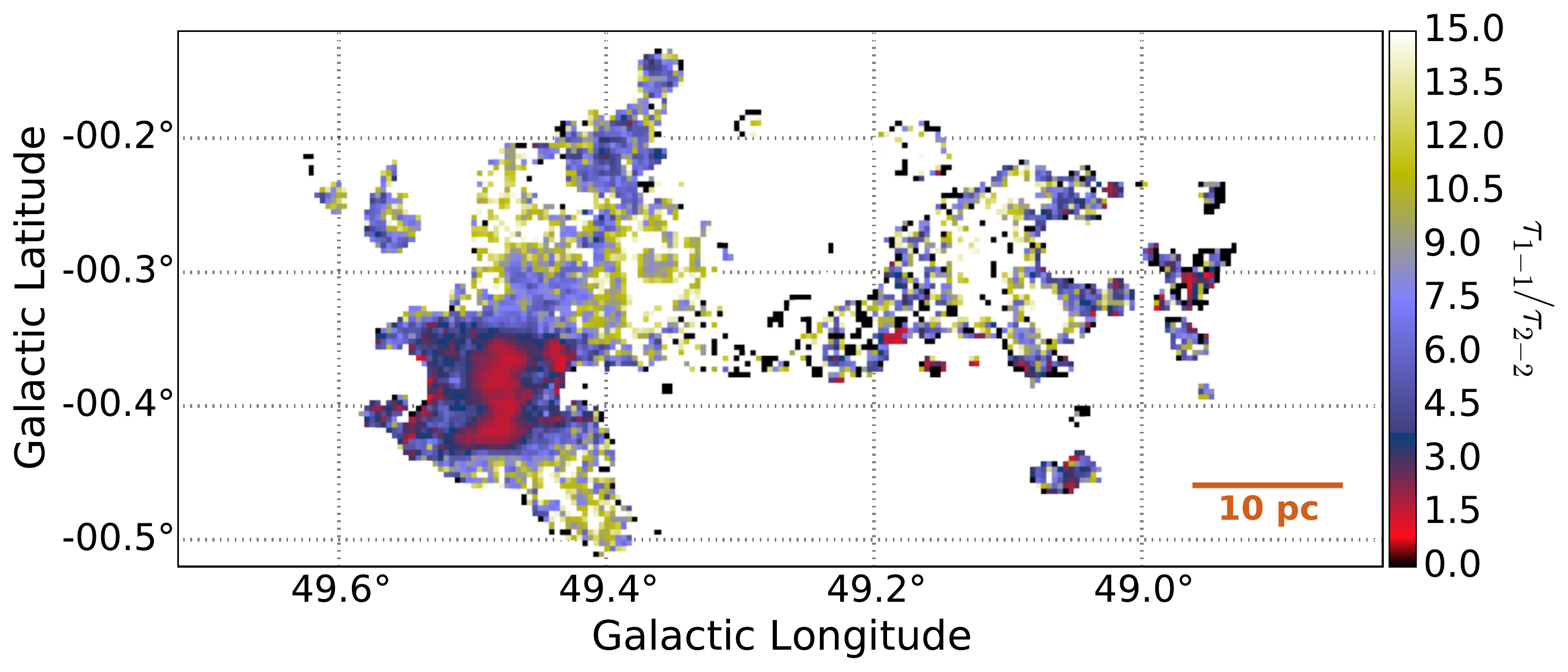}
{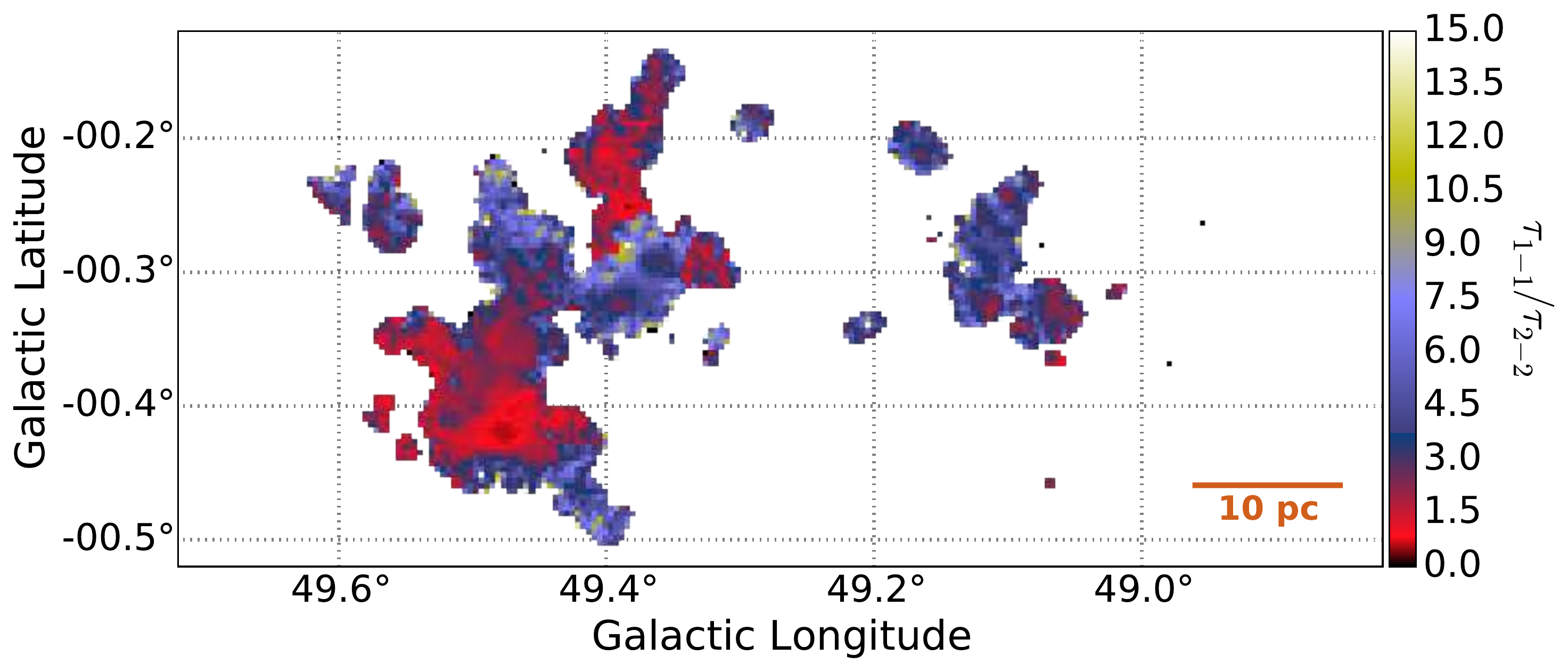}
{Same as Figure \ref{fig:peakoptdepth}, but limited to $62 > V_{LSR} > 40$ \kms.}
{fig:lowerpeakoptdepth}

\FigureFourPDF
{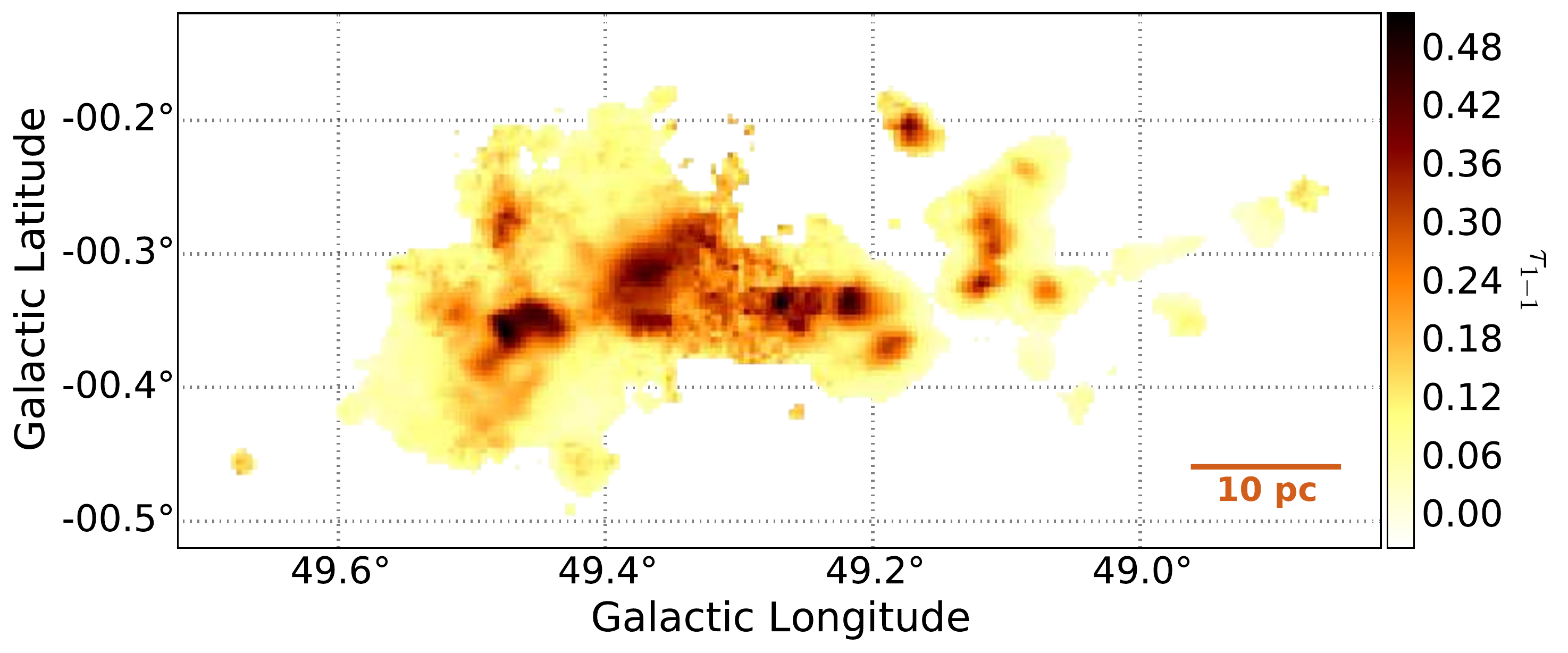}
{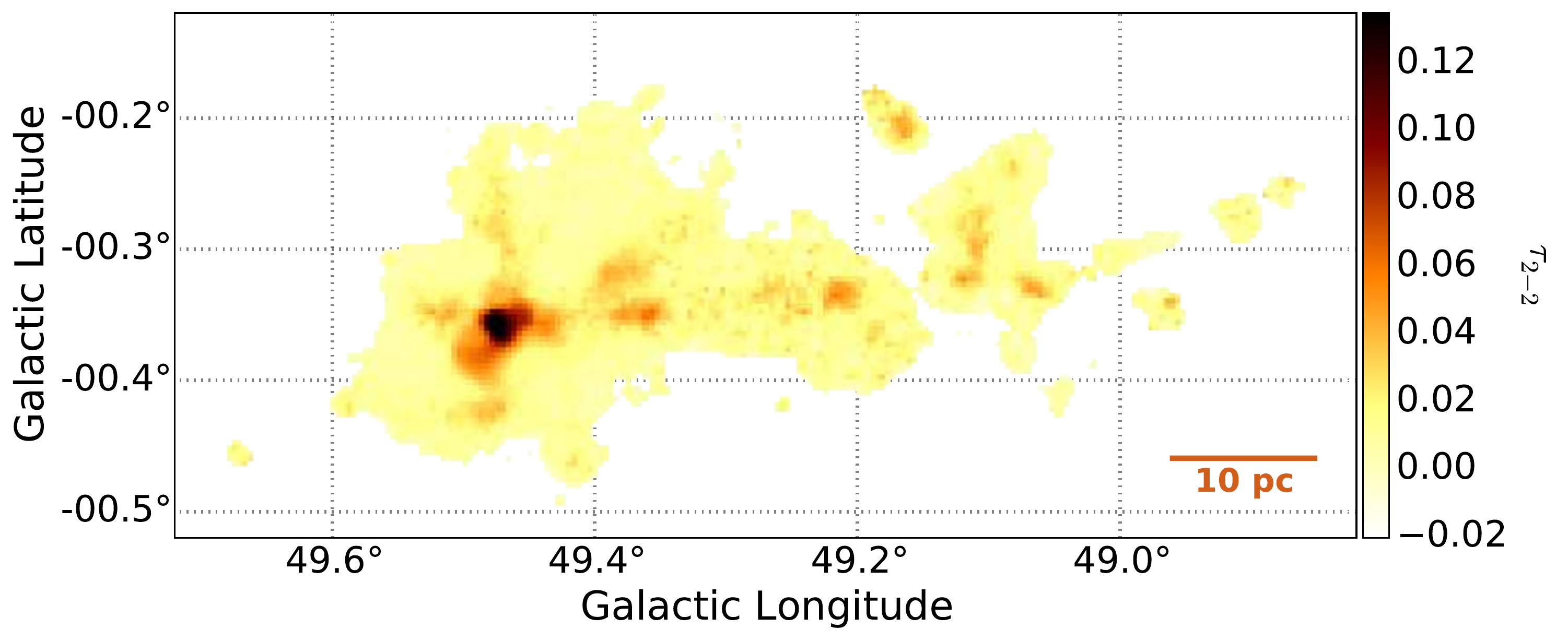}
{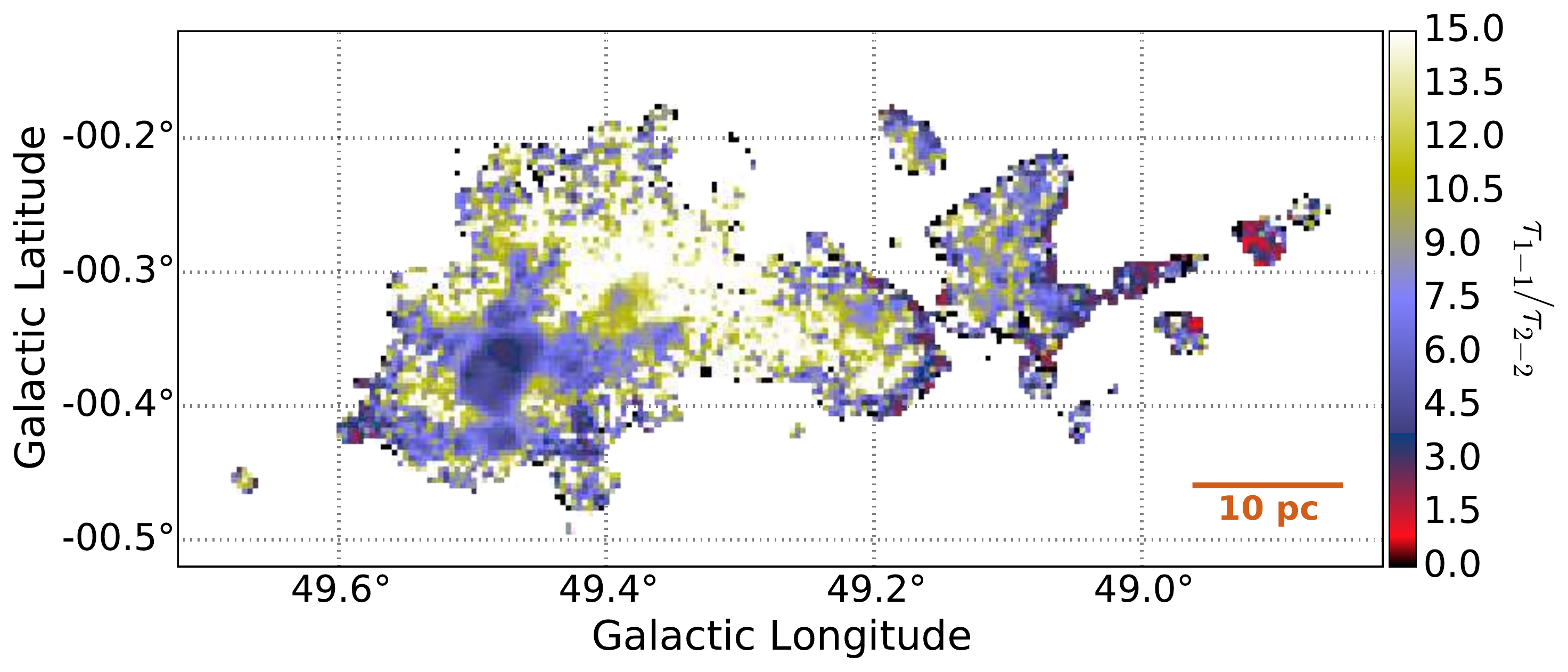}
{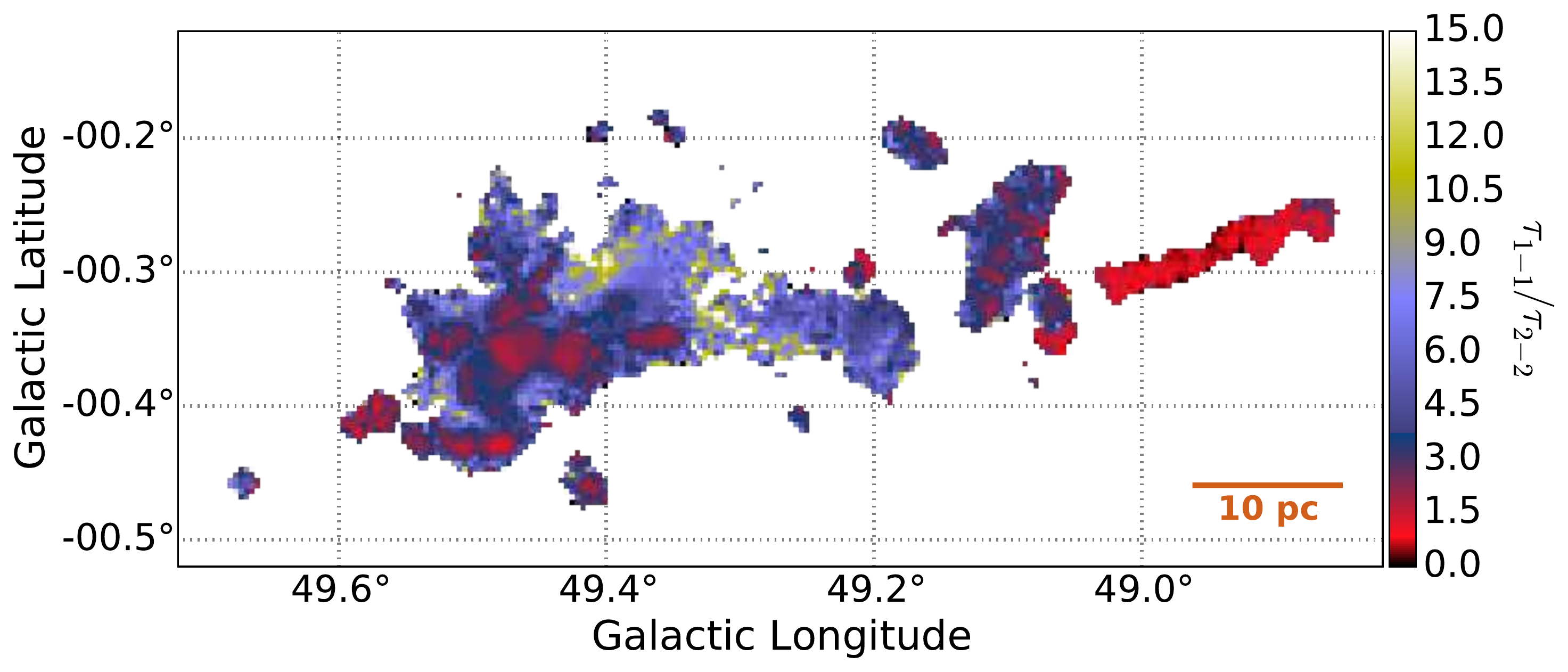}
{Same as Figure \ref{fig:peakoptdepth}, but limited to $75 > V_{LSR} > 62$ \kms.}
{fig:upperpeakoptdepth}

\subsection{Density Maps}
\label{sec:densmaps}
We computed the density in each voxel using the $\chi^2$ minimization technique
from \citet{Ginsburg2011a}.  We measure $\chi^2$ over the full 4D parameter
space (density, column density, gas temperature,
and ortho-to-para ratio [OPR])
\begin{equation}
    \begin{split}
    \label{eqn:chi2}
    \chi^2 =  \left( \frac{T_B(\oneone)-T_{model}(\oneone)}{\sigma(T_B,\oneone)}\right)^2 +\\
              \left( \frac{T_B(\twotwo)-T_{model}(\twotwo)}{\sigma(T_B,\twotwo)}\right)^2
    \end{split}
\end{equation}
The modeled brightness temperature $T_{model}$ is different for each spatial
pixel in order to account for the varying continuum background, although for
some pixels with significant
continuum detected, we still use the CMB as the background continuum because
the molecular gas is behind the other continuum emission; see Section
\ref{sec:geometry}.

We have not enforced any constraints on the column density,
temperature, or ortho-to-para ratio when fitting.  The best-fit value of each
of these parameters is taken to be the mean of those parameters over the range
$\Delta \chi^2 = \chi^2 - \chi^2_{min} < 1$, where $\chi^2_{min}$ is the
minimum $\chi^2$ value.

The gas temperatures returned from the fitting process are, as expected, purely
noise: the \formaldehyde \oneone/\twotwo ratio provides virtually no constraint
on the gas temperature and therefore leaving it as a free parameter has no
effect on the fitted density.  Similarly, the ortho-to-para ratio of \hh is
unconstrained in our data.  In principle, the \hh OPR has some effect on
\formaldehyde excitation \citep{Troscompt2009a}, but in the regime we have
modeled and observed, no effect is apparent.

The \ortho-column-weighted volume-density along each line of sight is shown in
Figures \ref{fig:wtdmeandens} and \ref{fig:wtdmeandensvrange}.  The former
shows the weighted density over all voxels and the latter shows the weighted
density over the two velocity ranges previously discussed.  These projections
include no information about the errors in the individual fits, which are
available from Equation \ref{eqn:chi2}, but by weighting by column density, we
have effectively selected the highest signal-to-noise points; the statistical
errors are therefore negligible relative to the systematic (i.e., those caused
by invalid assumptions about the single-zone nature of the models) in these
maps.  The maps are shown split into two velocity components,
$v<62$ \kms and $v>62$ \kms, which approximately separates out the 68 \kms
cloud from other components, though because the lines are quite broad the
separation is imperfect.  Figures \ref{fig:kinematics} and
\ref{fig:h2cokinematics} shows the velocity
separation in more detail.

The overall picture is of a central protocluster region (W51 Main and IRS2)
with most of the gas mass at a density $n\sim10^{5.5}$ \percc within a diameter
of $\sim3$ pc, surrounded by a rich cloud in which most of the mass is at a
density $\sim10^4$ \percc out to a diameter $d\sim14$ pc.

\Figure
{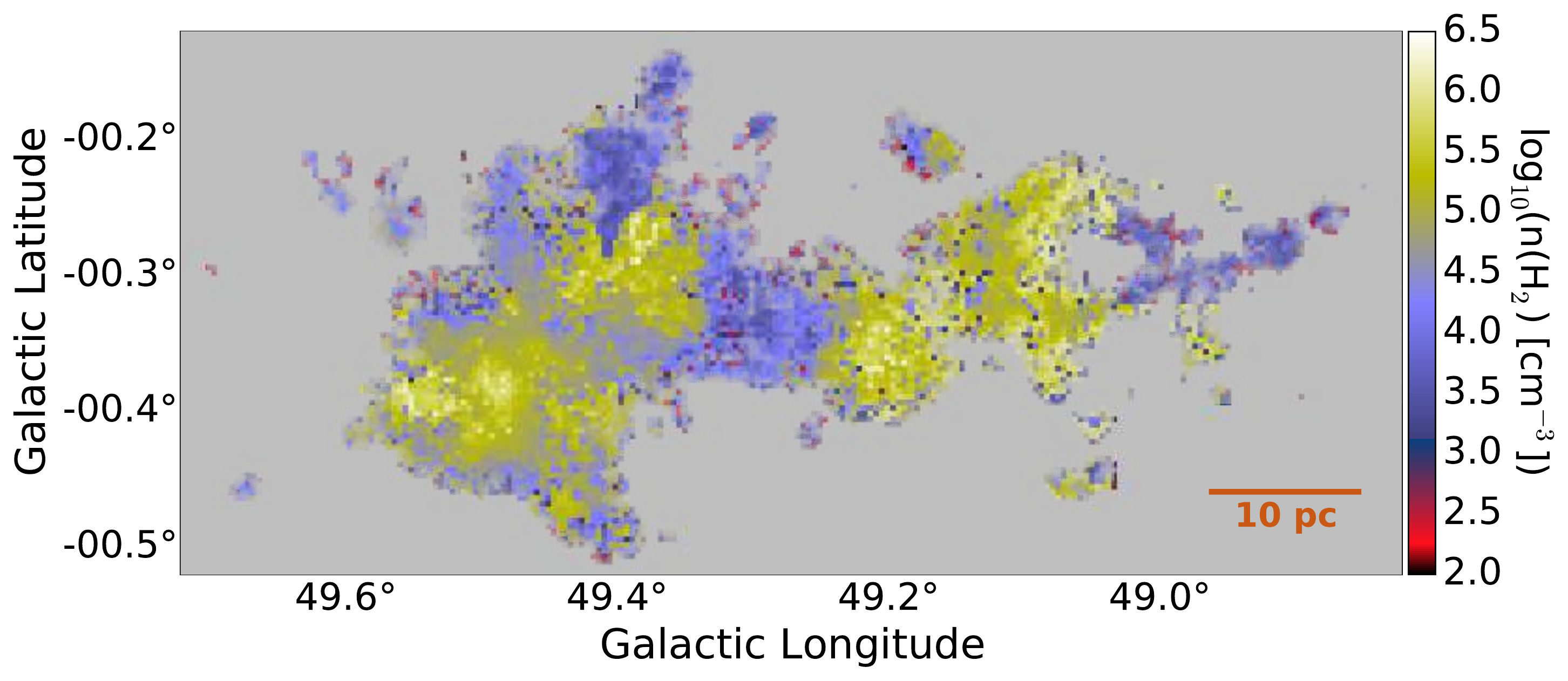}
{Map of the column-weighted volume density along the line of sight averaged
over all velocities.  The colors are `greyed out' where the
signal-to-noise ratio in the \oneone line is less than $\sim7$, with
lower-signal regions being progressively more gray.}
{fig:wtdmeandens}{0.5}{0}

\FigureTwoAA
{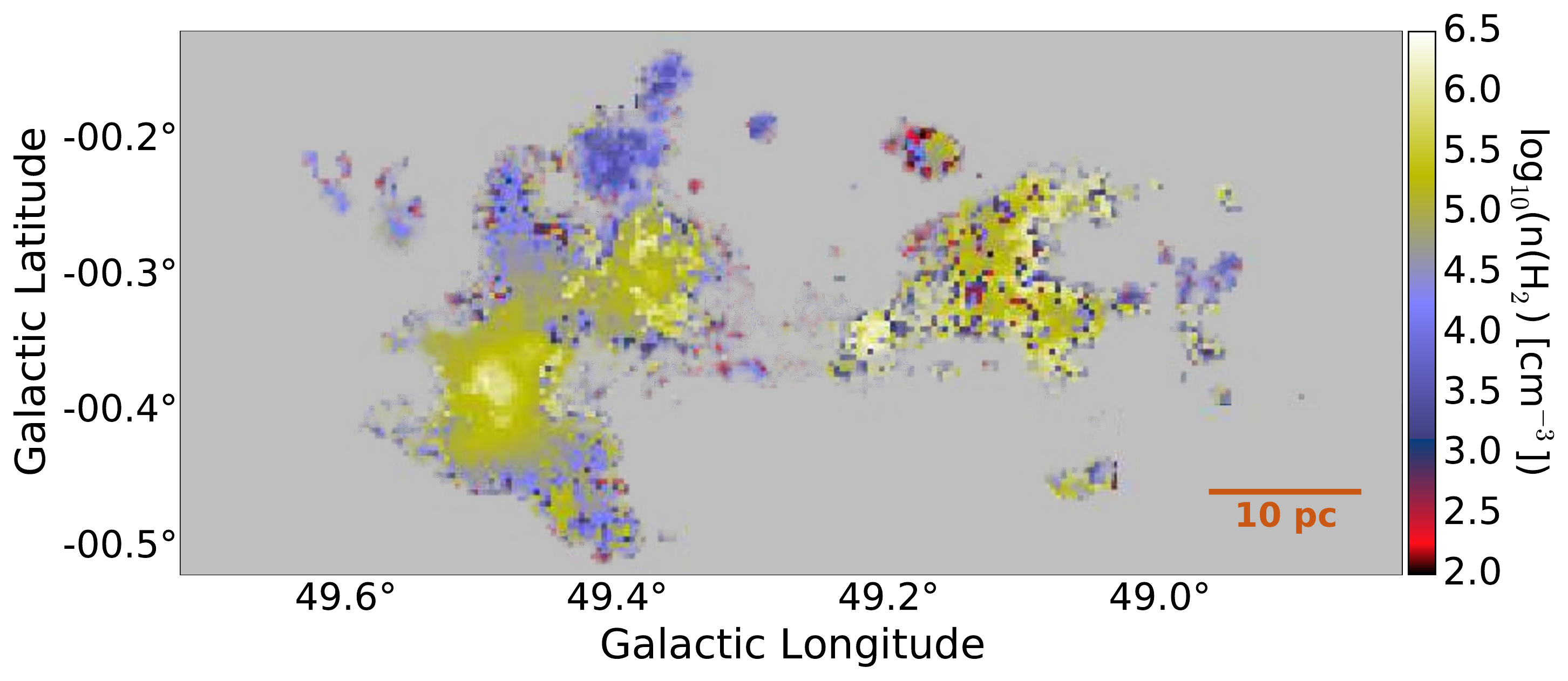}
{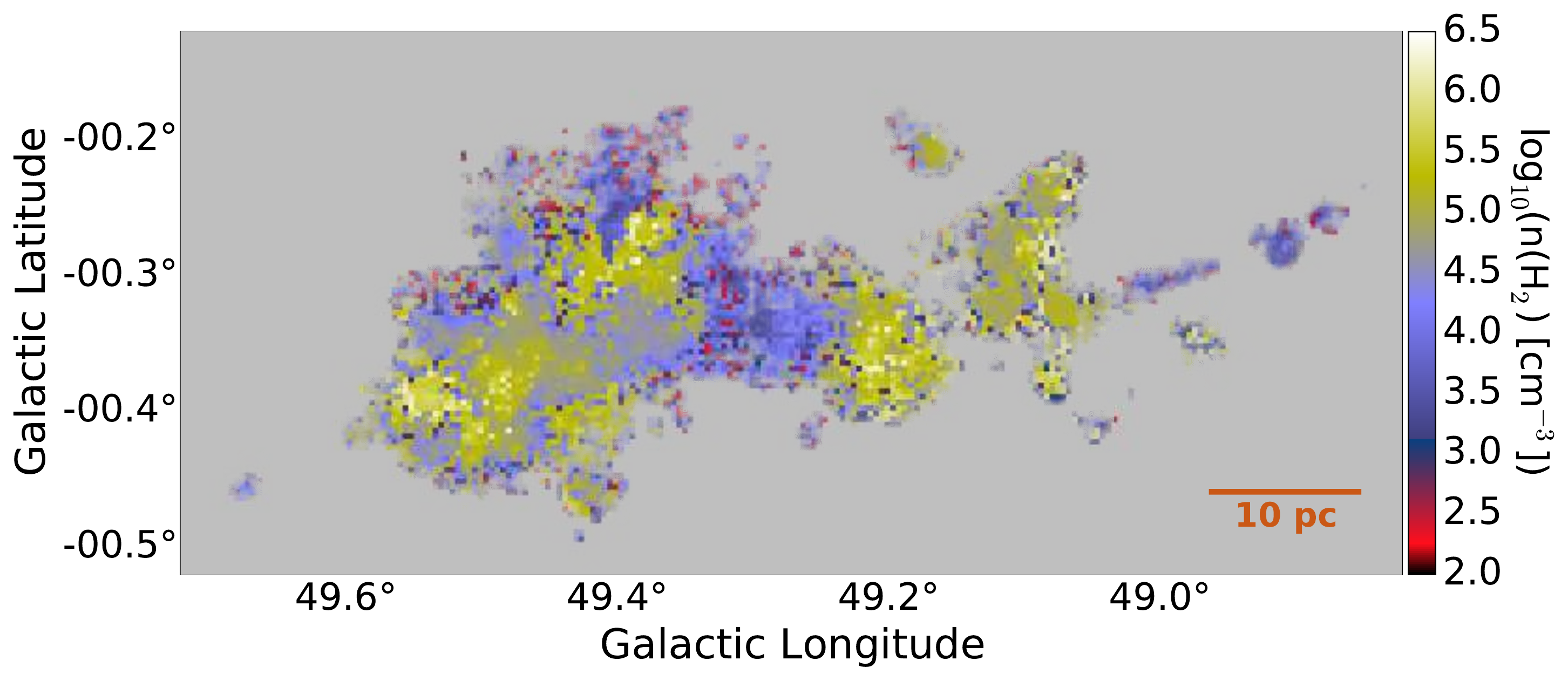}
{Map of the column-weighted volume density along the line of sight, split into 
(a) the 40-62 \kms component and (b) the 62 to 75 \kms component.  The
similarity between the two figures is due to large line widths; the cut at 62 \kms
is meant to highlight the low-density filament around $\ell=49$ and the clouds
surrounding W51 Main.}
{fig:wtdmeandensvrange}{1}{6.5in}

\subsection{Model fitting and geometry}
\label{sec:geometry}
Both \formaldehyde lines are seen only in absorption.  However, in some cases
the absorption is against a continuum background, while in others the
absorption may be only against the CMB.

We have fit the \formaldehyde lines constrained by the LVG models (Section
\ref{sec:models}) to spectra averaged over apertures with coherent molecular
absorption signatures (e.g., Figure \ref{fig:h2cofrontbackmodel}).  We compared
the $\chi^2$ values for fits with the observed continuum as the background to
those with the background fixed to $T_{BG} = T_{CMB}$.  We then selected the
better of the two fits as representative of the physical conditions.  Figures
\ref{fig:filament_pvslice} and \ref{fig:northsouthpv} show the qualitative
version of this analysis, highlighting CO-bright regions that lack the expected
\formaldehyde absorption signatures and therefore have a different geometry.
Regions with the continuum emission in front of the molecular absorption were
converted into masks that were then used to decide which models to use for the
per-pixel fitting process described in the previous section.  The geometric
analysis of each region is discussed in detail in Appendix
\ref{appendix:geometry}.

\Figure{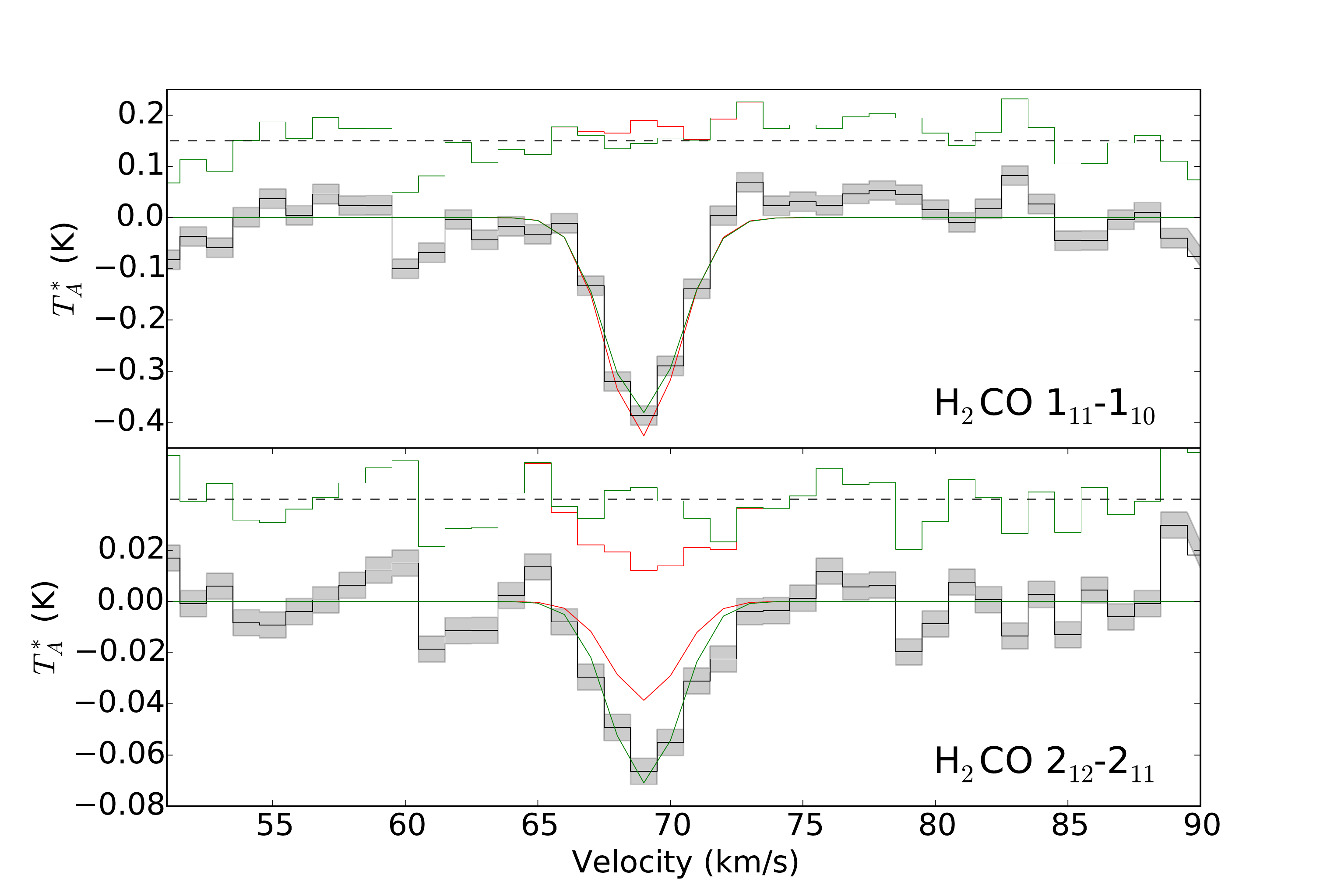}
{An example of the difference in models between a continuum source (red) and
the CMB (green) as the background.  The top plot shows the \oneone line and the
bottom shows the \twotwo line both with the continuum level set to zero in the
plot.  The residuals are shown
offset above the spectra, with the dashed line indicating the zero-residual
level.  The grey shaded regions show the 1$\sigma$ error bars on each pixel.
The model with the CMB as the only
background is able to reproduce the absorption line, while the model with the
\hii region in the background cannot account for the depth of the \twotwo line.
The reduced $\chi^2/n$ for the models are 14.1 (red) and 2.8 (green), evaluated
only over the pixels where the model is greater than the local RMS.}
{fig:h2cofrontbackmodel}{0.5}{0}

\FigureTwoAA
{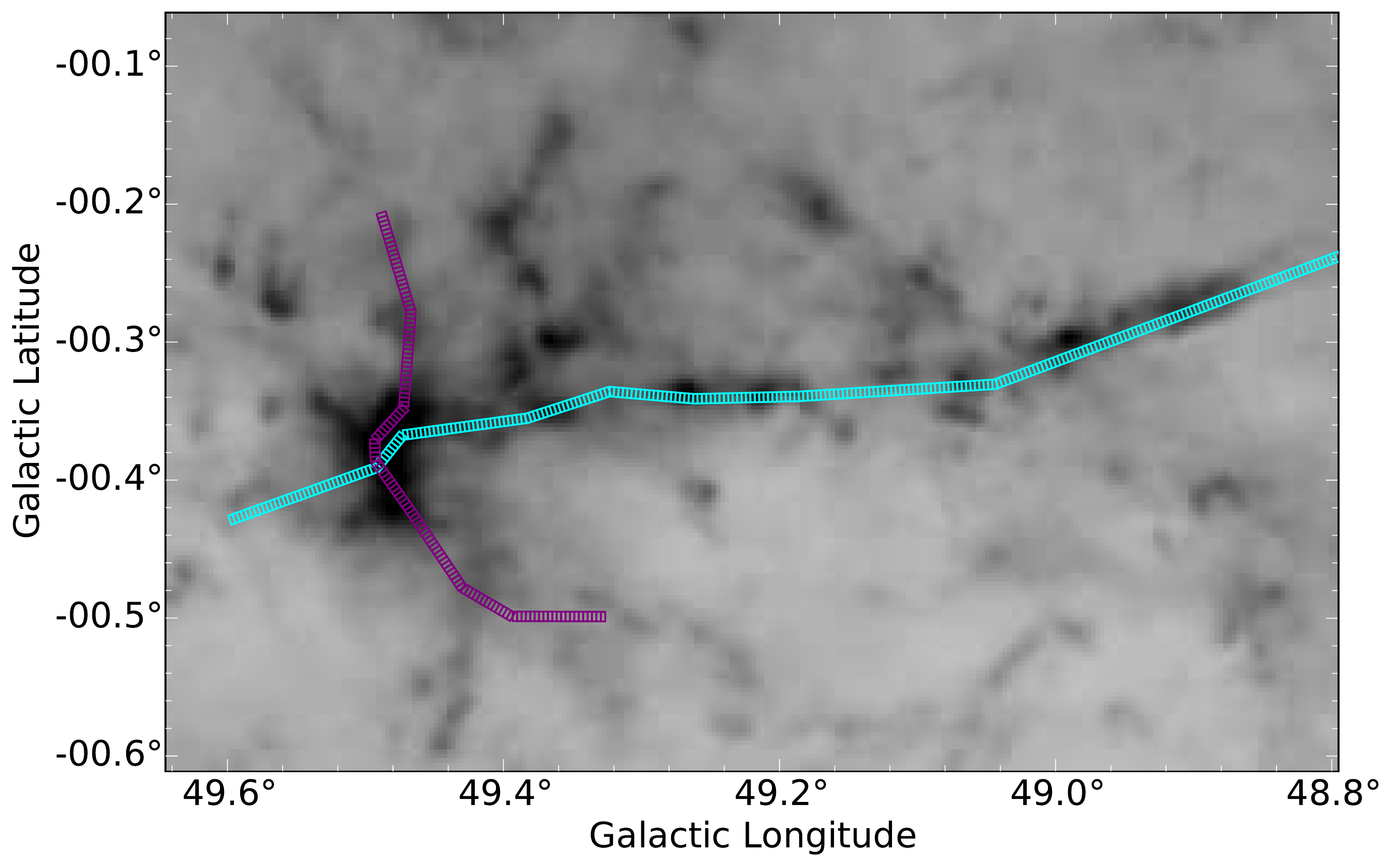}
{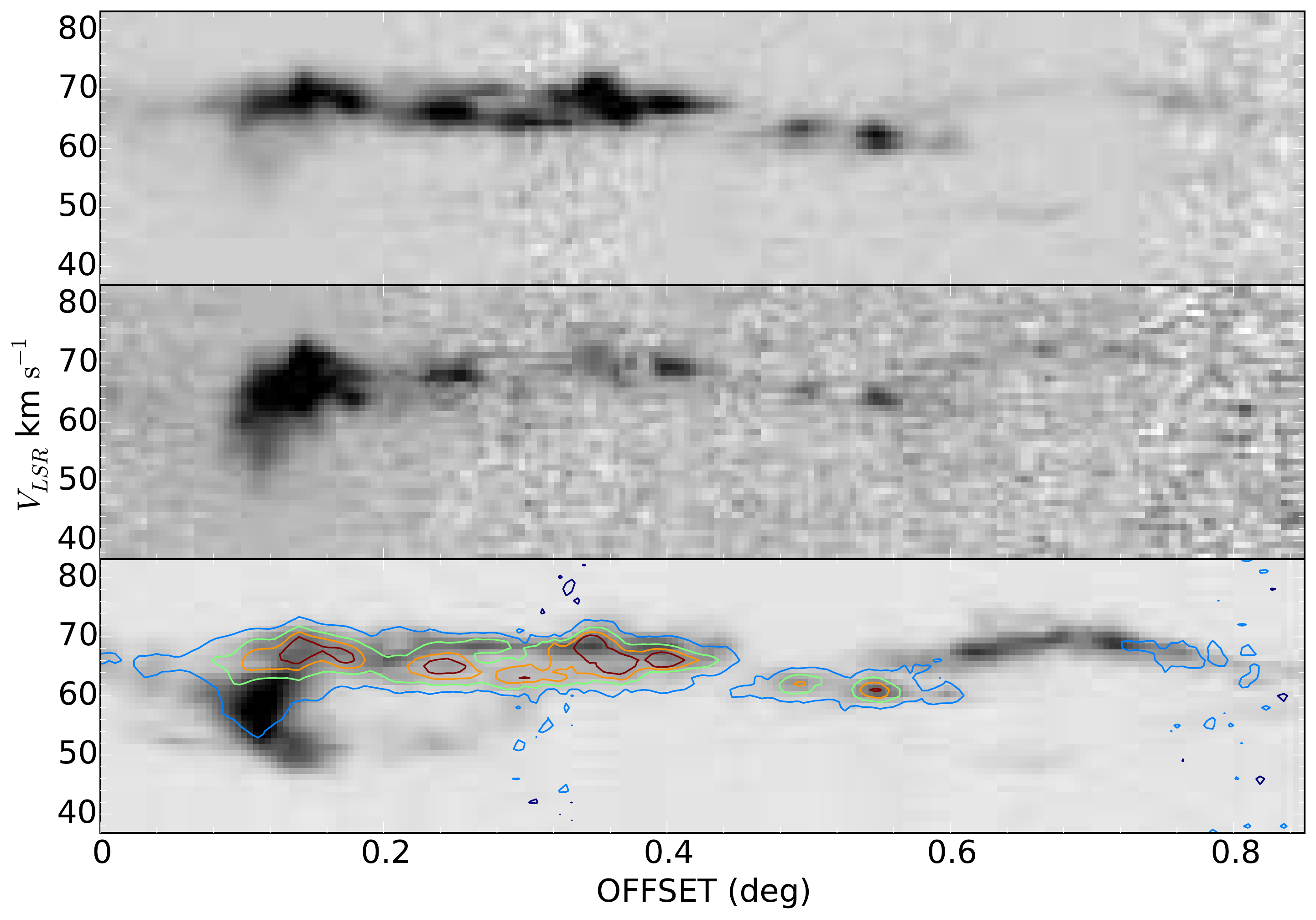}
{(top figure) A column density map fitted from the Herschel Hi-Gal data with two
filament extraction regions superposed in cyan and purple.
The purple extracted position-velocity diagram is shown in Figure
\ref{fig:northsouthpv}.
(bottom figure) A position-velocity slice of the 68 \kms cloud, shown in cyan in the
left figure, which includes an 8\um-dark cloud and the interaction region
with the W51C supernova remnant.
(bottom figure, top panel) \formaldehyde \oneone observed optical depth
(bottom figure, middle panel) \formaldehyde \twotwo observed optical depth
(bottom figure, bottom panel) \thirteenco 1-0 emission from the Galactic Ring Survey
\citep[GRS][]{Jackson2006a} with \formaldehyde \oneone
contours superposed.  The weakness of the \formaldehyde absorption on the right
half of the cloud corroborates the geometry inferred from comparison of the \oneone
and \twotwo lines in Figure \ref{fig:h2cofrontbackmodel}.
The \thirteenco emission without corresponding \formaldehyde absorption at offset
0.2 degrees is primarily background material in the 51 \kms cloud (see Figure
\ref{fig:geosketch}.
These figures were made using \texttt{wcsaxes} (\protect\url{http://wcsaxes.rtfd.org})
and \texttt{pvextractor} (\protect\url{http://pvextractor.rtfd.org}).
}
{fig:filament_pvslice}{1}{6.0in}

\Figure
{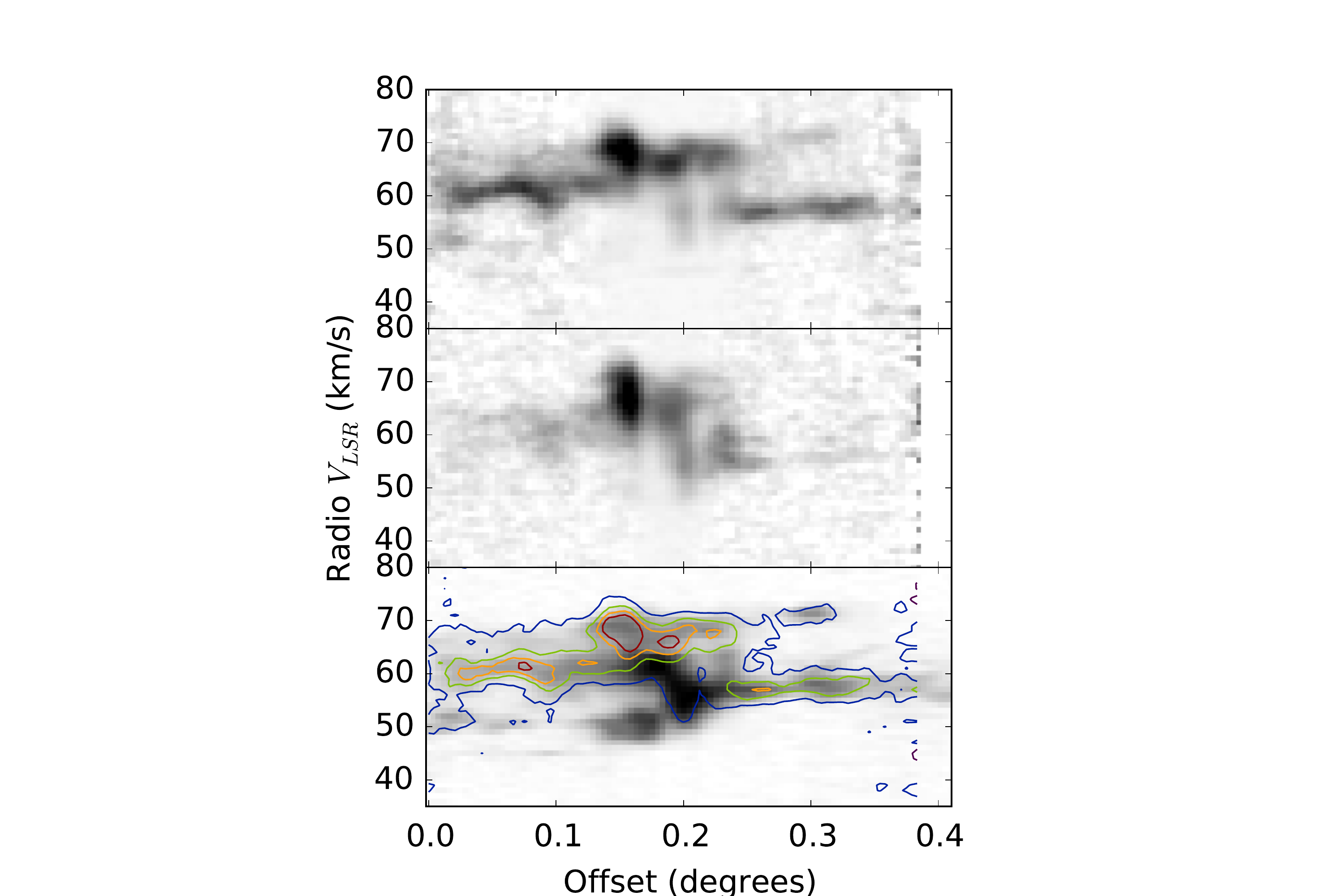}
{Position-velocity diagrams of filamentary structures to the north and south of
W51 Main. The panels are \formaldehyde \oneone, \twotwo, and \thirteenco
1-0 as in Figure \ref{fig:filament_pvslice}.  The 50 \kms component at offset
$\sim0.15$ degrees ($\ell\approx49.47, b\approx-0.42$) is in the background of
the \hii region.  The extracted region is identified in purple in Figure
\ref{fig:filament_pvslice}; the left side of the position-velocity diagram
corresponds to the $b=-0.5$ end of the region.}
{fig:northsouthpv}{0.5}{0}

It is possible that there are multiple continuum emitters along the line of
sight in many cases, with the absorbing molecular gas somewhere in the middle.
While this possibility adds uncertainty to the measurements, there are some
cases in which the dominant continuum can unambiguously be assigned a
foreground or background position.

Summary figures of our geometric analysis are shown in the cartoon Figure
\ref{fig:geosketch}, with an accompanying labeled on-sky map in Figure
\ref{fig:labeledzoom}.

\Figure{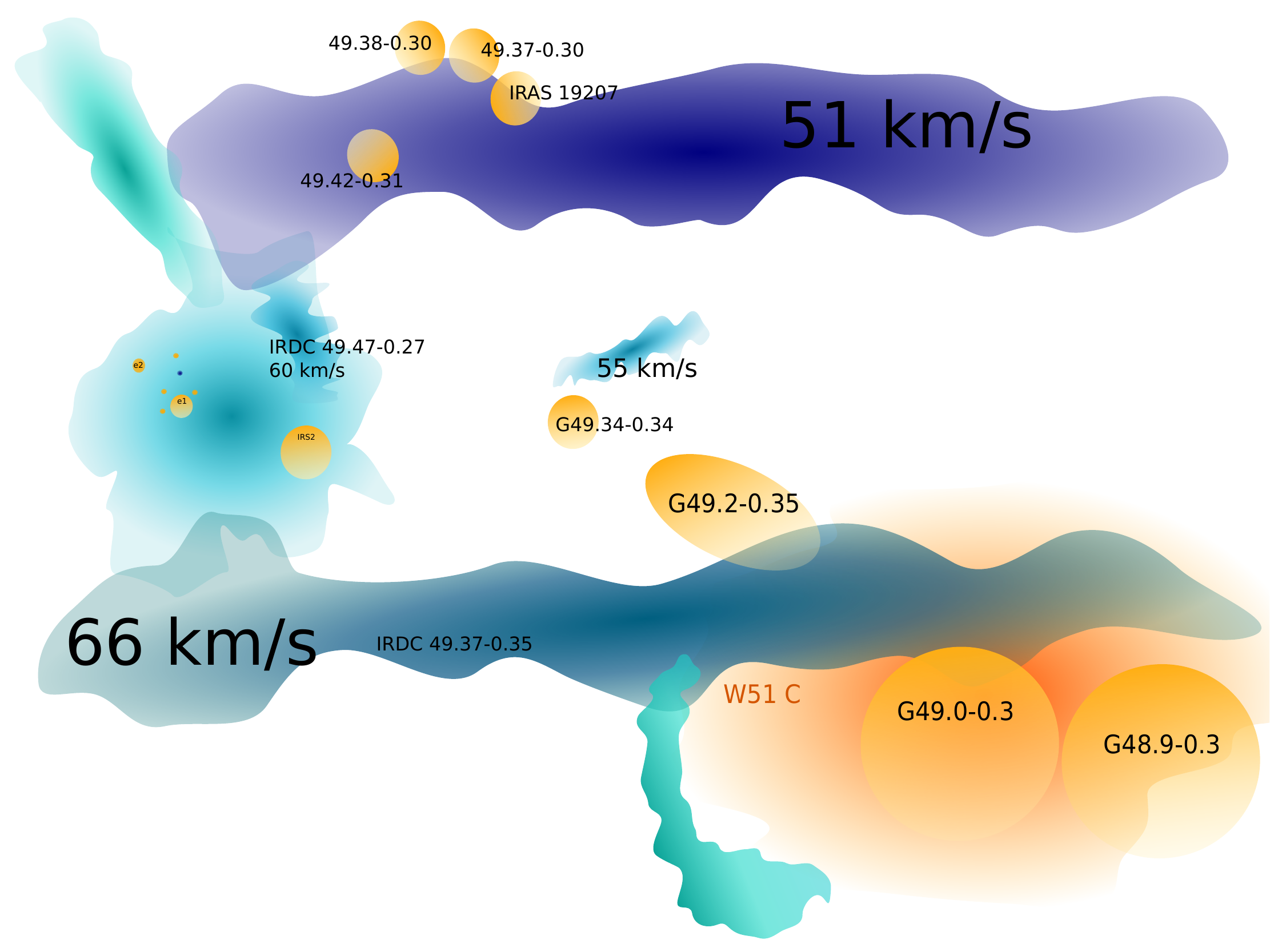}
{A sketched diagram of the W51 region as viewed from the Galactic north pole,
with the observer looking up the page from the bottom (i.e., W51C is the 
front-most labeled object along our line-of-sight).  There are a few
significant differences between this and Figure 29 of \citet{Kang2010a},
particularly the relative geometry of the cloud and the \hii regions in W51 B.
We also show a good deal more detail, revealing that there are \hii regions on
both front and back of many clouds.
The orange areas represent \hii regions and ionized gas (the W51 C SNR), while
purple/blue/cyan regions show molecular clouds.
The shapes of the clouds approximately reflect their shape on the sky, but these
shapes are only intended as mnemonics to help associate this face-on view with
the edge-on view of the real observations.
Figure \ref{fig:labeledzoom} shows the face-on view and can be compared
side-by-side with this figure to get an approximate 3D view of the region.
}
{fig:geosketch}{0.5}{0}

\Figure
{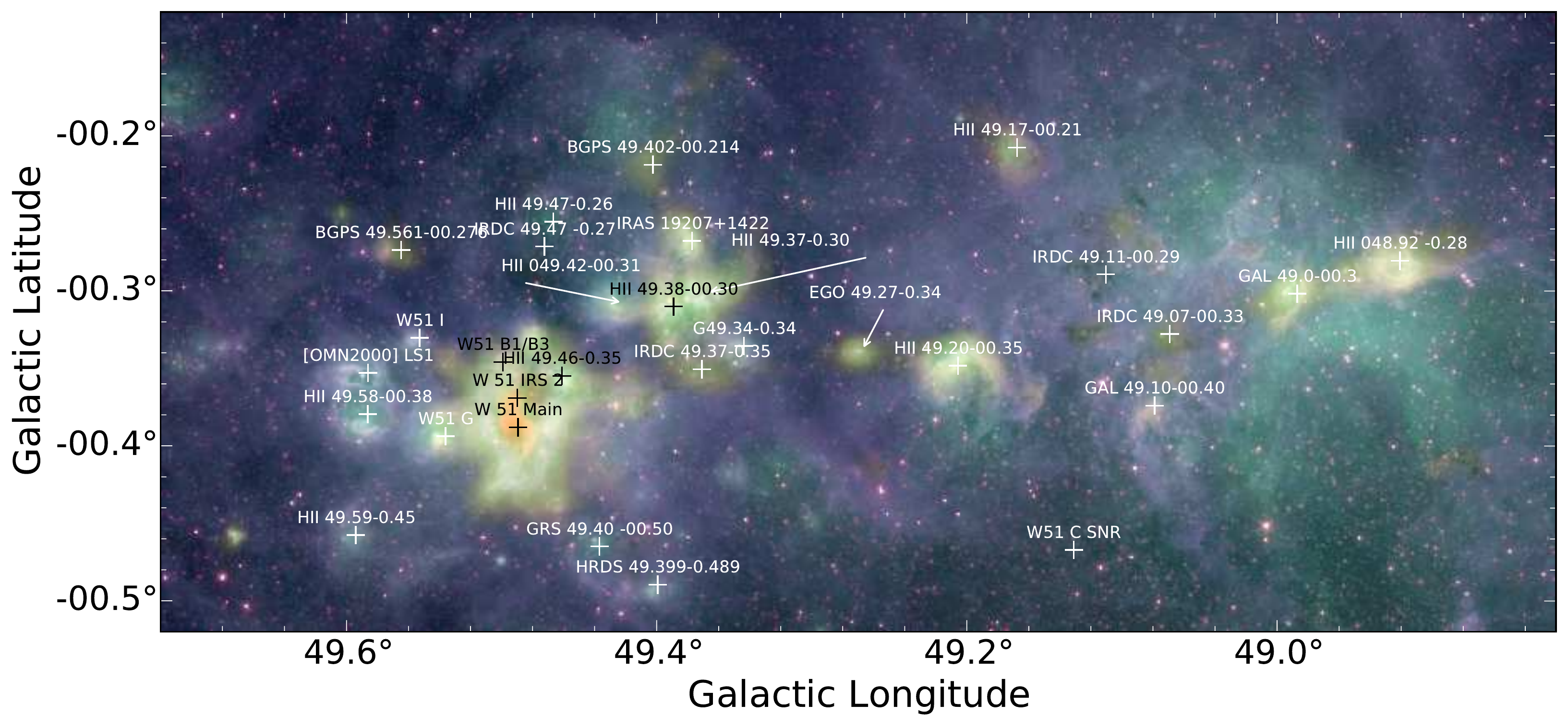}
{Labeled figures of the W51A and W51B regions, highlighting \hii regions and
8\um-dark clouds.   The colors are described in Figure \ref{fig:w51large}.
These labels can be compared to Figure \ref{fig:geosketch} to associate labeled
regions in the plane of the sky with their counterparts in the face-on view of
our Galaxy.}
{fig:labeledzoom}{0.5}{0}

\subsection{Dense Gas Mass Fractions}
Because the \formaldehyde densitometer yields a
mass-weighted\footnote{This is a simplification; other factors in the
weight include the optical depth for an optically thick line, the abundance of
\formaldehyde, and the geometry if there is a background source.  For most of
the W51 cloud, a fixed abundance and optically thin lines are not a bad
assumption, and we account for the geometry in the discussion.}
measurement of
the gas volume density, it is difficult to connect directly to the total gas
mass, which is the quantity of interest when determining bulk properties like
star forming efficiency.  However, because the \formaldehyde and CO chemically
trace the same gas, the \formaldehyde-derived density can be applied to the
total mass measured by CO.   We use the the FCRAO Galactic Ring Survey
of \thirteenco \citep{Jackson2006a} and assume that each \thirteenco PPV
`voxel' has a mass proportional to its integrated intensity and a density given
by the $n(\hh)$ delivered from the \formaldehyde densitometer.

The `dense gas mass fraction' (DGMF) is an oft-quoted measurement used to argue
about the speed of the star formation process, the existence of density
thresholds, and turbulent properties of the ISM \citep[e.g. Fig. 5
of][]{Krumholz2007a,Battisti2014a,Kainulainen2013a,Juneau2009a,Muraoka2009a,Hopkins2013e}.
However, these fractions are most often quoted as mass of gas at a
\emph{single} density divided by the total mass.  We present an improvement on
these measurements, showing the continuous distribution of the dense gas mass
fraction.

Figure \ref{fig:dgmf} shows the result of using our \formaldehyde PPV density
cubes to measure the DGMF from \thirteenco.  We use a range of density
thresholds from $\sim10^3$ to $\sim10^6$ \percc.  At each density, we identify
all voxels in the \thirteenco PPV cube above that density and integrate those.
We then divide by the total integrated \thirteenco brightness to get the mass
fraction: 
\begin{eqnarray}
    N(n>n_0) =  C \int T_{\thirteenco}(v,\ell,b) 
                G\left(n(v,\ell,b)\right) dv~d\ell~db \\
    G(n) = \left\{ \begin{array}{ll}
        1 & : n >= n_0 \\
        0 & : n < n_0 \\
    \end{array} \right.\\
    N(total) = C \int T_{\thirteenco}(v, \ell, b) dv~d\ell~db\\
    DGMF = N(n>n_0) / N(total)
\end{eqnarray}
where $N$ is the column of \hh and $C=8.1\ee{20} \persc /
(\Kkms)$ is a constant used to convert \thirteenco brightness to
\hh column; it cancels in the DGMF equation.

Figure \ref{fig:dgmf_regions} shows the same results, but for two individual
regions: the W51 Main/IRS2 protoclusters and the W51 B region.  Within about 10 pc
of W51 Main, around half of the mass is at density $n>10^4$ \percc.  By
contrast, the rest of the molecular cloud shows a consistent fraction
$f(n>10^4~\percc) \sim10\%$.

These DGMF plots are similar to cumulative distribution functions of the column
density \citep[e.g.][Figure 6]{Battersby2014a}, but with the added advantage of
assigning a density to each resolution element in both velocity and position.

\Figure{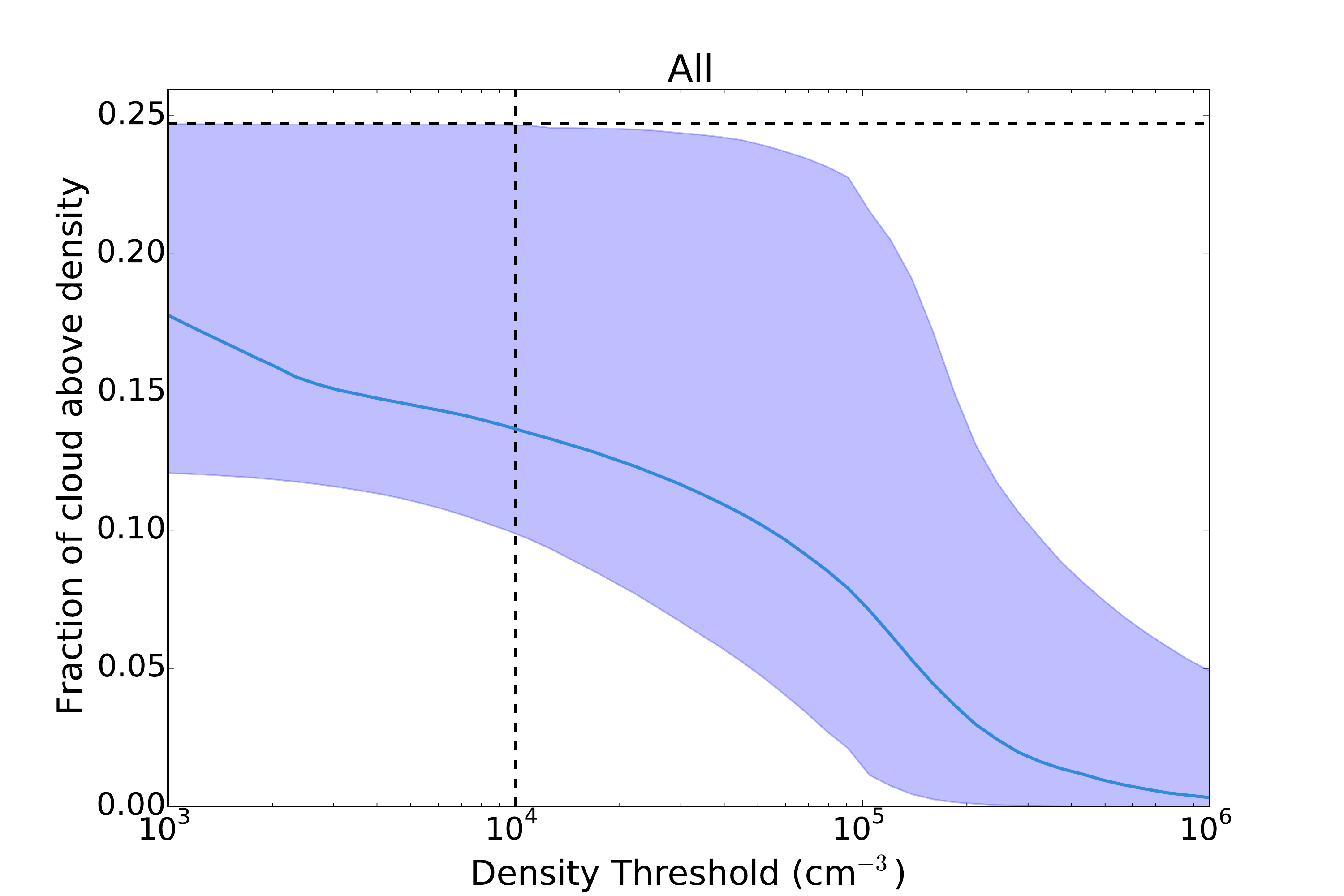}
{The `dense gas mass fraction' as a function of volume density threshold
$n(\hh)$ \percc.  The $y$-axis shows the sum of \thirteenco pixels from the GRS
cube with
\formaldehyde-derived density above the value shown on the $x$-axis divided by
the total.  Both values are computed over the velocity range $40~\kms < v_{LSR}
< 75~\kms$.  The solid line represents the fraction of \thirteenco emitting
voxels above the mean measured \formaldehyde density as described in Section
\ref{sec:densmaps}.
The blue shaded region shows the extent of plausible model fits at
each density: effectively, this is the $\sim1\sigma$ error region.  The
vertical line at $n=10^4$ \percc indicates the approximate completeness limit.
The horizontal line shows the fraction of \thirteenco flux in pixels that had
corresponding detections at
$>2\sigma$ in both the \formaldehyde \oneone and \twotwo lines: it represents
the upper limit of what could have been detected if, e.g., all \formaldehyde
detections were
toward regions with $n>10^4$ \percc.  The failure to converge to a fraction
$f\rightarrow f_{max}$ indicates that there are some real detections of
low-density gas.}
{fig:dgmf}{0.5}{0}

\FigureTwoAA
{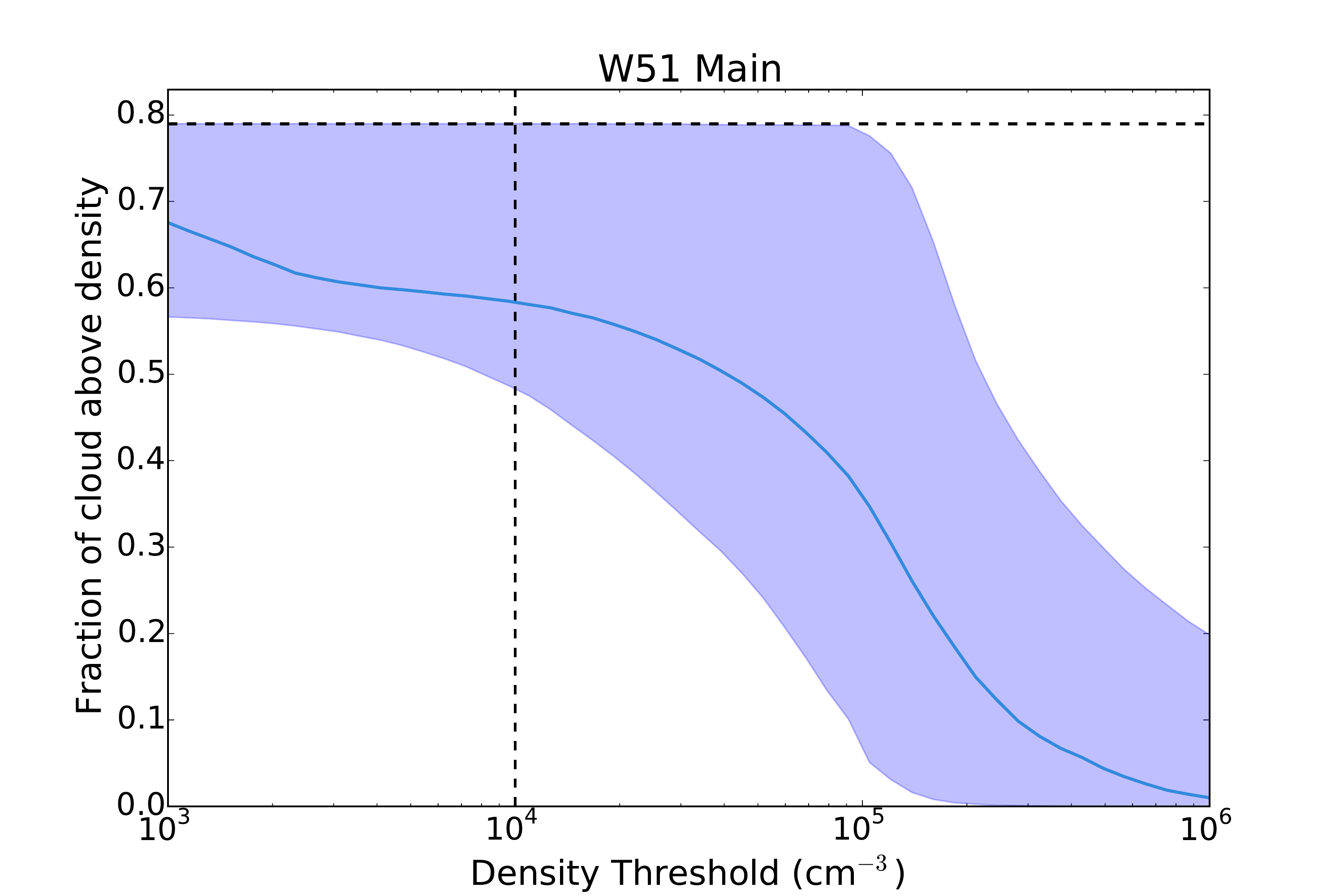}
{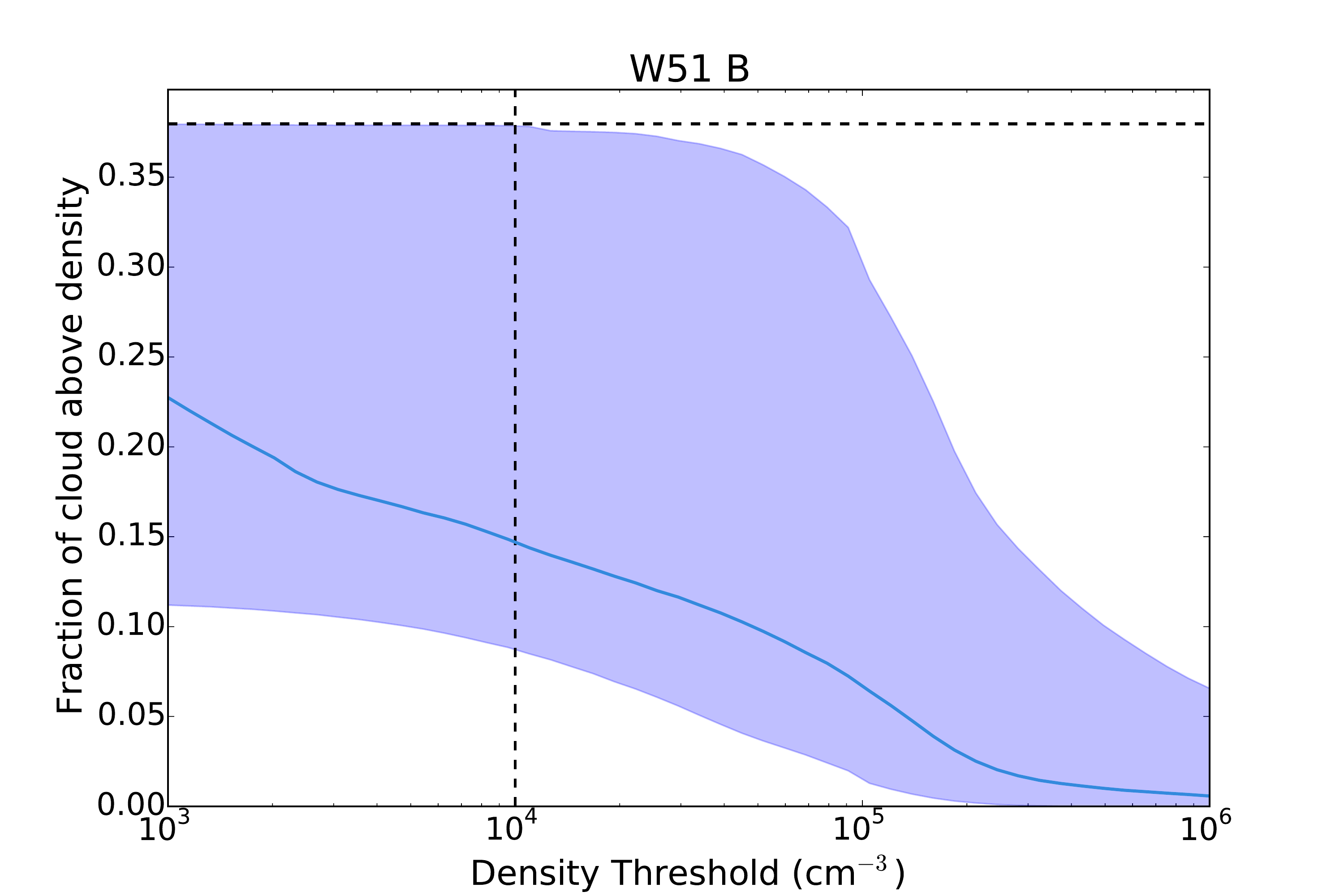}
{Same as Figure \ref{fig:dgmf}, but for two individual regions: W51 Main
($a$), the region $49.4\arcdeg < \ell < 49.6\arcdeg$, $-0.5\arcdeg < b <
-0.3\arcdeg$, and W51 B ($b$) with $48.8\arcdeg \ell < 49.4\arcdeg$ and
$-0.5\arcdeg < b < -0.1\arcdeg$.  Note that the $y$ axes have different
ranges.  The area covered by the W51 B cutout is $6\times$ larger than
the W51 A cutout, so its effect on Figure \ref{fig:dgmf} is greater. The total
area covered in Figure \ref{fig:dgmf} includes some regions with no detected
\formaldehyde, which is why the peak fraction in that figure is lower.}
{fig:dgmf_regions}{1}{6.5in}

\FigureTwoAA
{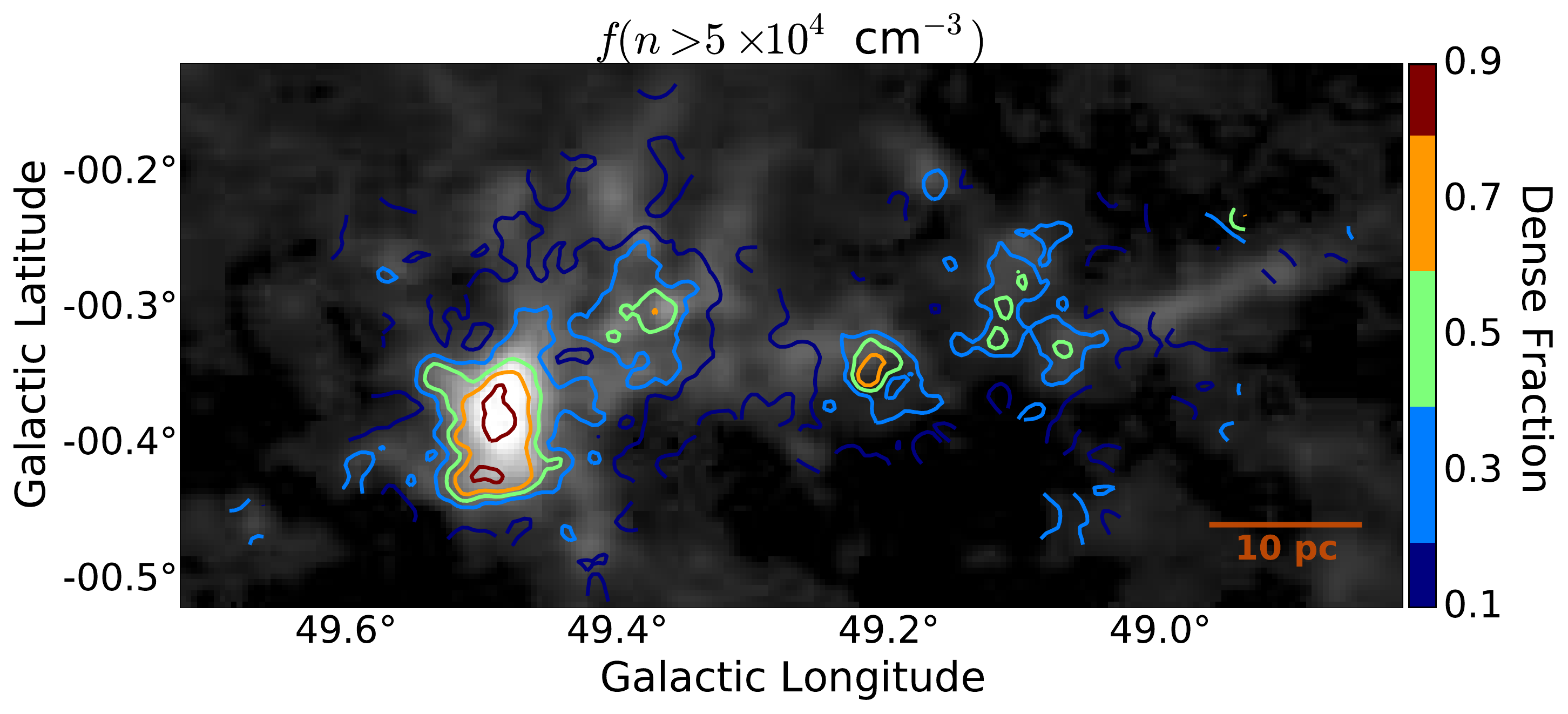}
{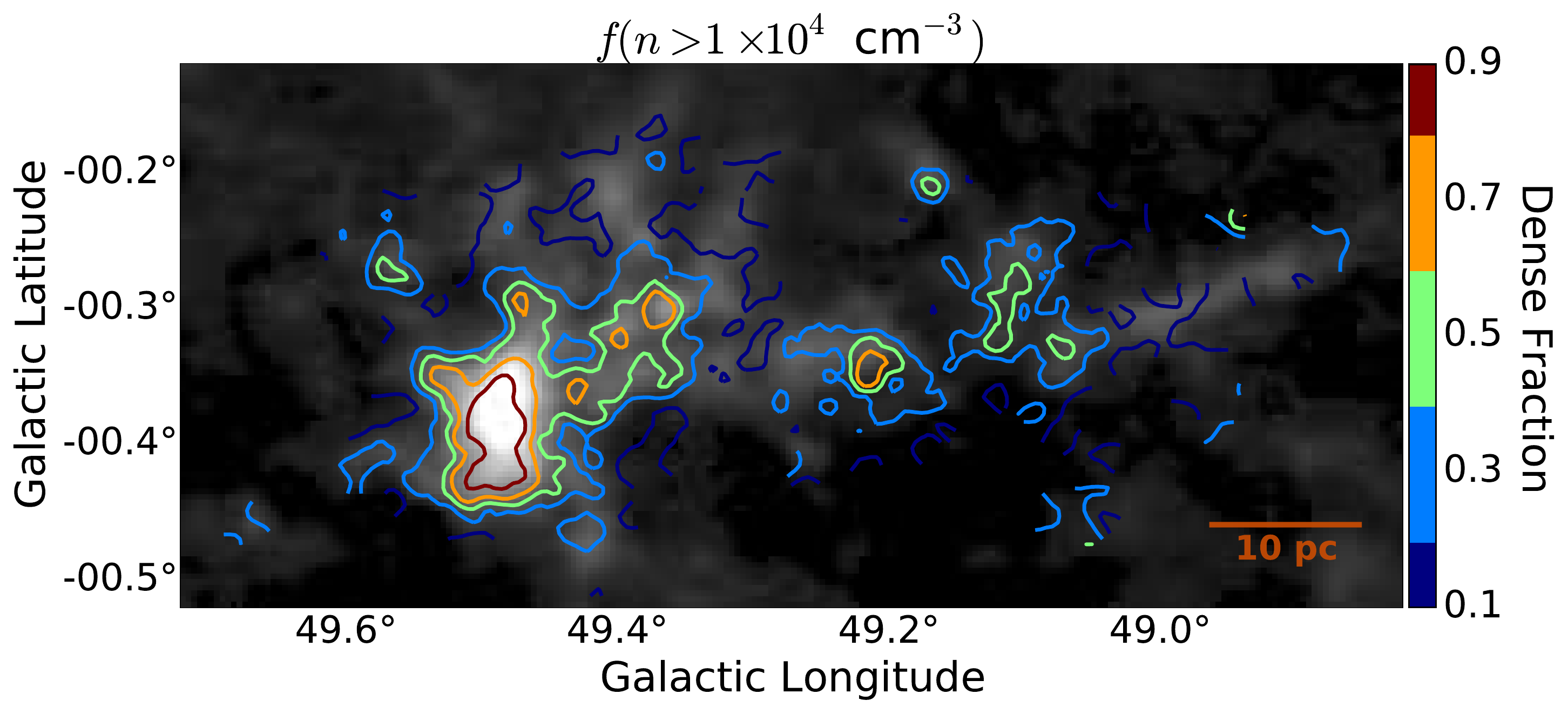}
{Contours of the dense gas mass fraction using two different thresholds
overlaid on the integrated \thirteenco map.  The regions with fraction $f>0.5$
should be rapidly forming stars.  The background image in both frames is the
GRS \thirteenco image integrated over the range $40~\kms < v_{lsr} < 75~\kms$,
masked to include only pixels with $T_B>0.5$ K.}
{fig:dgmf_contours}{1.0}{6.5in}

The multi-density DGMF presented here can be compared to models of `global'
collapse in which progressively more gas should be observed in denser structures
over time.  They are effectively a gas density cumulative distribution function.
However, to understand the systematic effects of line-of-sight stacking of
different velocity components (and corresponding radiative transfer issues),
similar analysis should be performed on hydrodynamic simulations.

\subsubsection{Dense Gas Fraction assumptions and caveats}

The DGMF analysis relies on the \thirteenco being optically thin and thermally
excited, both of which are generally good assumptions for the majority of the
mass.  The molecular cloud probably includes no more than $\sim20\%$ of its mass
in \twelveco-dark gas \citep{Pineda2013b,Langer2014b,Smith2014b}, which adds little to
the overall uncertainty.  Similarly, we expect that there is little 
\thirteenco-dark molecular gas \citep[the ratio of \twelveco to \thirteenco
should not vary significantly in molecular clouds;][]{Visser2009a}.

In Section \ref{sec:densmaps}, we discussed the various caveats and issues
related to \formaldehyde density fitting.  To account for the full range of
errors in that analysis, we have plotted the DGMF calculated using the minimum
and maximum values of the \formaldehyde-derived density consistent with the
data at the $1\sigma$ level in Figure \ref{fig:dgmf} and
\ref{fig:dgmf_regions}.

We have assigned \emph{all} of the mass associated with a given PPV voxel with
a single, fixed density in this analysis.  There is certainly some mass at a
lower density in each PPV pixel associated with each voxel along that line of
sight.  This additional mass biases the measured DGMF upwards,
but probably only by a small amount at each
threshold.  This systematic bias can be better characterized by performing a
similar analysis on molecular cloud simulations projected into PPV space
\citep[as demonstrated for other analysis techniques by][]{Beaumont2013a}.

\section{Discussion}
\label{sec:discussion}
The W51 cloud complex includes a full range of star forming conditions.  In the
west, W51 B, there is an older generation of stars including at least one
supernova remnant.  In the east, there is a pair of forming, still-embedded
massive clusters.  We have described the geometry of these regions and features
of the cloud structures, now we speculate on the broader implications of these
observations.

The gas in the W51 B region, while clearly impacted by the expanding W51 C
supernova, is less dense than most of the gas in the W51 A region.  The
supernova feedback is, if anything, destructive; a `collect-and-collapse'
scenario does not fit the observed gas structure since there is less dense
gas in the vicinity of the SNR.  Given the large radius of the W51 C
SNR, $\gtrsim100$ pc, collection and the early stages of collapse should
have happened by now if they are to happen at all.

The proximity of the 68 \kms filamentary `high velocity stream' and the W51
Main protocluster and their relative line-of-sight positions have been
presented as evidence for a cloud-cloud collision \citep{Kang2010a}.
Examination of the \formaldehyde line ratios has shown that the protocluster is
embedded in the $\sim55$ \kms molecular cloud (e.g. Figure
\ref{fig:northsouthpv}).  The velocity difference and their relative positions
along the line of sight unambiguously indicate that these
components are approaching each other, which is consistent with the cloud-cloud
collision hypothesis, though the distinction of these velocity components
as individual clouds is somewhat arbitrary since they are components of the same
hierarchical medium.

It is possible that the 68 \kms cloud is streaming in to a spiral arm, while
the 50 \kms cloud is slowed down as it exits the spiral arm on the far side.
In this scenario, W51 Main is in the deepest part of the spiral arm potential.
Gas is accumulating at W51 Main,  becoming compressed and undergoing a
``mini-starburst".  The W51 B/C region, with its mature HII
regions and SNR, represents a slightly older generation than W51 Main.

The line-of-sight length of the W51 complex is still uncertain, despite our
constraints on the relative geometry of different regions.  The best prospect
for resolving the line-of-sight structure of the region is via precise
constraints on distances to the individual regions.  Spectrophotometric surveys
of the individual stellar sub-clusters may be able to provide this and should
be undertaken.  Maser parallax observations of different zones may also provide
differential distance estimates.

\subsection{Gas density and its relation to star formation}
To assess the relation between gas density and star formation rate or
efficiency, one can multiply the DGMF by the total mass to get a naïve
measurement of the total mass directly involved in star formation (Figure
\ref{fig:dgmf_contours} transformed to Figure \ref{fig:sfmassmap}).  We have
compared this map to the distribution of Class I and flat-spectrum Spitzer YSO
candidates from \citet{Kang2009a} in Figure \ref{fig:sfmassmap}.  While there
are some regions of decent agreement between the YSO density and the
star-forming gas mass, e.g.  in the W51 B cloud and some of the more diffuse
filaments, the densest pockets of star-forming gas contain few or no YSOs.
Spitzer mid-infrared sources are either too confused or obscured to be detected
in these regions.  The Herschel images are also too crowded to identify a full
sample of individual YSOs.

We also compare the star-forming gas surface density to a 21+24
\um-derived star formation rate surface density in Figure \ref{fig:sfmassmap}
using standard extragalactic SFR calibrations
\citep{Price2001a,Carey2009a,Rieke2009a,Kennicutt2012a}\footnote{The same
exercise was performed with the 2 cm radio continuum data yielding broadly
consistent results.}.  The map was created
by filling the saturated regions of the 24 \um MIPS map with empirically scaled 
21 \um values.
The star formation rate density is only well-correlated with the dense gas
surface density in the central portion of W51 A.  At lower gas surface
densities, the SFR and gas are nearly anticorrelated.  This offset is due to an
age difference: the gas is tracing star formation yet to begin up to $\sim1$
Myr, while the infrared emission traces older \citep[0-100 Myr, with a peak at
5 Myr;][]{Kennicutt2012a} star formation.  The small-scale anticorrelation
is consistent with the 5-20\arcmin scales required to recover a star formation
`law' from Galactic plane survey observations \citep{Vutisalchavakul2014a}
because of the \citet{Kruijssen2014a} `uncertainty principle for star
formation'.
The improved correlation at the highest densities also implies that the massive
star formation threshold is closer to $10^5$ \percc than the $10^4$ \percc
often measured for nearby, low-mass star-forming regions.

In order to evaluate star formation efficiency as a
function of dense gas fraction, a more complete assessment of the present-day
or very recent, $t<5$ Myr, star formation in the W51 clouds is needed.

\FigureTwoAA
{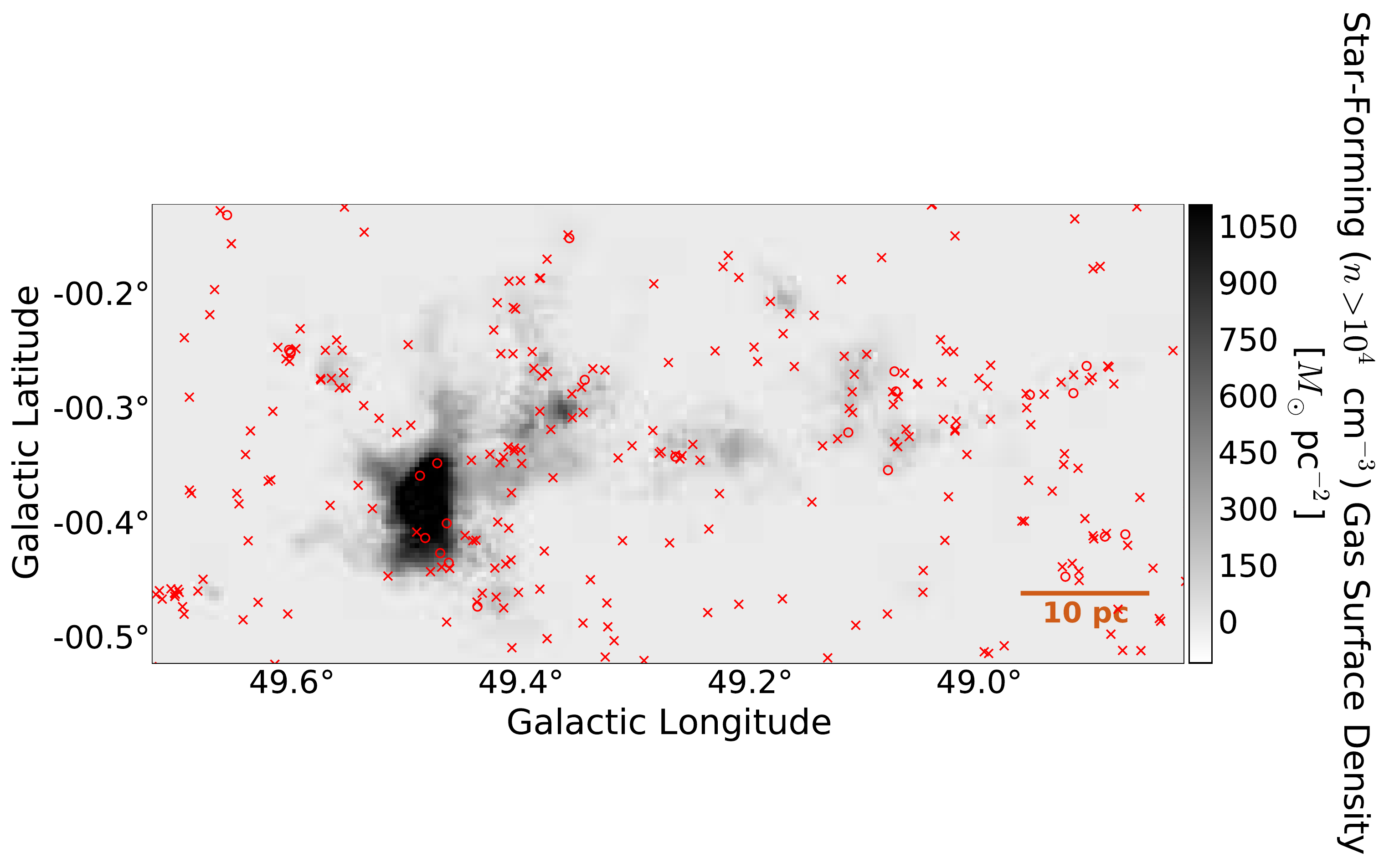}
{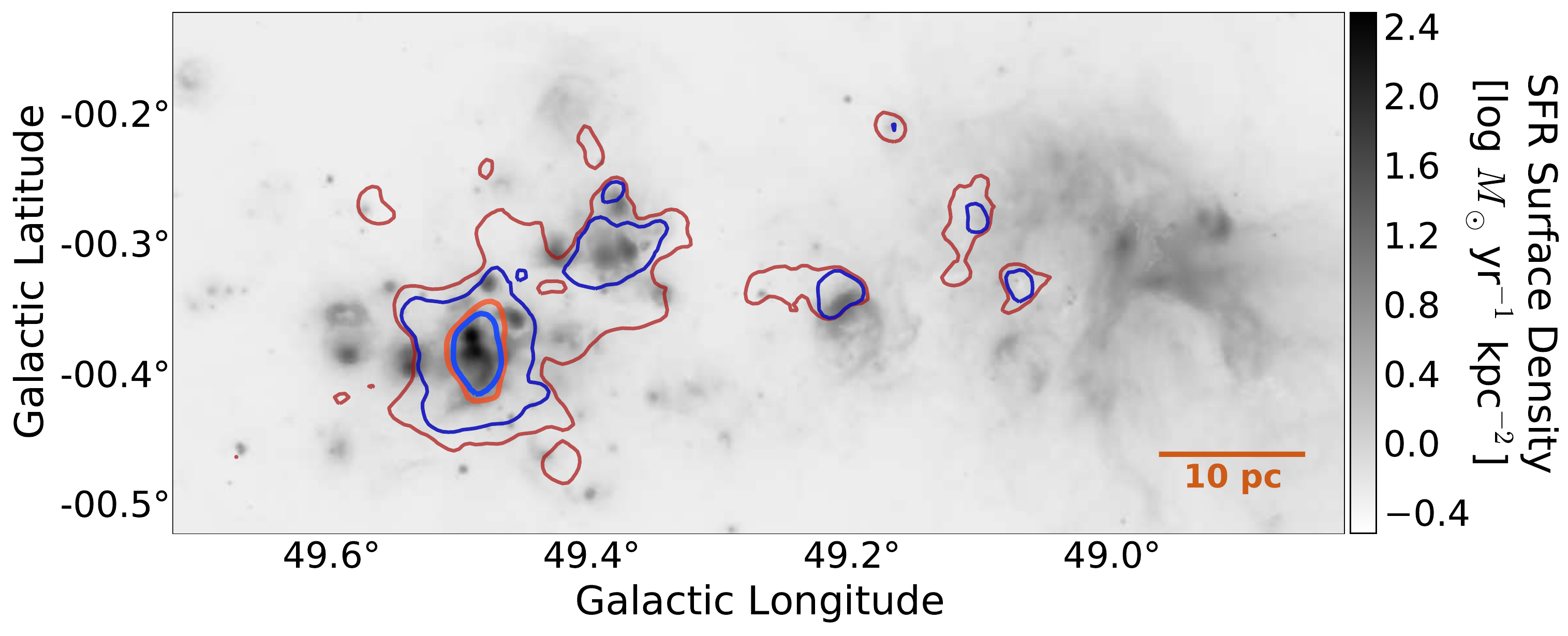}
{(top) A `star-forming gas mass' map, created by multiplying the dense gas
fraction by the integrated \thirteenco mass surface density.  The red
symbols are Class I and flat-spectrum YSOs from \citet{Kang2009a} with
$M<10 \msun$ ($\times$'s) and $M>10 \msun$ (circles).  They are
absent from the highest density regions.
(bottom) A star formation rate surface density map computed from the MSX
21 cm and MIPS 24 \um \citep{Price2001a,Carey2009a} data with a star formation
rate computed using the \citet{Rieke2009a} calibration \citep{Kennicutt2012a}.
The contours show the star-forming gas surface density above a threshold
$n>10^4$ \percc (red) and $n>10^5$ \percc (blue) at levels of [100,1000] \msun
kpc$^{-2}$.  At low star-forming gas surface density,
the star formation surface density is essentially anticorrelated with the gas
surface density.  The two converge at the highest surface densities and at the highest
volume densities.
}
{fig:sfmassmap}{0.5}{6.5in}

\subsection{The future evolution of W51}
\label{sec:futureev}
The low DGMF associated with the W51 B cloud indicates that it has a
low star formation potential despite containing significant mass
($M\gtrsim1\ee{5}$ \msun).  The presence of a supernova remnant and old,
diffuse \hii regions indicates that the cloud did previously (and recently) form
stars, but is now being destroyed.  The fact that this cloud contains $10-30\%$
of the total mass of W51, but has only a tiny fraction of its total mass above
the purported star forming thresholds supports this story.

This cloud is therefore a good region to examine the effects of different
feedback mechanisms (radiation, ionization, and supernovae) in parallel.  It
may also be a good location to examine how star formation comes to an end in
the presence of massive star feedback.

By contrast, the W51 Main region has a dense gas fraction $\sim1$ in its
center, or at least $\sim50\%$ out to nearly 5 pc.  It has presumably only
formed a small fraction
of its total potential.  It exceeds all of the various star formation and
massive star formation thresholds
\citep[e.g.][]{Lada2010a,Krumholz2008a,Kauffmann2010a}, and therefore is
expected to form additional stars efficiently, up to $\sim10^4-10^6$
\msun.

\subsection{Implications for extragalactic observations of \formaldehyde}
\label{sec:exgal}
Although W51 is one of the most massive and active GMCs in the galaxy,
containing 7\% of the present-day massive star formation galaxy-wide
\citep{Urquhart2014a}, its star-forming gas mass is predominantly at a moderate
density, $n\sim5\ee{4}$ \percc; there is very little gas above $10^6$ \percc even
in W51 Main.  There were \emph{no detections} of \formaldehyde emission on the
$\sim1.25$ pc (50\arcsec) scales observed.  

The nondetection is not just a geometric effect,
as the only gas within W51 that is capable of emitting is tightly associated
with the deeply embedded \uchii regions within W51 Main.  The dense gas is
spatially compact and very likely to reside along the line of sight toward the
continuum sources independent of viewing angle.  This geometry is confirmed by
existing interferometric observations that also show only absorption
\citep{Martin-Pintado1985a,Arnal1985a} and observations in other tracers
showing that the dense gas is associated on $\sim0.1$ pc scales with the
\uchii regions \citep{Zhang1997a,Zhang1998a}.

By contrast, in extragalactic observations of starburst galaxies, there have
been detections of \formaldehyde emission on $\sim100$ pc scales.
\citet{Mangum2013a} report
detections of \formaldehyde \oneone emission in NGC 3079, IC 860, IR
15107+0724, and Arp 220 on $\sim10$ kpc scales.  The implied local column
densities from their analysis are modest, but the densities are extreme: their
observations imply that the local-scale \emph{chemical} conditions are
comparable to W51, but the densities are different.

The only location in which densities $n\gtrsim5\ee{5}$ \percc (comparable to
$n(\textrm{Arp } 220)$, etc.) are observed in W51 are in the central W51
Main region.  We do not observe emission because of the bright
continuum background source.  It is therefore not possible to explain a
\formaldehyde-emission galaxy by constructing it from collections of \uchii
regions; such a galaxy would be continuum-bright and show only \formaldehyde
absorption.  Instead, they must be assembled from huge quantities of
high-density, non-star-forming gas.  The presence of this gas in turn
implies that the density threshold for star formation must be higher in
starburst galaxies.
This result is in contradiction to the
idea that Giant \hii Regions are the `building blocks' of starburst galaxies
\citep[e.g.][]{Miura2014a}.

One interpretation of the difference between the Galactic and
extragalactic \formaldehyde is that the \hii regions that form in starburst
galaxies have their radio luminosity significantly suppressed.  The most
straightforward explanation for a lower luminosity is that they are smaller and
optically thick, compressed by a much more massive and overpressured
ISM.  Tightly squeezed \hii regions such as these are locations where radiation
pressure is likely the dominant form of feedback, with multiple photon
scatterings transferring additional momentum to the surrounding medium as
described in \citet{Murray2010b}; this scenario does not occur in the optically
thin \hii regions in the Galaxy where ionized gas pressure dominates.

\section{Conclusion}
\label{sec:conclusion}
We have presented maps of the \formaldehyde \oneone and \twotwo and H77$\alpha$
and H110$\alpha$ lines covering the W51 star forming complex and used these maps
to examine the geometry and density structure of the complex.  For the
recombination lines, see Appendix \ref{sec:contrrldata}.

The \formaldehyde \oneone/\twotwo line ratio was used to measure gas volume
densities and dense gas mass fractions.  The W51 protoclusters have the majority
($>70\%$) of their mass in gas with density $n(\hh)>5\ee{4}$ \percc.  The rest of
the cloud has a small dense gas fraction, with $f(n>1\ee{4}~\percc) \sim 10\%$.
The W51 B cloud therefore appears to be at the end of its star-forming
lifetime, while W51 A will continue to form stars efficiently in the
future.  The relative weakness of present and future star formation
in the W51 B/C region suggests that the `collect-and-collapse' mechanism
is operating inefficiently or not at all.

Present-day high-mass star formation is associated only with W51 A, in
which most of the molecular gas has density $n\gtrsim1\ee{5}$ \percc.  The
highest gas density is closely associated with bright mid-infrared emission.
W51 B and the outskirts of the W51 cloud have some gas with $n>10^4$ \percc,
but these regions exhibit limited star formation and show infrared emission
anticorrelated with the dense gas.  The tighter correlation with massive-star
driven star formation indicators at high densities suggests that the density
threshold for high-mass star formation is higher than that for low-mass star
formation.

The \formaldehyde lines and their ratios have also been used to constrain the
geometry of the W51 GMC and the associated \hii regions.  The Galactic-face-on
view of W51 is presented in more detail than has previously been possible.
Analysis of our \formaldehyde data lead to the following conclusions about the
structure of the GMC:
\begin{itemize}
    \item The high velocity 68 \kms cloud is in front of the 51 \kms cloud
        and the rest of the W51 GMC complex.
    \item The most luminous clusters and associated HII regions are in between
        the 51 \kms and 68 \kms clouds.
    \item There is molecular absorption associated with W51 B both in front of
        and behind the W51 C supernova remnant; W51 C is therefore within the
        W51 B cloud.
\end{itemize}
It is possible that the foreground 68 \kms cloud is falling into the spiral
potential from the near side, interacting with the 51 \kms cloud exiting the
potential on the far side to produce the W51 protoclusters.  

Finally, we did not detect any \formaldehyde emission throughout the entire W51
GMC.  The nondetection on scales from $\sim1-100$ pc implies that detections of
\formaldehyde \oneone emission in other galaxies comes from gas that does not
surround bright \hii regions.  These galaxies may therefore have interstellar
media dominated by very high-density gas ($n(\hh)>10^{5.5}$ \percc) that is not
presently forming stars.
The density threshold for star formation in these galaxies must
therefore be larger than in the Galactic disk, confirming earlier empirical
\citep{Longmore2013b} and theoretical
\citep{Krumholz2005a,Hennebelle2013a,Padoan2011b,Federrath2012a} results
that such a threshold cannot
be universal. These results are inconsistent with a `universal' density
threshold for star formation observed in studies of nearby clouds
\citep{Lada2010a,Lada2012a,Andre2013c}.

The data are made available in FITS cubes and images hosted at the CfA
dataverse \url{doi:10.7910/DVN/26818},
\url{http://thedata.harvard.edu/dvn/dv/W51_H2CO}.
The entire reduction and analysis process and all associated code and scripts
are made available via a git
repository hosted on github:
\url{https://github.com/keflavich/w51_singledish_h2co_maps}, with a
snapshot of the publication version available from zenodo
\url{http://dx.doi.org/10.5281/zenodo.11737}.

\textbf{Acknowledgements}:
We thank Xiaohui Sun for providing the Urumqi 6 cm Stokes I image prior to its
availability on the survey website.  The paper benefitted from discussions
with Jonathan Tan, Neal Evans, and Diederik Kruijssen.
We thank our referee Jeff Mangum for a helpful and rapid referee report.
We are grateful to the editor, Malcolm Walmsley, for additional comments
and for catching many small but significant errors.
ER is supported by a Discovery Grant from NSERC of Canada.
This research has made use of the VizieR catalogue access tool, CDS,
Strasbourg, France. The original description of the VizieR service was
published in \citep{Ochsenbein2000a}.
This publication makes use of molecular line data from the Boston
University-FCRAO Galactic Ring Survey (GRS). The GRS is a joint project of
Boston University and Five College Radio Astronomy Observatory, funded by the
National Science Foundation under grants AST-9800334, AST-0098562, AST-0100793,
AST-0228993, \& AST-0507657.
This research has made use of the NASA/ IPAC Infrared Science Archive, which is
operated by the Jet Propulsion Laboratory, California Institute of Technology,
under contract with the National Aeronautics and Space Administration.

\textbf{Code Bibliography:} 
\begin{itemize}
    \item The GBT KFPA Pipeline \url{https://safe.nrao.edu/wiki/bin/view/Kbandfpa/ObserverGuide}
    \item aoIDL \url{http://www.naic.edu/~phil/download/aoIdl.tar.gz}
    \item gbtidl \url{http://gbtidl.nrao.edu/}
    \item astropy \url{www.astropy.org} \citep{Astropy-Collaboration2013a}
    \item astroquery \url{astroquery.readthedocs.org} (http://dx.doi.org/10.5281/zenodo.11656)
    \item sdpy \url{https://github.com/keflavich/sdpy}
    \item \texttt{FITS\_tools} \url{https://github.com/keflavich/FITS_tools}
    \item aplpy \url{http://aplpy.github.io}
    \item image-registration \url{http://image-registration.rtfd.org}
    \item wcsaxes \url{wcsaxes.rtfd.org} ($>=$0.3.dev409)
    \item pvextractor \url{pvextractor.rtfd.org}
    \item agpy \url{https://code.google.com/p/agpy/}
    \item pyspeckit \url{pyspeckit.bitbucket.org} \citep{Ginsburg2011c}
    \item ipython \url{http://ipython.org/} \citep{Perez2007a}
\end{itemize}

\clearpage
\appendix
\section{Continuum and RRL Data}
\label{sec:contrrldata}
We compare various data sets to assess calibration uncertainties and
provide details of reduction for archival purposes.  The data presented in this
section are suitable for comparisons of Galactic to extragalactic star
formation rate measurements, for example, since they are among the largest
angular scale maps of radio recombination lines available with sub-parsec
resolution.

\subsection{Comparison between GBT and GPA data}
\label{sec:gpacompare}
The Galactic Plane ``A'' survey \citep{Langston2000a} covered the Galactic
plane at 14.35 GHz using the Green Bank Earth Station (GBES) 13.7m telescope,
with a reported FWHM beam size of 6.6\arcmin. 
The published images were released with a
FWHM resolution of 8\arcmin.  We compared our GBT
continuum observations to theirs in order to determine whether a significant DC
component is missing from our data.
Because the GPA used $10\deg$ long scans in Galactic latitude, it should fully
recover all diffuse Galactic Plane emission.  In the released brightness temperature
maps, brightness down to a scale of $1.5\deg$ is recovered.  However, because
the GPA data undersampled the sky (its 5\arcmin steps between scans were
larger than the Nyquist sampling scale of the 14.35 GHz beam), point source
fluxes in the GPA are underestimated by 19\% and flux on small angular scales
may be unreliable.

We resampled the GPA image onto the GBT grid using cubic spline interpolation,
then smoothed both data sets to 9.5\arcmin.  There are image artifacts
(particularly vertical streaking) in the GPA data that are diminished by this
large smoothing kernel.

We compared the surface brightness in the GPA and GBT data, and found that the
GPA data was $\sim0.2$ K brighter than the GBT in the diffuse portion of the
W51 Main region; the offset is not consistent with a purely multiplicative
offset (Figure \ref{fig:2cmcompare}).  The GBT observed the W51 Main peak to be
moderately brighter, which is likely a result of the sparse sampling in the
GPA.  The morphological agreement between the maps is imperfect, perhaps in
part because of the small area mapped in our GBT data, though there also
appears to be vertical (along a line of constant longitude) stretching of the
W51 Main source in the unsmoothed GPA data that is not consistent with the GBT
observations.

\Figure{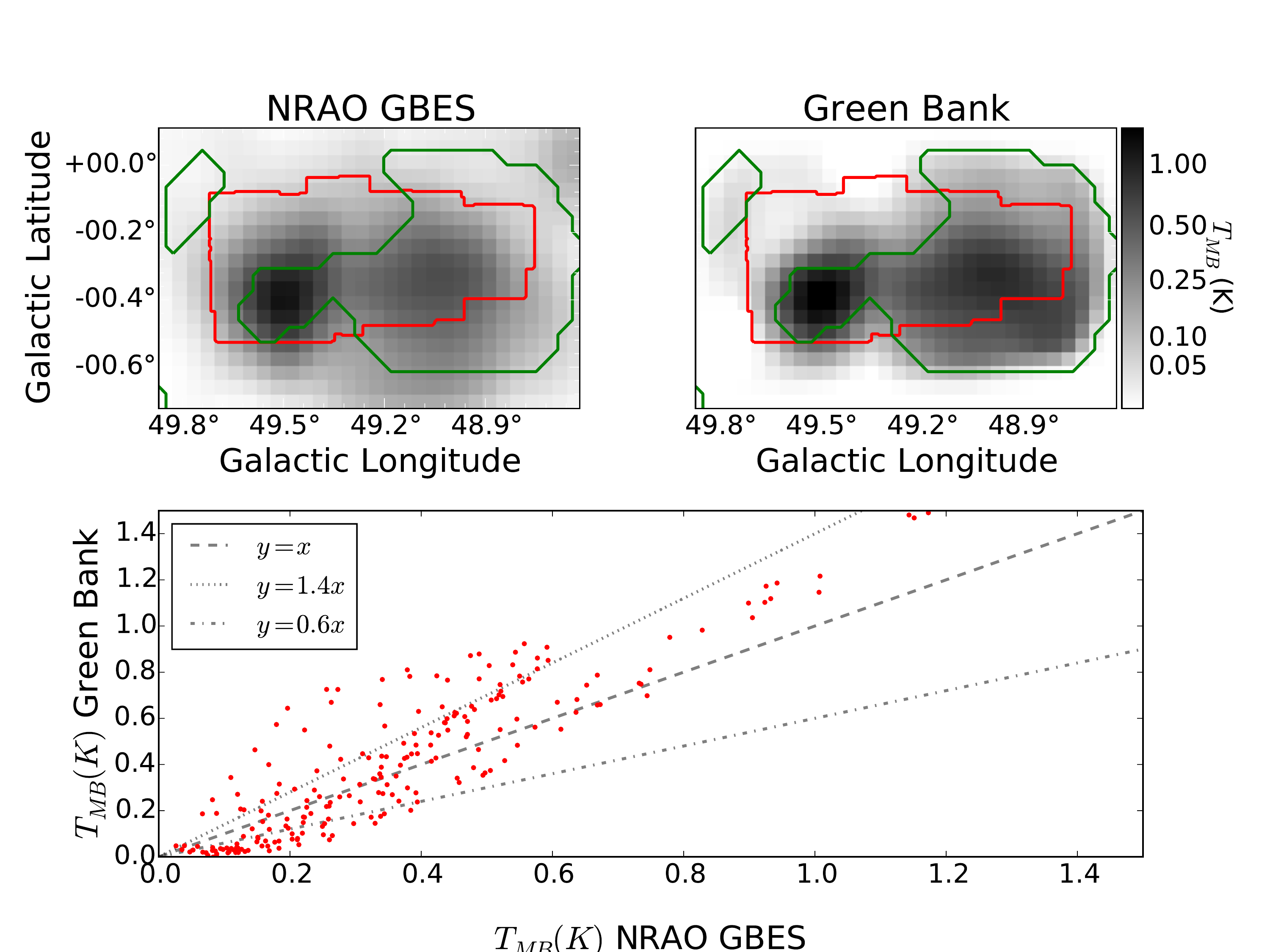}
{Comparison between the GBT and NRAO GBES \citep{Langston2000a} data.
(top left) NRAO GBES 2 cm map
(top right) GBT 2 cm map of the same region smoothed to about 8.9\arcmin. 
The colorbar applies to both figures,
showing brightness temperature units in K.  The red contours in both figures
show the region observed by Green Bank; flux outside of those boundaries is
extrapolated with the smoothing kernel.  The green contours show the region
where $T_B$(GBT)$>T_B$(GBES).  (bottom) Plot of the GBT vs the GBES surface
brightness measurements.
The large red dots show the region within the red contours.  
}
{fig:2cmcompare}{0.5}{0}

\subsubsection{Comparison between Arecibo and Urumqi data}
We compare the 6 cm continuum to the Urumqi 25m data from \citet{Sun2007a} and
\citet{Sun2011a}.  Figure \ref{fig:6cmcompare} shows the comparison of the
Urumqi data and the Arecibo continuum data smoothed to 9.5\arcmin resolution.
The Arecibo and Urumqi data agree well as long as the main beam
efficiencies of the respective telescopes ($0.5$ and $0.67$) are accounted for.

\Figure{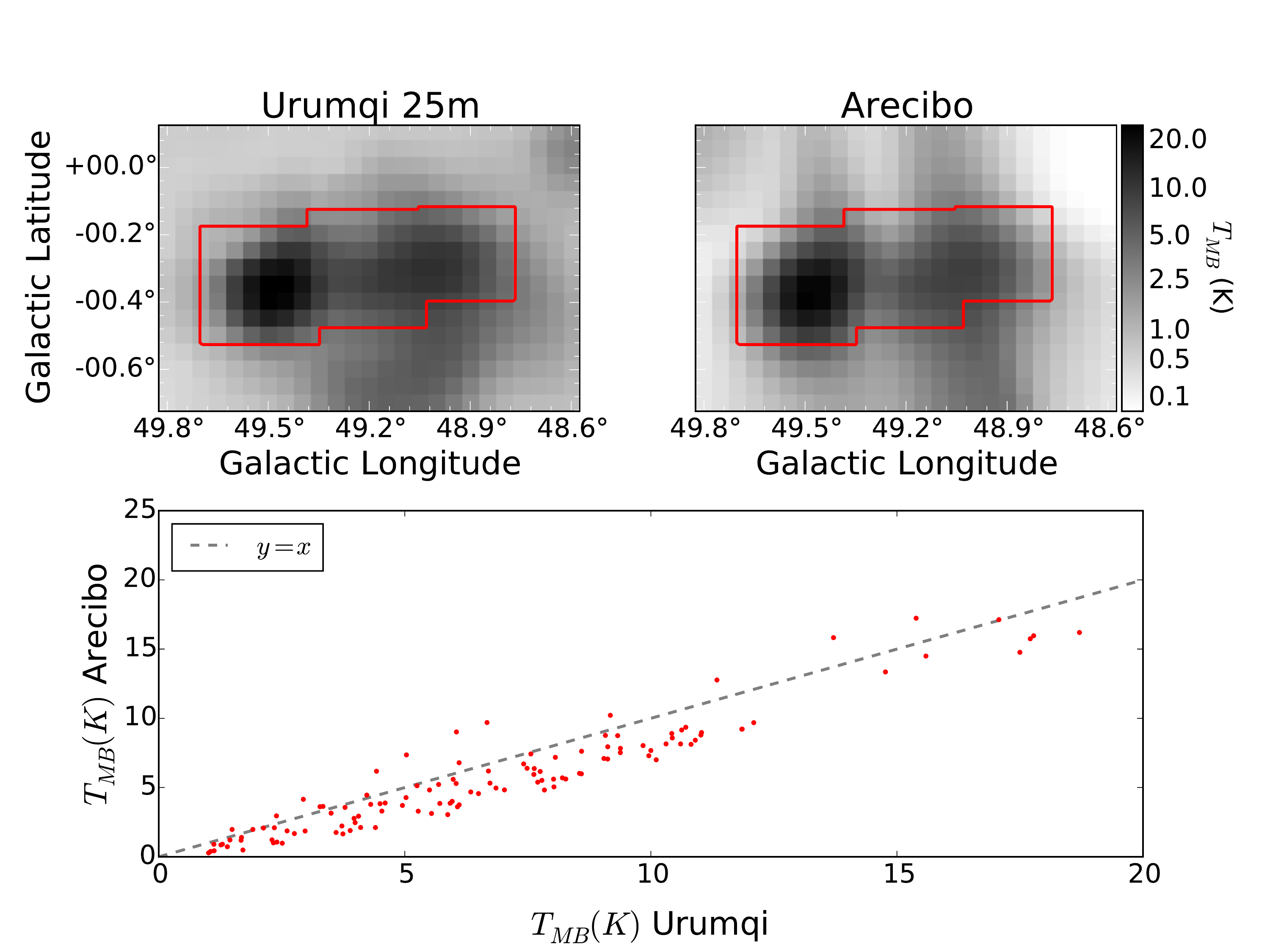}
{Comparison between the Arecibo and Urumqi 25m \citep{Sun2011b} data.
(top left) Urumqi 6 cm map of the W51 region.
(top right) Arecibo 6 cm map of the same region smoothed to the 9.5\arcmin
resolution of the Urumqi data set.  The colorbar applies to both figures,
showing brightness temperature units in K.  The red contours in both figures
show the region observed by Arecibo; flux outside of those boundaries is
extrapolated with the smoothing kernel. 
(bottom) Plot of the Arecibo vs the Urumqi surface brightness measurements.
The red dots show the region within the red contours.
}
{fig:6cmcompare}{0.5}{0}

\subsubsection{Comparison of GBT and Arecibo data}
In order to compare the Green Bank and Arecibo continuum data, we converted
the brightness temperature maps to Janskys assuming a beam FWHM of 50\arcsec
for both surveys and central frequencies of 4.8 and 14.5 GHz for Arecibo and
Green Bank respectively.  Measured beam widths for both telescopes were
$\sim49-54\arcsec$, so the relative error from assuming the same beam size
should be $\lesssim10\%$.
In this section, the target frequencies are referred
to as $S_{5 GHz}$ and $S_{15 GHz}$ for brevity.

The data are well-correlated, with $S_{5 GHz} \sim 1.4 S_{15 GHz}$ ($S_{15 GHz}
\sim 0.7 S_{5 GHz}$; Figure \ref{fig:gbtaocontcompare}), consistent with a
spectral index $\alpha_\nu=-0.3$
slightly steeper than usually observed for optically thin brehmsstrahlung and
consistent with there being some contribution from synchrotron emission.  The
lower-brightness regions have a lower $S_{15 GHz}/S_{5 GHz}$, indicating that
these regions are more affected by synchrotron.  In Figure
\ref{fig:contrrlcompare}a, a great deal of structure in the $S_{15 GHz}/S_{5
GHz}$ ratio is evident in the vicinity of W51 Main: the ratio is higher towards
the continuum peaks, indicating that the peaks have higher free-free optical
depths, or lower relative contributions from synchrotron emission, than their
envelopes.

We additionally compare the radio recombination lines observed simultaneously
with the continuum and \formaldehyde.  Hydrogen RRLs are often extremely
well-correlated with the continuum and are therefore good indicators of the
calibration quality.

In Figure \ref{fig:contrrlcompare}, we show the ratios between the two
frequencies in RRLs and continuum and the line-to-continuum ratios at both
frequencies.  The `line' values are the integrated flux densities over the
range 20 to 100 \kms, which includes all H$\alpha$ emission but no He$\alpha$.

The ratios between the $x$ and $y$ axis in each plot in Figure
\ref{fig:gbtaocontcompare} are fitted using a total least squares approach with
uniform errors for each data point.   The line-to-continuum ratio is
$L/C(H77\alpha)\sim0.15$ and $L/C(H112\alpha)\sim0.04$; in both cases there is
little evidence for deviation from a linear relationship.

\subsubsection{Comparison of the RRL and continuum data}
\label{sec:rrlvscont}
Radio recombination lines are generally observed to be well-correlated with the
corresponding radio continuum, particularly at low frequencies.  At 5 and 15
GHz, the population level departure coefficients are close to 1, $b_n > 0.95$
\citep{Wilson2009a,Walmsley1990a}.

While radio recombination lines are purely thermal in nature, the large-scale
continuum may include a contribution from synchrotron emission.  The
morphological similarity between the 90 cm and 4 m (meter - i.e., 74 MHz)
images presented by \citet{Brogan2013a} and our 6 and 2 cm data hint that
synchrotron emission could be significant.  However, the high degree of
correlation between the 2 and 6 cm described below suggest that synchrotron
`contamination' is minor at both wavelengths.

Figure \ref{fig:gbtaocontcompare} shows a comparison between the integrated RRL
surface brightness and radio continuum at both 2 and 6 cm\footnote{The H107,
108, 109, and 111$\alpha$ data were affected by missing (corrupted) data in one
segment of the map.  H107 and 108$\alpha$ were also affected by RFI.  We
therefore used the average of the H110 and H112$\alpha$ lines for the 6 cm line
ananylsis.}.  The figure shows
the total least squares best-fit slopes to the data assuming uniform error,
which yield a measurement of the line-to-continuum ratio.

We use the line-to-continuum ratio in both bands to measure the electron
temperature using Equation 14.58 of \citet{Wilson2009a}, which assumes a
plane-parallel, optically-thin emission region with lines formed in local
thermodynamic equilibrium (the $*$ in $T_e^*$ is meant to indicate these three
assumptions are made).  The two lines yield consistent measurements, with mean
$T_e^*\sim7000-8000$ K; these measurements are consistent with smaller-scale
measurements using the VLA with H92$\alpha$ \citep{Mehringer1994a}.  There is
little structure in the $T_e^*$ maps, with a hint of higher temperatures
around G49.1-0.4, coincident with the W51C supernova remnant.  Other structures
are most likely due to the limited S/N.

Finally, we fit a single-component Gaussian to each pixel to produce velocity
maps.  These are discussed in Section \ref{sec:kinematics}.

\subsection{Carbon and Helium RRLs}
Helium RRLs were prevalent and reasonably well-correlated with the hydrogen
RRLs, but we did not examine them in detail.  He77$\alpha$ is detected at much
higher signal-to-noise than than He107-112$\alpha$.  There were no clear
detections of C77$\alpha$ or C107-112$\alpha$, though there is a possible
C77$\alpha$ signal at G49.366-0.304 with $v_{lsr} \approx 55$ \kms and a
possible detection toward W51 Main along the wing of the He77$\alpha$ line.
The He77$\alpha$ line detections are associated with regions of high $H_n\alpha$
but not regions of different $T_e^*$.

\Figure{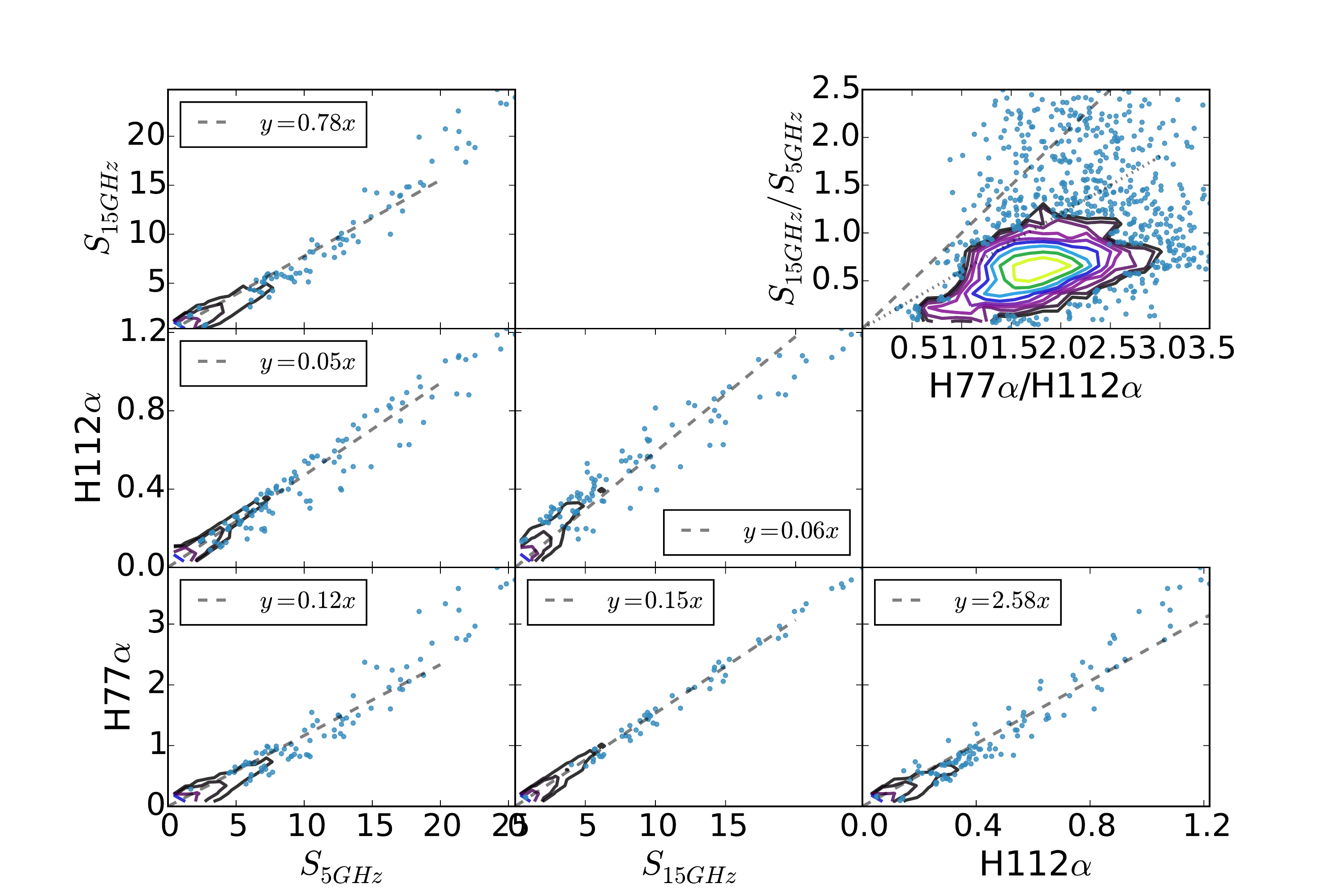}
{Plots of the 5 GHz and 15 GHz continuum and RRL flux densities against one
another; all units are in Jy.  The dashed lines show the total least squares
best fit line with the slope shown in the legend.  Wherever the density of
points is too high to display, the points have been replaced with a contour
plot showing the density of data points.  The upper-right panel shows a
comparison of the continuum ratio to the RRL ratio.  The dashed line in the
upper-right plot has slope 1, and the dotted line has slope 0.6.}
{fig:gbtaocontcompare}{0.5}{0}

\FigureFourPDF
{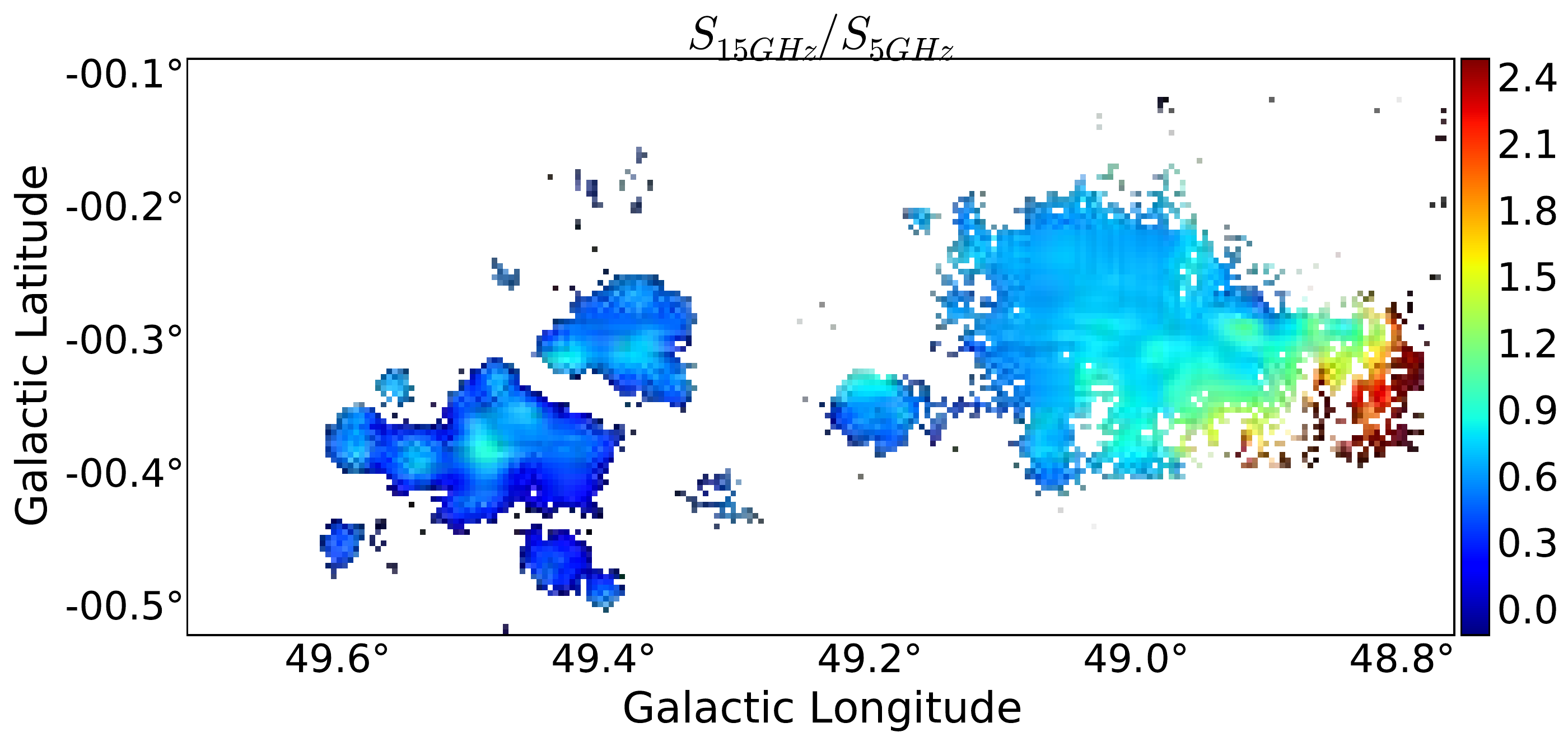}
{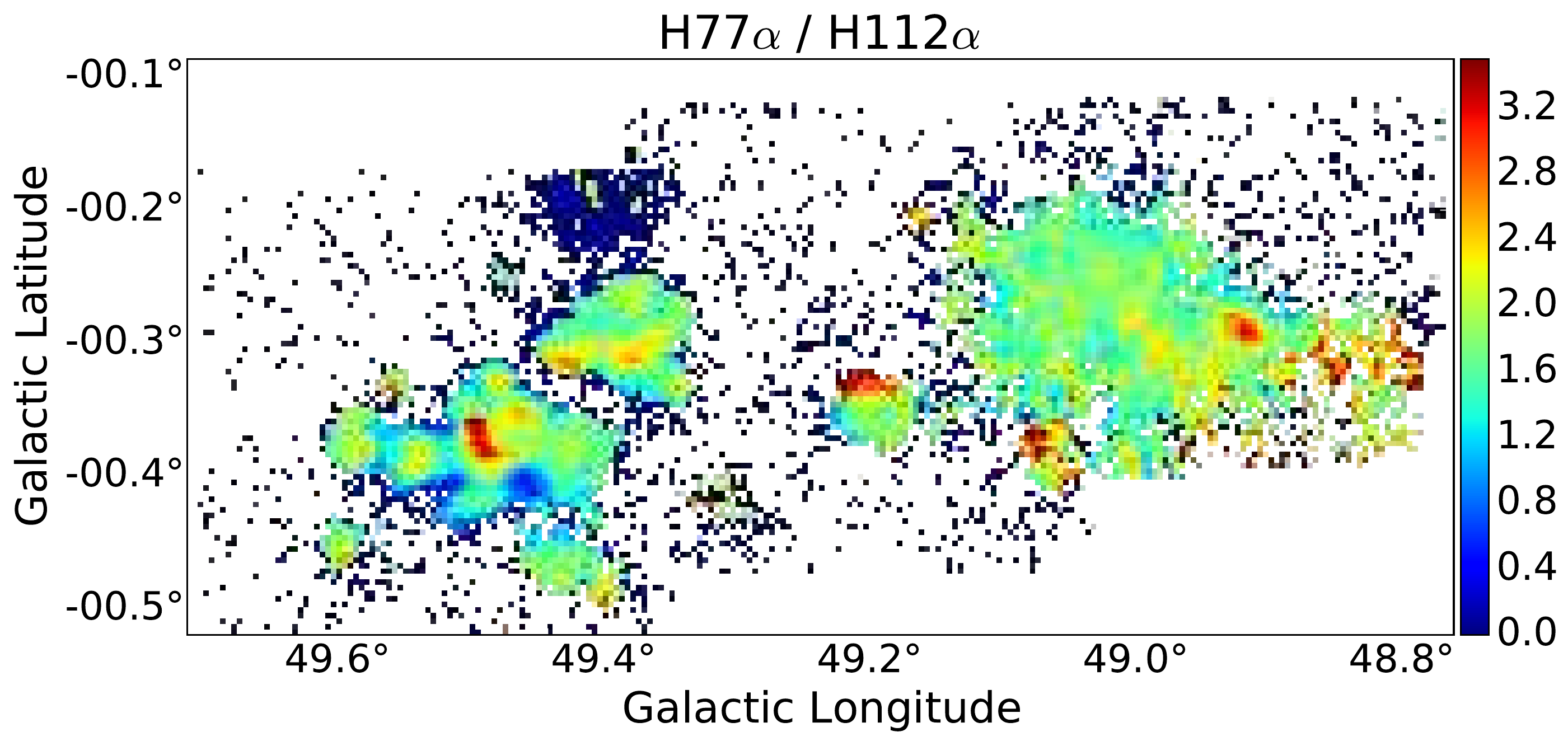}
{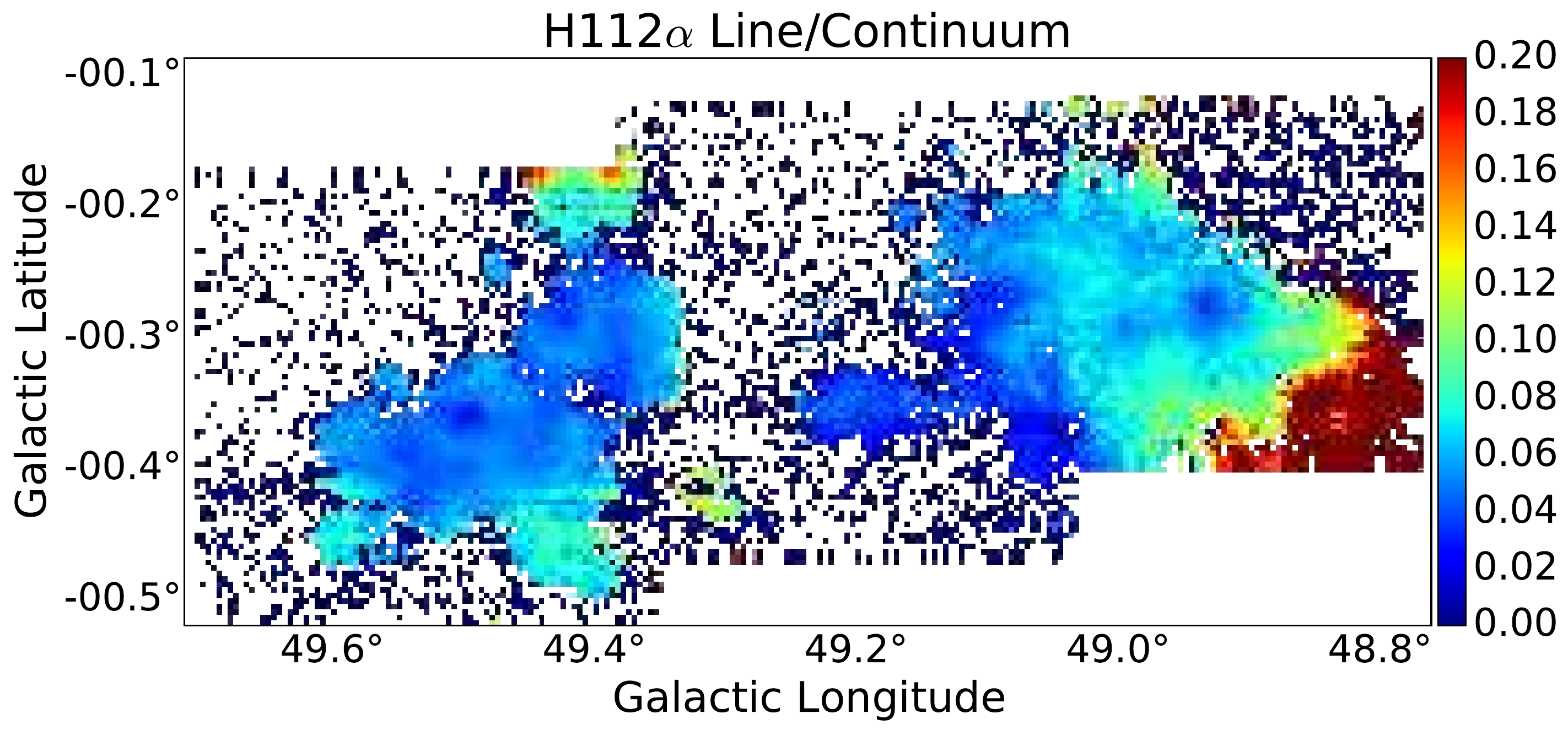}
{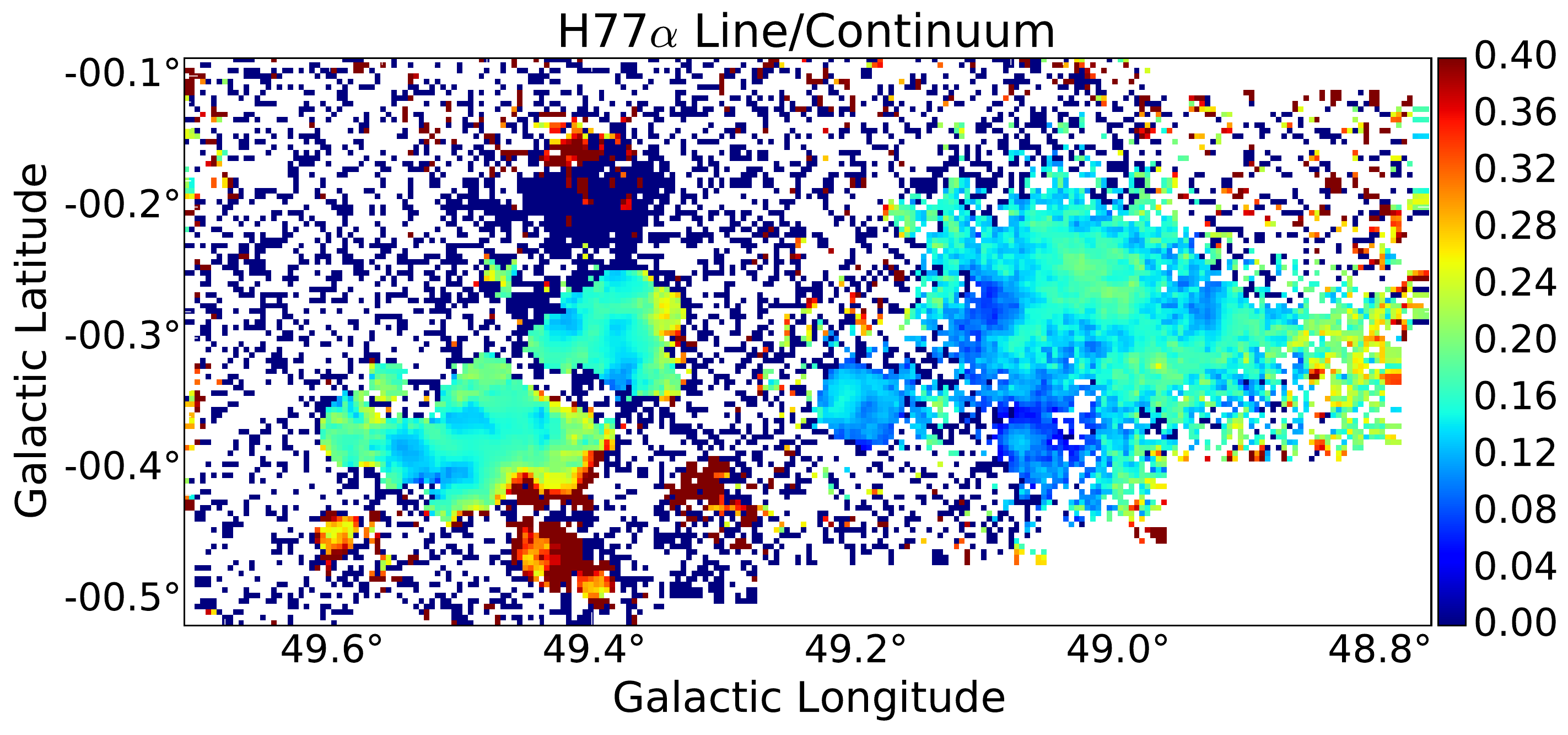}
{Ratio maps of the ionized gas in W51.  
(a) Continuum ratio $S_{15 GHz} / S_{5 GHz}$.  For $\alpha=-0.1$, an optically
thin free-free source, the ratio is 0.9, while for $\alpha=2$, an optically thick source,
the ratio is 9.
(b) The ratio of the H77$\alpha$ peak to the H112$\alpha$ peak.
(c) The line-to-continuum ratio H112$\alpha$ / $S_{5 GHz}$
(d) The line-to-continuum ratio H77$\alpha$ / $S_{15 GHz}$
}
{fig:contrrlcompare}

\FigureFourPDF
{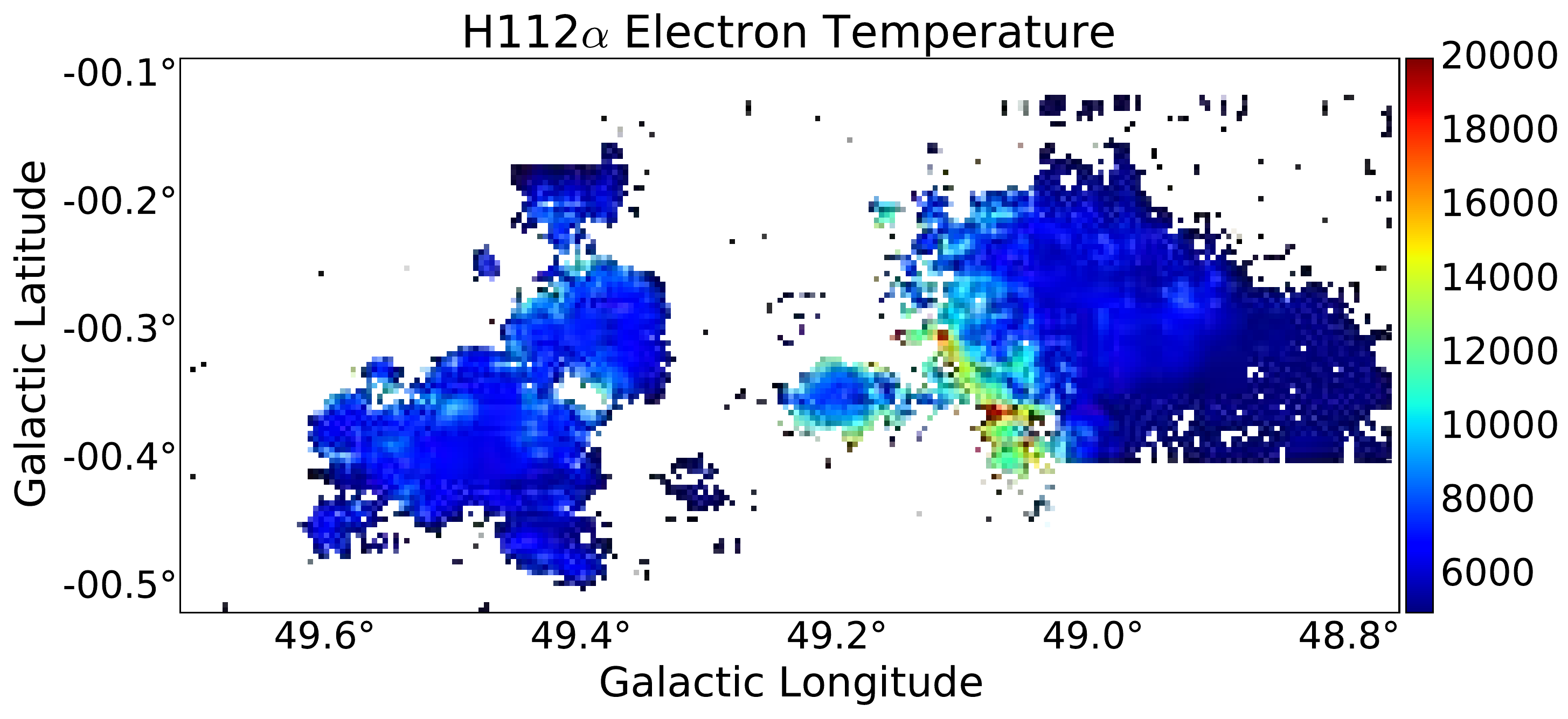}
{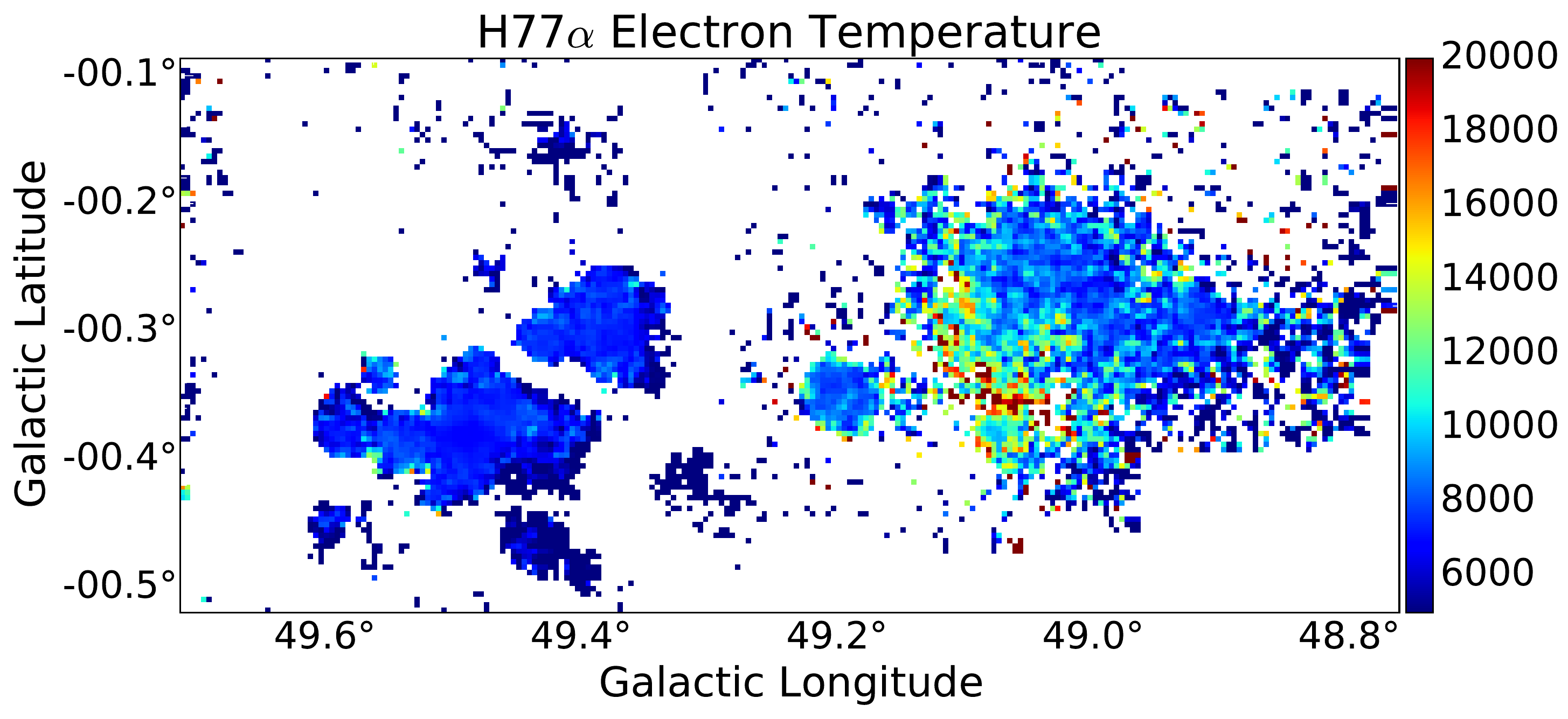}
{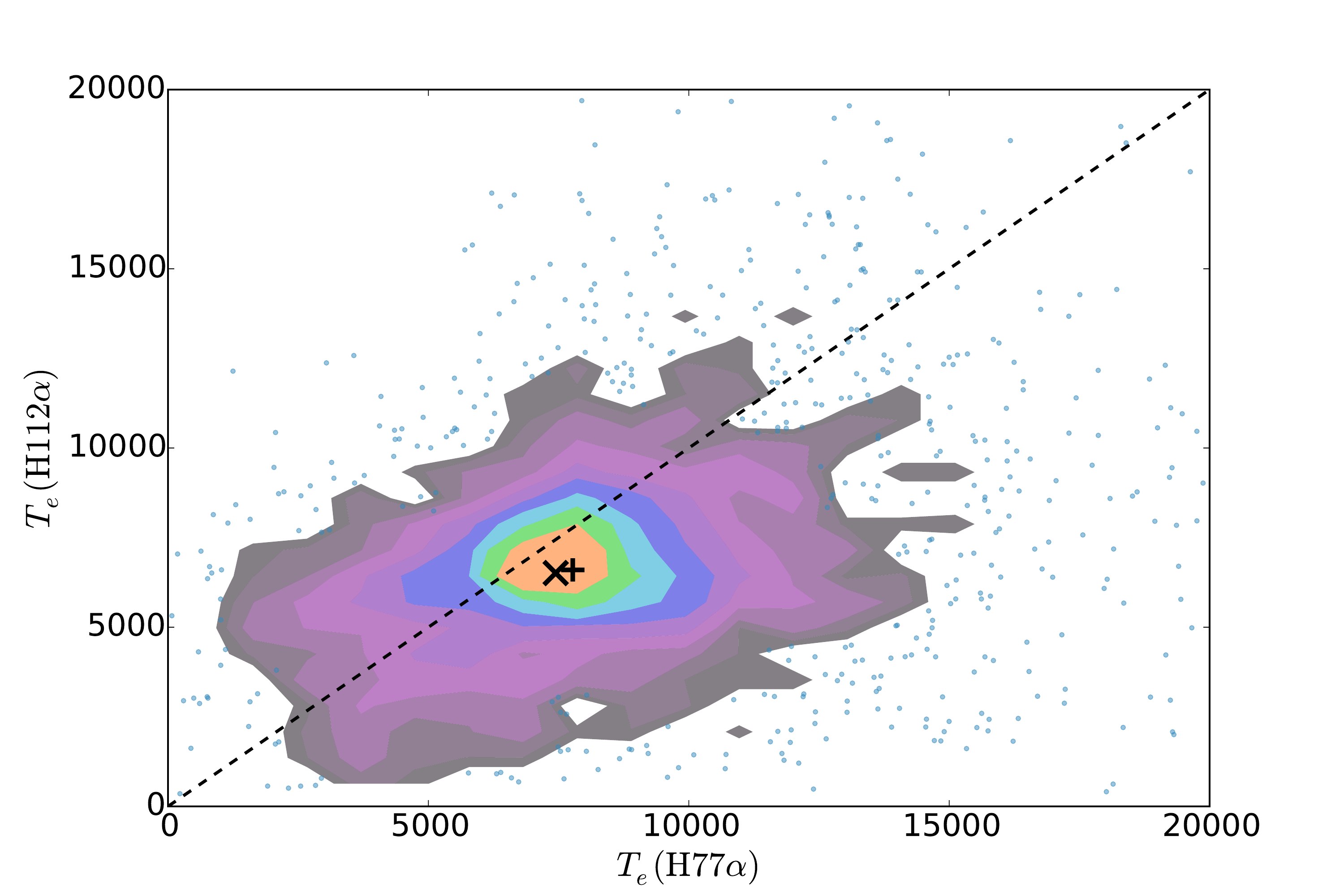}
{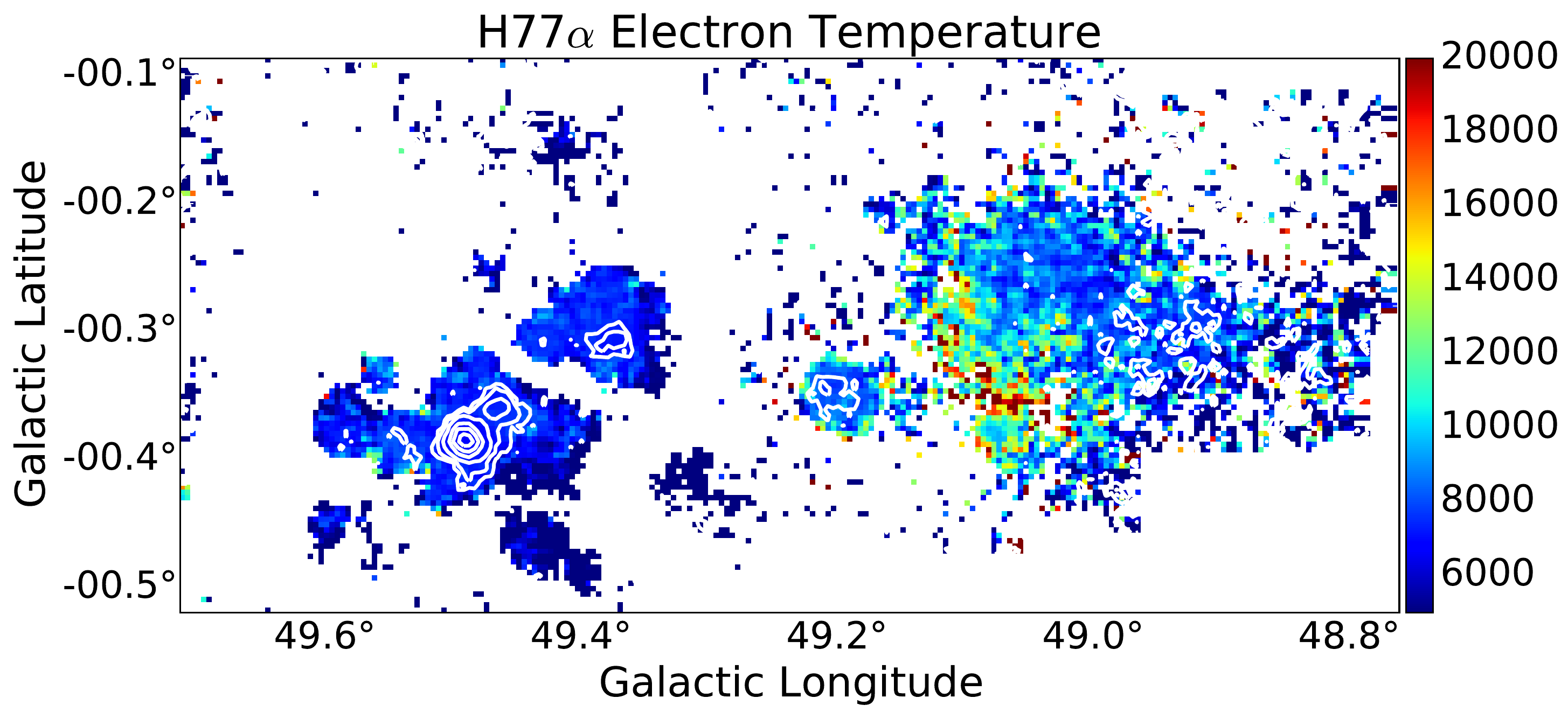}
{(a) The H112$\alpha$ electron temperature map showing $T_e^*$ in K. 
(b) The H77$\alpha$ electron temperature map showing $T_e^*$ in K.
(c) The measured electron temperature in the 6 cm vs the 2 cm band at each spatial
pixel with significant detected RRL emission.  The contours show regions of
increasing pixel density.  The $x$ marks the median and the $+$ marks the mean
over all valid pixels.
(d) Same as (b), but with integrated He77$\alpha$ contours at levels [0.0125,
0.025, 0.05, 0.1, 0.15, 0.2] K \kms overlaid.  The contours on the right side
($\ell<49$) most likely trace noise, since the noise in that region is higher.
}
{fig:tevste}

\section{Geometry of Individual Regions}
\label{appendix:geometry}
\subsection{W51 Main \& W51 IRS 2}
The W51 Main and IRS 2 spectra show that both have ionized gas components at
$v_{LSR}\sim 55$ \kms.  This velocity approximately coincides with the peak of
the \thirteenco emission.

The \formaldehyde \oneone spectra are deepest at $\sim68$ \kms, while the
\twotwo have depths approximately equal between the $\sim58$ \kms and $\sim 68
\kms$ components.  The 55-60 \kms components are too deep to be entirely behind
the \hii regions.  This indicates that the 55 \kms ionized gas must be embedded
within the molecular cloud, with molecular gas on \emph{both} sides of the
ionized gas along the line of sight.

Because these are well-studied regions, the low spatial resolution
\formaldehyde spectra we present here add little new information about the gas
kinematics.  However, all of the velocity components observed in the W51 region
are apparently kinematically connected to the W51 clusters.

\subsection{The W51 B Filament}
\label{sec:w51b}
The W51 B filament (right side of Figure \ref{fig:geosketch} at $\sim68$ \kms),
exhibits bright CO emission ($T_A^*\sim30-50$ K in the \citet{Parsons2012a} CO
3-2 data) but has relatively weak \formaldehyde absorption.
The absorption models are inconsistent with the molecular gas being in front of
the continuum emission, so Figure \ref{fig:geosketch} shows the continuum sources
in front of the cloud at lower $\ell$.  Figure \ref{fig:h2cofrontbackmodel}
shows an example model fit with the continuum assumed to be in front and in
back, illustrating that the best-fit model parameters with continuum in the
back do not reproduce the data.
The relative positioning of the molecular gas behind the \hii regions suggests
that the molecular gas is also behind the W51 C supernova remnant.

\subsection{The edge of W51 C}
W51 C is a supernova remnant that spatially overlaps with the W51 B star
forming region.  \citet{Brogan2013a} argue that the supernova remnant must be
in front of the \hii region G49.20-0.35 because the \hii-region has not absorbed
all of the 4m (74 MHz) nonthermal emission.  The G49.1-0.4, G49.0-0.3, and G48.9-0.3
regions, however, show 4m absorption signatures and may be in the foreground.
There are clumps aligned along the 68 \kms filamentary cloud with very high CO
and \hi velocities \citep{Koo1997b,Koo1997c,Brogan2013a}, indicating that the
SNR is interacting with the molecular gas.

The clumps at G49.1-0.3, $\sim68$ \kms are either lower density ($n<1.5\ee{4}$
\percc) and in the background of the \hii region or high density ($n>1.5\ee{5}$
\percc), low-column density and in the foreground.  The $62$ \kms clumps have
densities a few times higher, $n\sim4\ee{4}$ \percc, and are clearly in the
foreground of the continuum emission because their absorption depths are
$\sim2.5$ K, which cannot occur for absorption against the CMB.  Figure
\ref{fig:contbetween63kms} shows a model spectrum fitted assuming the continuum
lies between the two molecular velocity components.  The relative strength of
the \thirteenco and the \formaldehyde also suggests that the 68 \kms component
is behind the continuum.

We are seeing molecular gas both in front of and behind the supernova.  This
geometry can be readily confirmed by looking for molecular absorption at much
lower frequencies where the SN synchrotron emission dominates over the \hii
region free-free emission, i.e. the 335 and 71 MHz \para lines.

\Figure{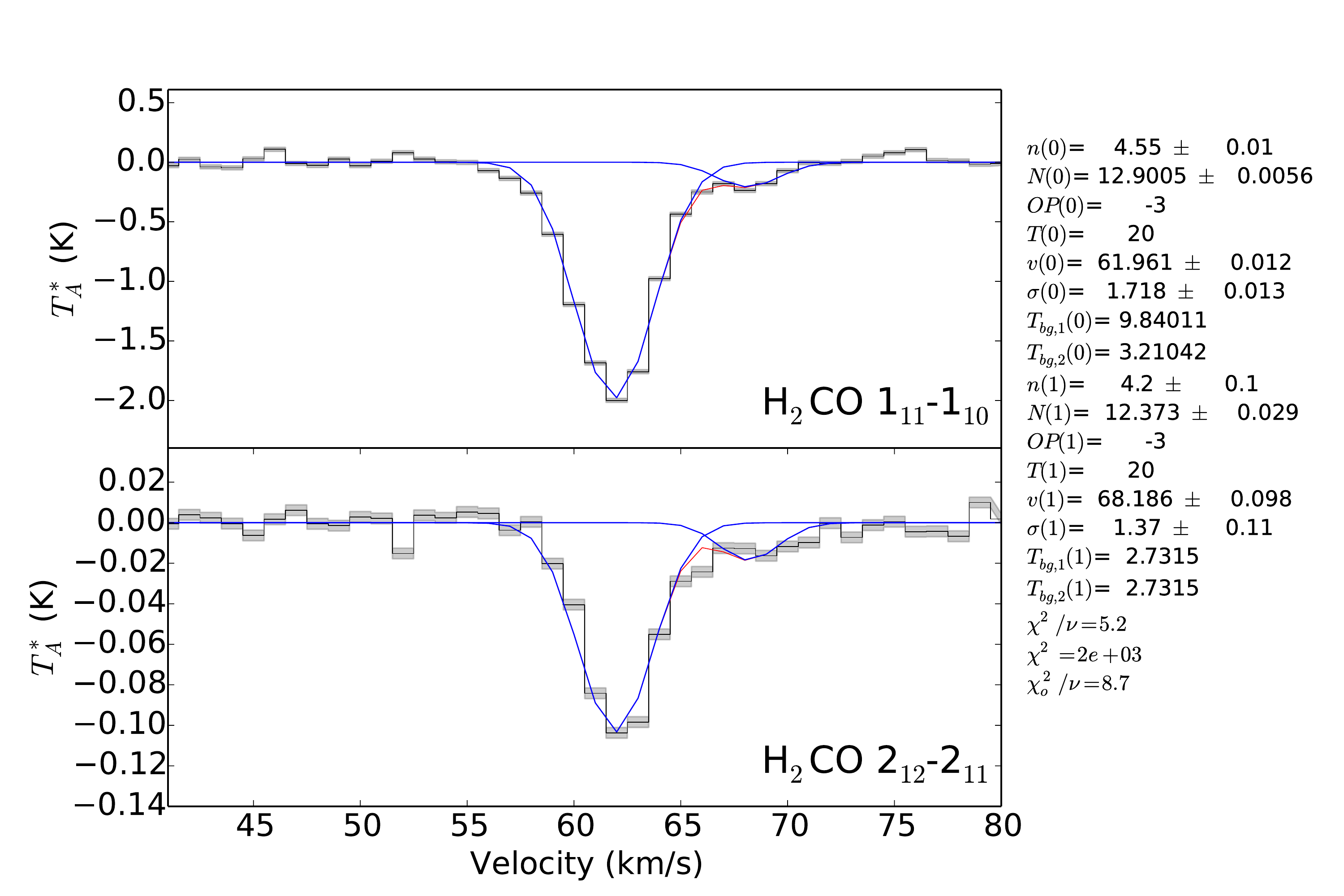}
{The spectrum extracted from G49.119-0.277 in a 55\arcsec radius aperture,
showing a model in which the continuum is \emph{behind} the 63 \kms component
but in front of the 68 \kms component.  The legend gives the fit parameters
along with $1\sigma$ error bars.  The parameters with no errors indicated
(OPR, $T$, $T_{BG}$) are assumed or independently measured values.}
{fig:contbetween63kms}{0.5}{0}

\subsection{G49.20-0.35 and G49.1-0.4}
\citet{Tian2013a} focus on the \hii regions G49.20-0.35 and G49.10-0.40 (called
G49.10-0.38 in their work) to determine the relative geometry
of the W51 C SNR and the W51 B \hii/star-forming region.  They observe that the
high-velocity \hi is not detected toward either of these sources, indicating
that the \hii regions must be behind the high-velocity \hi features.

We detect \formaldehyde \oneone at $\sim58$ and $\sim63$ \kms toward
G49.10-0.40, with line ratios that are consistent with the \hii region being
behind the molecular cloud complex.  It also has an extreme RRL velocity,
$v_{110\alpha} \approx 72~\kms$, the most redshifted seen in the entire W51 region
(see Figures \ref{fig:kinematics} and \ref{fig:hiirrlspec}).

G49.20-0.35 is also clearly behind the molecular cloud, as evidenced both by
\formaldehyde absorption depth and the IRDC absorption in the foreground.
It has an RRL velocity $v_{110\alpha} \approx 70$ \kms.

Because both \hii regions are extremely redshifted, they are most likely
associated with the W51 B cloud complex, contrary to the interpretation by
\citet{Tian2013a} in which they are unrelated background clouds.  The Galactic
rotation curve doesn't allow for velocities red of $\sim60$ \kms, and almost
none of the molecular gas exceeds $\sim70$ \kms even on the wings.  The \hii
regions are therefore probably shooting out the back side of the molecular
cloud in a `champagne flow,' perhaps accelerating ionized gas from the $\sim
68$ \kms component
further to the red.

\FigureTwoAA
{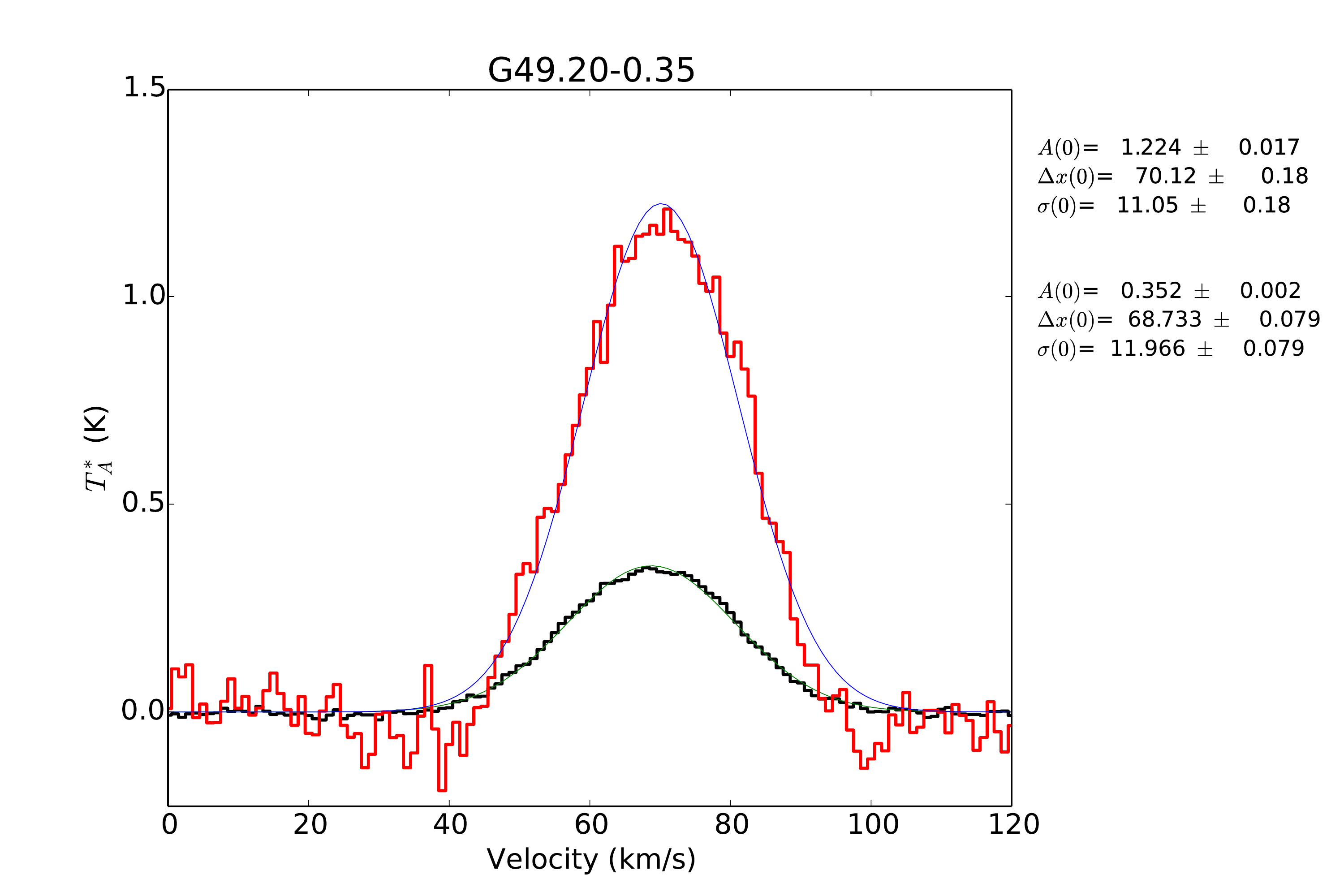}
{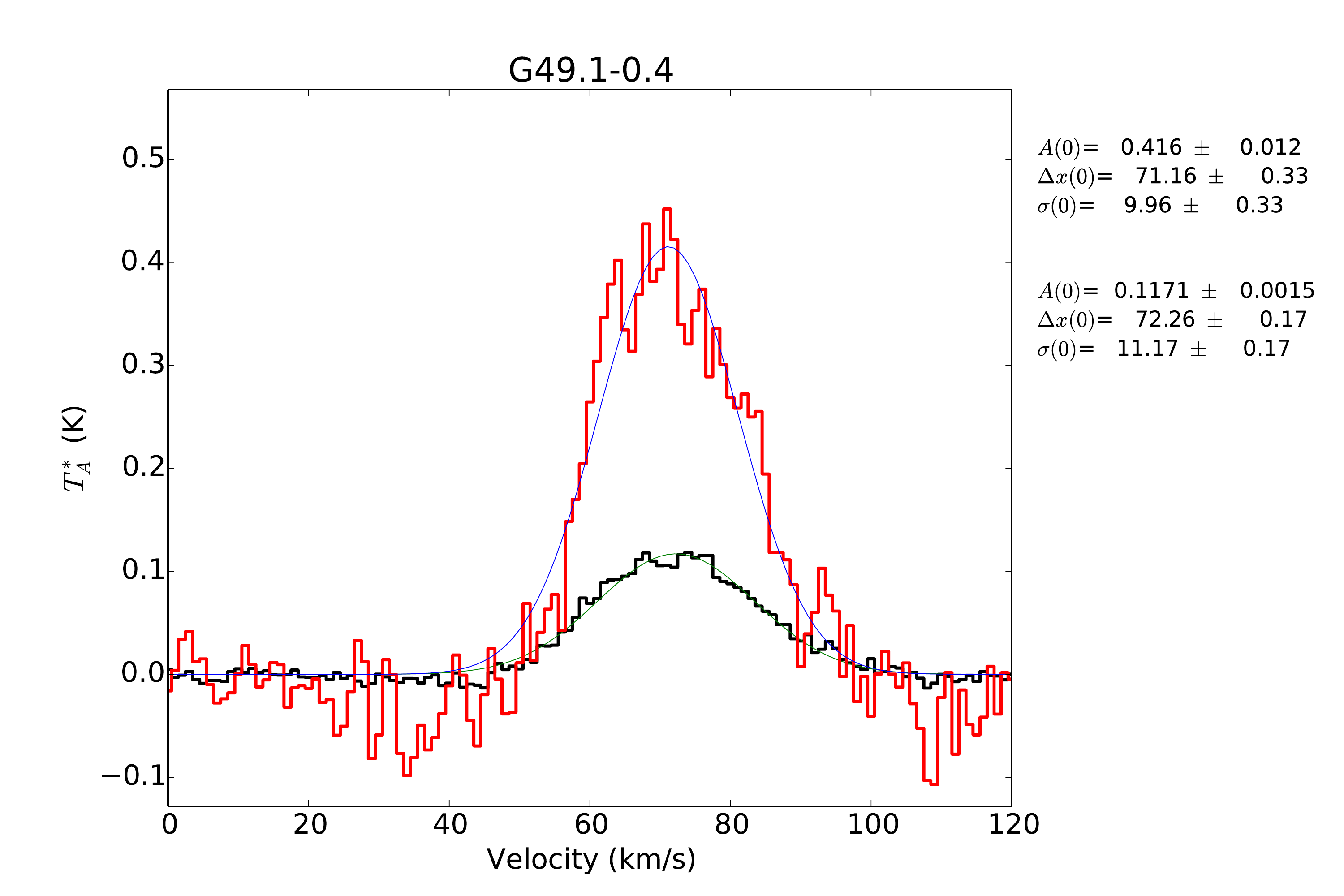}
{Fitted H110$\alpha$ (red) and H77$\alpha$ (black) spectra extracted from
55\arcsec apertures centered on G49.20-0.35 (left) and G49.1-0.4 (right).
The best-fit Gaussian parameters are shown in the legends, with the lower
legend corresponding to H77$\alpha$.}
{fig:hiirrlspec}{1}{6.5in}

\subsection{The 66 \kms 8 \um dark cloud}
Between W51 A and W51 B, there is a component of the 68 \kms cloud that is
filamentary and in the foreground of all of the free-free emission.
This cloud component is evident as an IRDC in the Spitzer GLIMPSE images from
$\ell=49.393$, $b=-0.357$ to $\ell=49.207$, $b=-0.338$; it is labeled as IRDC
G49.37-0.35 in Figures \ref{fig:geosketch} and \ref{fig:labeledzoom}.

The \hii region G49.20-0.35 is clearly behind the IRDC, though there are strong
morphological hints that it is interacting with and truncated by the cloud.

\subsection{G49.27-0.34}
The \uchii region G49.27-0.34, which was considered a candidate extended green
object (EGO) and subsequently rejected for lack of \hh emission
\citep{De-Buizer2010a,Lee2013a}, exhibits a second velocity component at $\sim68
\kms$, slightly but clearly redshifted of the rest of the IRDC.  The dust
component contains a gas mass $\sim2\ee{3} \msun$ based on the BGPS flux and
using the assumptions
outlined in \citet{Aguirre2011a}, suggesting that the high velocity could be
due to infall or virialized gas within a deep potential.  The virialized
velocity width, given the radius and mass from the BGPS data, is
$\sigma_{vir}=8.8 \kms$, while the measured \formaldehyde linewidth is
$FWHM(\formaldehyde) = 7.2 \kms$, wider than in any other part of the cloud
except W51 Main.

Both radio continuum and RRLs are detected toward this source.  The H77$\alpha$
RRL velocity is $\sim58$ \kms, significantly blueshifted from the molecular
gas.  The \formaldehyde lines do not independently distinguish between the
continuum source being in the front or back of the cloud, but the mean density
from the BGPS mass and radius $n\sim2.5\ee{4}$ \percc is within a factor of 2
of the \formaldehyde-derived density, $n\sim1.4\ee{4}$ \percc, if the continuum
source is behind the gas, while the \formaldehyde-derived density is too low,
$n\sim2\ee{3}$ \percc if the continuum source is in front.

The implied geometry therefore has the \hii region behind the molecular gas,
plowing toward it at a velocity difference $\Delta v \sim 10$ \kms.  Such a
high velocity difference may indicate that the \hii region is confined by the
molecular gas and on a plunging orbit into the cloud.

\subsection{G49.4-0.3f, aka G49.34-0.34, aka IRAS 19209+1418}
The \hii region centered at 49.34-0.34 was identified by \citet{Mehringer1994a}
as part of the G49.4-0.3 complex.  There are 3 distinct \formaldehyde line
components at 51, 63.70, and 68.47 \kms.  The 51 \kms component is behind the \hii
region; the \thirteenco line is detected at comparable brightness at 51 \kms
and 63 \kms, while the \formaldehyde \oneone line is $\sim10\times$ deeper at
63 \kms.  The RRLs associated with this source are at $v_{LSR}=58 \pm 1$ \kms.

The \formaldehyde lines are moderately well-fit by the two-velocity-component
model, but there is a relative excess of \twotwo absorption at 66 \kms
(associated with the 68 \kms component).  The extra absorption may indicate
that there is a high-density, low-column
component at this velocity.

The 8 \um GLIMPSE image shows that the 68 \kms IRDC crosses in front of this
source.  Herschel Hi-Gal 70\um images reveal a ring structure that is hinted at
in the 8 \um image.  There is no evidence for interaction between the ring
feature and the IRDC.  This intriguing feature will likely be difficult to
study in detail because the dusty, molecular gas feature lies in front of it.

\subsection{G49.4-0.3}
\label{sec:maus}
The collection of \hii regions around G49.4-0.3 vaguely resembles a cartoon
mouse.  As noted in \citet{Carpenter1998a}, the molecular gas in this region is
separated into two distinct components, one at 51 \kms and the other at 64
\kms.  The 64 \kms component is in the foreground, while the 51 \kms is in the
background of most of the \hii regions.  

Both cloud components are in the foreground of the central \hii regions at
G49.36-0.31, the `eyes' of the mouse.  The density of the 51 \kms component is
an order of magnitude higher than that in the 64 \kms component in this region,
suggesting that the gas is being compressed by the \hii region.
The clean separation between the 64 and 51 \kms cloud components suggests that
they are not interacting at this location.  

Based on the absorption line depths, the G49.38-0.30, IRAS 19207+1422, and
G49.37-0.30 \hii regions are behind the 51 \kms cloud.  The 8 \um absorption
features are associated with the 64 \kms cloud and are in front of all of the
\hii regions.

The 8 \um morphology of G49.42-0.31 is bubble-like, so it is plausible that the
\hii region is neither in front nor behind the 51 \kms cloud but embedded
within it, blowing a hole in the cloud.

\subsection{8\um Dark Cloud G49.47-0.27}
The cloud to the north of W51 Main/IRS2 appears as a dark feature in Spitzer
GLIMPSE 8 \um maps.  It is detected in \formaldehyde from 54 to 64
\kms.  Throughout, it has a high \oneone/\twotwo ratio, $\gtrsim7$ in
most voxels, indicating a low density $n\lesssim10^3$ \percc.

Centered at 60.6 \kms, the region has a line FWHM 5 - 7 \kms, indicating that
it is quite turbulent, with 3D Mach number in the range $10 < \mathcal{M} < 20$
for an assumed $10 < T < 20$ K.  At its centroid velocity, it is connected
to the W51 Main cloud.

There is a previously unreported bubble HII region in the north part of this
cloud, which we designate G49.47-0.26, with radius $\sim70\arcsec$ (1.7 pc).
The \hii has RRL velocities $v_{lsr}\approx50$ \kms.  Because it is not detected
in Brackett $\gamma$ emission \citep[from the UWISH2
survey:][]{Froebrich2011a}, it is most likely behind the cloud.

The \citet{Kang2009a} Spitzer survey of YSOs in the region indicates that there
are no YSOs within the boundaries of this cloud; it is very likely
non-star-forming at present (see also Figure \ref{fig:sfmassmap}).

Because the cloud is continuous with the W51 Main region in velocity and is
8\um-dark, it is most likely at the same distance as W51 Main and
associated indirectly with the massive cluster forming region.

\subsection{The 40 \kms clouds}
There are clouds observed at 40 \kms that show only weak \formaldehyde
absorption spread across nearly the entire region.  These molecular clouds are
behind nearly all of the \hii regions in the W51 complex.  There are additional
40 \kms clouds clearly seen in \hi absorption \citep{Stil2006a} that are not
associated with these molecular clouds, but instead represent a foreground
population of neutral atomic medium clouds.

\subsection{Kinematic Maps}
\label{sec:kinematics}
Maps showing the overall kinematics of the region are shown in Figures
\ref{fig:kinematics} and \ref{fig:h2cokinematics}.  Figure \ref{fig:kinematics}
shows the velocity at peak absorption of the \formaldehyde \oneone line and the
fitted radio recombination line centroid velocity.  Figure
\ref{fig:h2cokinematics} shows the best simultaneous fit to the \formaldehyde
\oneone and \twotwo absorption features over two different velocity ranges.
The \oneone absorption velocity in Figure \ref{fig:kinematics}a approximately
shows the velocity of the front-most molecular clouds along the line of sight
at each position.

\FigureTwoAA
{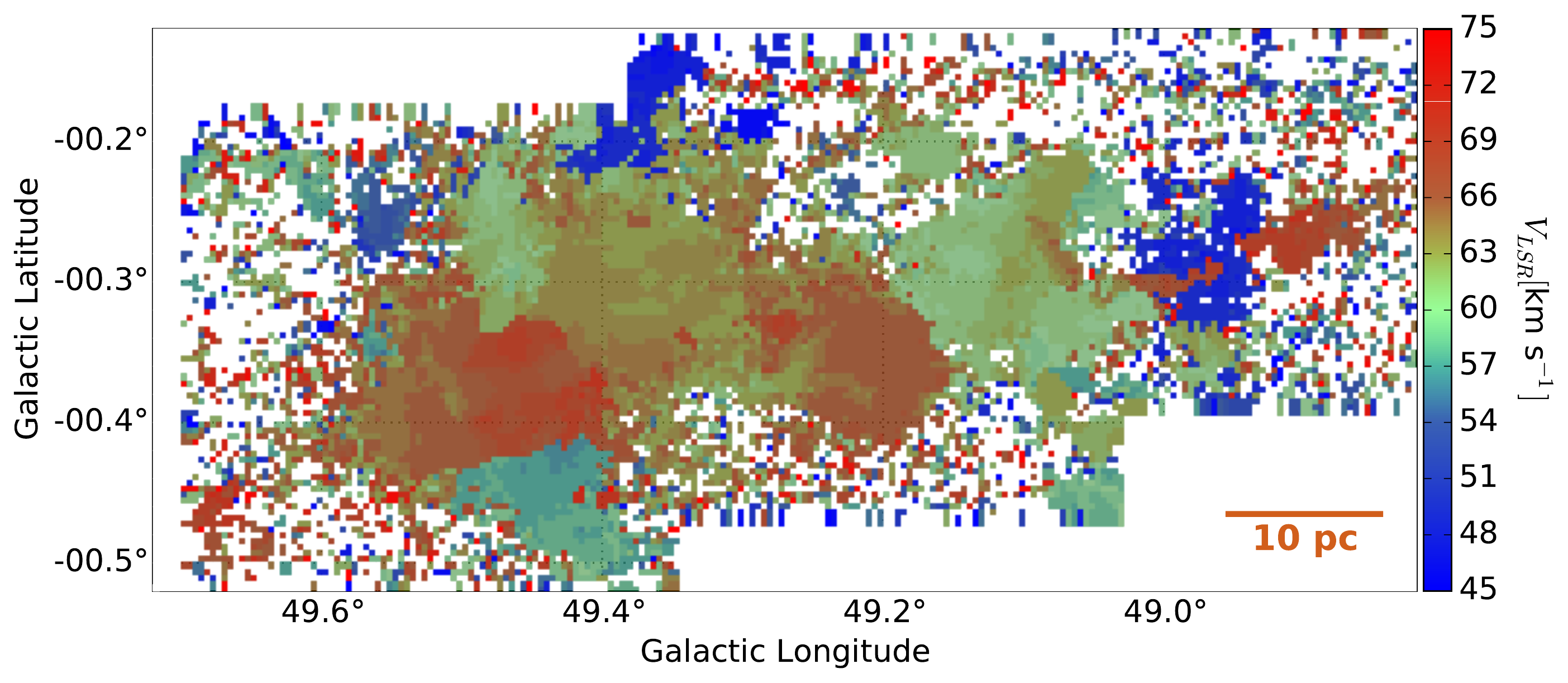}
{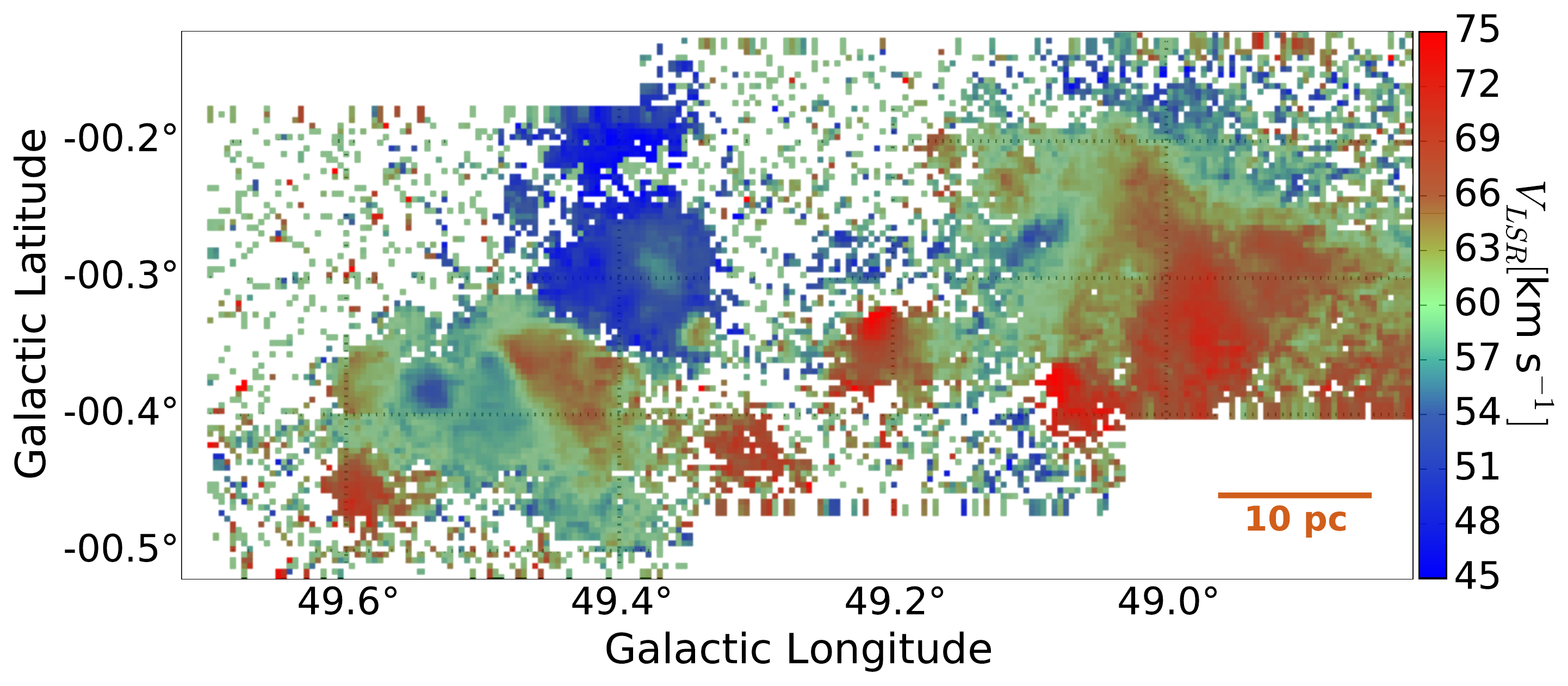}
{(a) Velocity of the peak \formaldehyde \oneone signal (deepest absorption) at
1 \kms resolution
(b) Velocity of the peak H110$\alpha$ emission as derived from Gaussian fits
to each spectrum.}
{fig:kinematics}{1}{6.5in}

\FigureTwoAA
{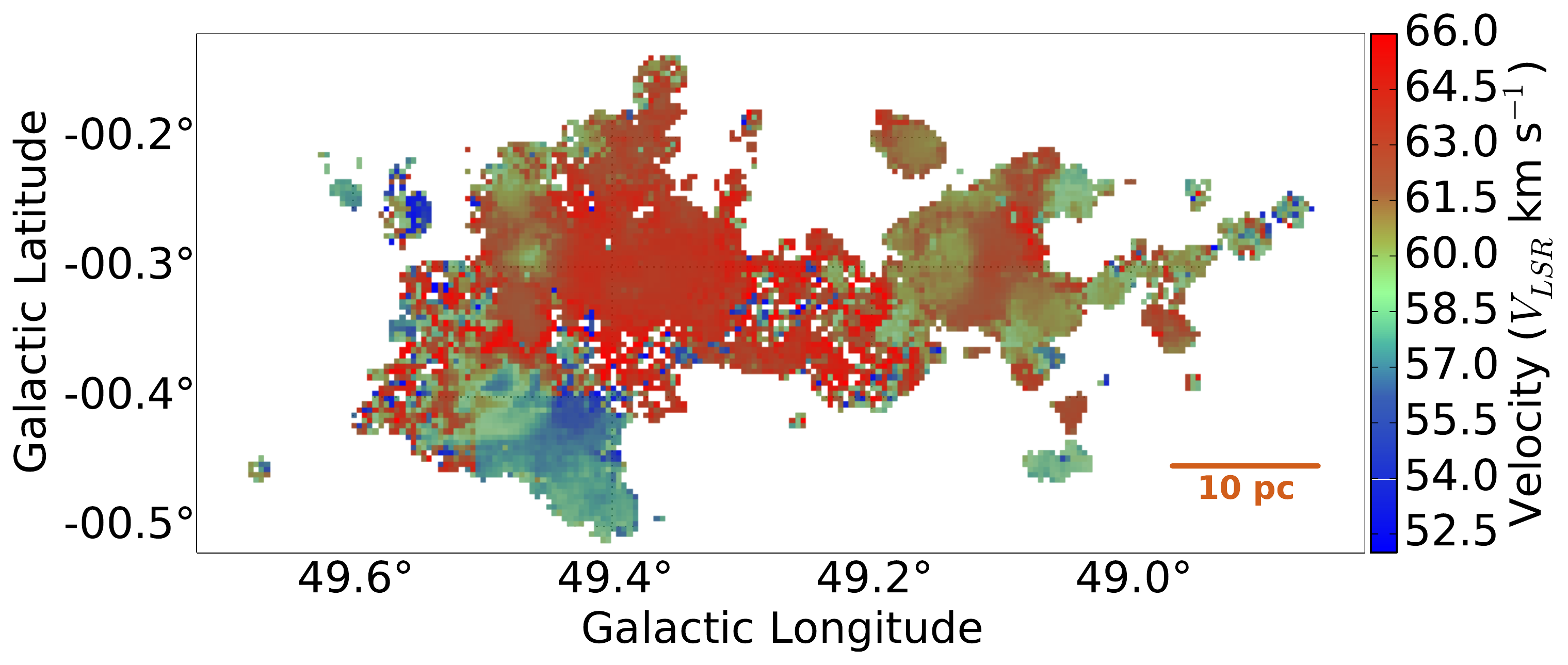}
{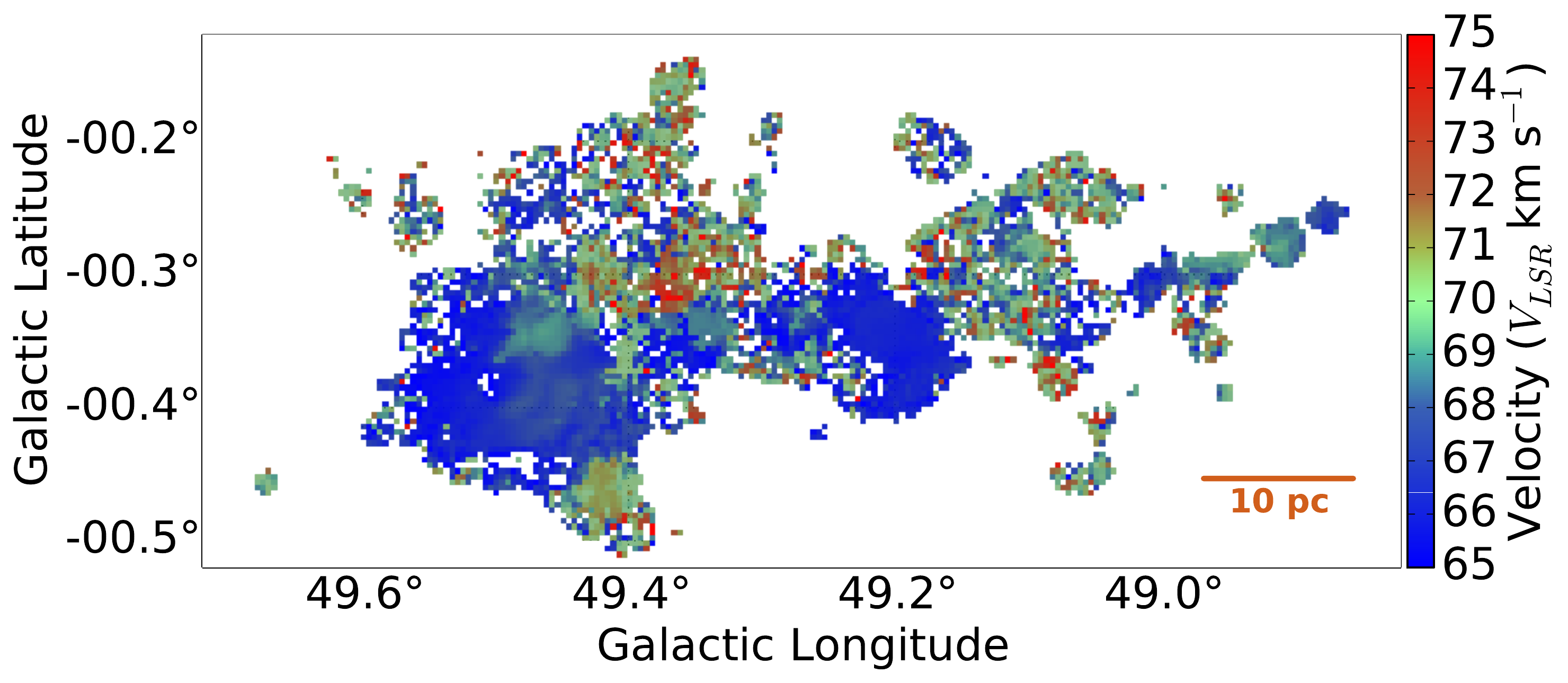}
{Maps of the fitted \formaldehyde velocity components over the range $40 <
v_{LSR} < 66$ \kms (left) and $66 < v_{LSR} < 75$ \kms (right).  The regions
that appear noisy have ambiguous multi-component decompositions.  }
{fig:h2cokinematics}{1}{6.5in}

\subsection{Abundances}
\label{sec:abundance}
The LVG modeling also yields measurements of abundance that are degenerate with
the assumed velocity gradient.  The abundance within the LVG model is defined
as 
\begin{equation}
    X = \frac{N(\ortho) [\persc / (\kms~\perpc)]}{n(\hh) [\percc]}
    \cdot \left(\frac{1~\kms~\perpc}{\mathrm{pc}}\right)
\end{equation}
We show the distribution of fitted abundances under this definition in Figure
\ref{fig:abundances}.  The histogram shows the average abundance along each
line of sight derived from the likelihood-weighted density divided by the
likelihood-weighted column.  We also show a two-gaussian fit to the 
abundance distribution: 30\% of the area is consistent with an abundance
$X(\ortho) = 10^{-8.4 \pm 0.5}$ and 70\% with $X(\ortho) = 10^{-9.9\pm0.6}$.
We caution that these abundance measurements are highly uncertain and are
contingent on both the backlighting source brightness and the assumed velocity
gradient.  The individual abundances going in to the histogram are generally
not well-constrained.  A more accurate abundance measurement could be obtained
by measuring the millimeter lines of \ortho at 140 and 150 GHz.

\Figure
{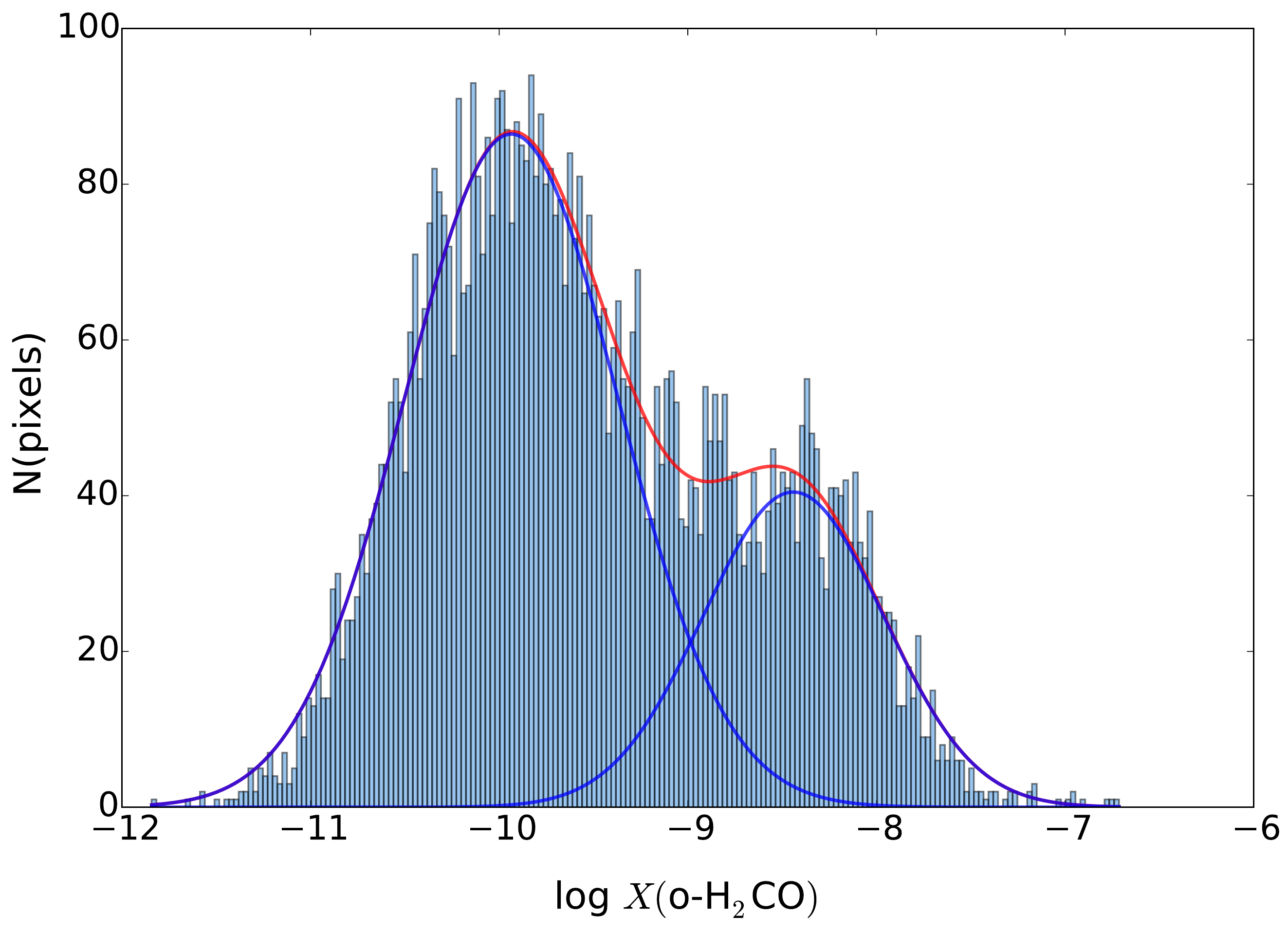}
{A histogram of the abundances derived for each spatial pixel using
the LVG grid fit.  The abundance shown assumes a velocity gradient 1 \kms
\perpc.  The overlaid fits show that 30\% of the area of W51 is consistent with
an abundance $X(\ortho) = 10^{-8.4 \pm 0.5}$ and 70\% with $X(\ortho) =
10^{-9.9\pm0.6}$.
}
{fig:abundances}{0.5}{0}

\subsection{Supplementary Picture}
We include a large-scale WISE + BGPS composite image of the entire W51 cloud
and its surroundings.  The figure highlights the W51 C supernova remnant.

\Figure{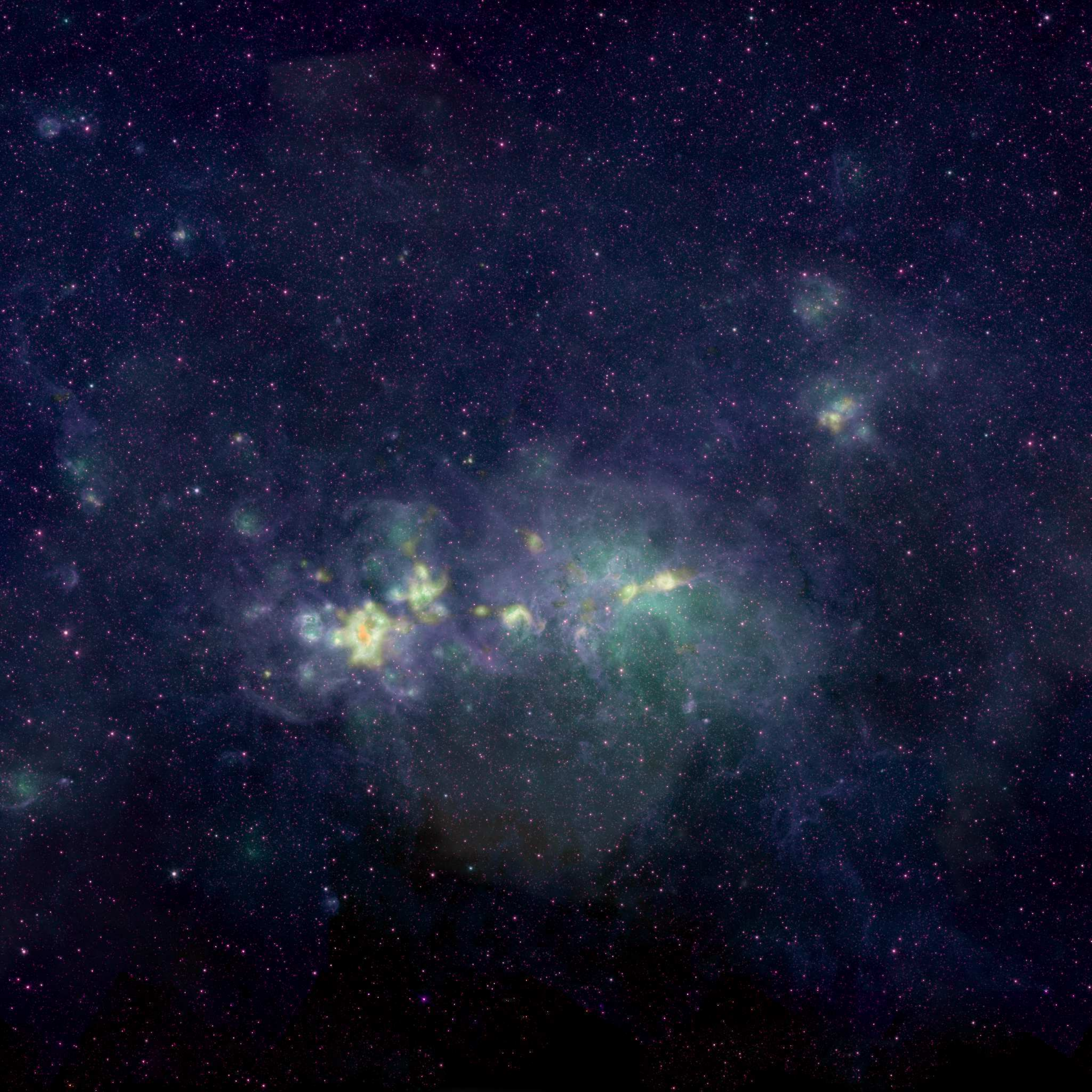}
{A large-scale color composite that highlights the W51 C supernova remnant as a
white haze using 90 cm data from \citet{Brogan2013a}.
The colors are the same as in Figure \ref{fig:w51large}:
The blue, green, and red colors are WISE bands 1, 3, and 4 (3.4, 12, and 22
\um) respectively.  The yellow-orange semitransparent layer is from the Bolocam
1.1 mm Galactic Plane Survey data \citep{Aguirre2011a,Ginsburg2013a}.
}{fig:w51huge}{0.2}{0}

\end{document}